\documentclass[final,onecolumn,titlepage]{jfm}

\usepackage{graphicx}
\usepackage{epstopdf,epsfig}
\usepackage{newtxtext}
\usepackage{newtxmath}
\usepackage{natbib}
\usepackage{hyperref}
\hypersetup{
    colorlinks = true,
    urlcolor   = blue,
    citecolor  = blue,
}

\newcommand{\RomanNumeralCaps}[1]
\linenumbers

\usepackage{nicefrac}
\usepackage{siunitx}
\usepackage{float}
\usepackage[caption=false]{subfig}
\renewcommand{\vec}[1]{\mbox{\boldmath$#1$}}

\title{The global flow state in a precessing cylinder}
\author{Andr\'e Giesecke\aff{1}
\corresp{\email{a.giesecke@hzdr.de}},
Tobias Vogt\aff{1},
Federico Pizzi\aff{2},
Vivaswat Kumar\aff{1},
Fernando Garcia Gonzalez\aff{2,3},
Thomas Gundrum\aff{1},
\and
Frank Stefani\aff{1}}

\affiliation{\aff{1}Institute of Fluid Dynamics, Helmholtz-Zentrum
  Dresden-Rossendorf, Bautzner Landstrasse 400, D-01328 Dresden, Germany
\aff{2}Departament de Mec{\`a}nica de Fluids, Universitat Polit{\`e}cnica de
  Catalunya-Barcelona Tech, Av. V{\'i}ctor Balaguer 1, 08800 Vilanova i la
  Geltr{\'u}, Barcelona, Spain}
\begin{document}
\maketitle
%\date{\today}

\begin{abstract}
We examine the fluid flow forced by precession of a rotating
cylindrical container using numerical simulations and experimental
flow measurements with ultrasonic Doppler velocimetry (UDV).  The
analysis is based on the decomposition of the flow field into
contributions with distinct azimuthal symmetry or analytically known
inertial modes and the corresponding calculation of their amplitudes.
We show that the predominant fraction of the kinetic energy of the
precession-driven fluid flow is contained only within a few
large-scale modes. 

The most striking observation shown by simulations and experiments is
the transition from a flow dominated by large-scale structures to a
more turbulent behaviour with the small-scale fluctuations becoming
increasingly important.  At a fixed rotation frequency (parametrized
by the Reynolds number, ${\rm{Re}}$) this transition occurs when a
critical precession ratio is exceeded and consists of a two-stage
collapse of the directly driven flow going along with a massive
modification of the azimuthal circulation (the zonal flow) and the
appearance of an axisymmetric double-roll mode limited to a narrow
range of precession ratios.  A similar behaviour is found in
experiments which make it possible to follow the transition up to
Reynolds numbers of ${\rm{Re}}\approx 2\times 10^6$.  We find that the
critical precession ratio decreases with rotation, initially showing a
particular scaling $\propto {\rm{Re}}^{-\nicefrac{1}{5}}$ but
developing an asymptotic behaviour for ${\rm{Re}}\gtrsim 10^5$ which
might be explained by the onset of turbulence in boundary layers.
\end{abstract}

%%%%%%%%%%%%%%%%%%%%%%%%%%%%%%%%%%%%%%%%%%%%%%%%%%%%%%%%%%%%%%%%%%%%%%%%%%%

%\begin{keywords}
%Geodynamo,Internal Waves,Quasi-geostrophic flows, Waves in rotating
%fluids, Transition to turbulence, Bifurcation, Chaos, Rotating
%turbulence, Shear Layer turbulence, Turbulent boundary layers,
%Turbulence simulation, Turbulent transition, Wave-turbulence
%interactions, Vortex breakdown, Vortex instability
%\end{keywords}

%%%%%%%%%%%%%%%%%%%%%%%%%%%%%%%%%%%%%%%%%%%%%%%%%%%%%%%%%%%%%%%%%%%%%%%%%%%

\section{Introduction}

A precessing body is a rotating object whose axis of rotation
undergoes a periodic change of its orientation.  In case of a
precessing container filled with a liquid, the force caused by
precession directly acts on the liquid, thus driving a complex
three-dimensional flow that is unstable and usually ends in a
turbulent state
\citep{mcewan1970,vanyo1971,vanyo1973b,vanyo1991,goto2007,goto2014,lebars2015,lebars2016,barker2016}.
Precessing flows are applied in technical devices because of their
efficiency in mixing aimed at the homogenization of viscous fluids
\citep{meunier2020}.  Mixing inside a liquid goes along with a
redistribution of the internal angular momentum and/or torque, which
in the case of freely precessing bodies can lead to surprising changes in
the orientation of the inertial axes so that, in turn, the stability of
trajectories of flying and rotating bodies with liquid payload can be
impacted \citep{vanyo1973a,rogers2013}.  On a larger scale, precession
also affects the stability of vortices in the atmosphere, for example
the formation and decay of hurricanes \citep{reasor2004} or the fluid
flow in the liquid interior of planets, moons or asteroids
\citep{lebars2016}.  Indeed, precessional forcing is a promising
additional source to explain the strong magnetic field of the ancient
moon three to four billion years ago with a field strength comparable
to that of the Earth's magnetic field today
\citep{dwyer2011,noir2013,weiss2014,tikoo2017,cebron2019,tikoo2022}.
Another example is the flow in the liquid core of the Earth, for which
the forcing is rather strong due to the rather fast precession
time scale of $\sim 26000 \mbox{ yrs}$ (compared with other planets in
the solar system) and the large angle between rotation axis and
precession axis ($\sim23.5^{\circ}$).  For many years it has been a
subject of (still ongoing) discussions as to whether precession would be
suitable to impact the electrically conducting fluid in the liquid
interior of the Earth in a way such that a large-scale magnetic field can
be generated \citep{stewartson1963,malkus1968,vanyo1991,lebars2015},
although it is generally assumed that the forcing mechanism for the
geodynamo is thermal or chemical convection.  In particular, for the
early geodynamo, which already functioned shortly after the formation
of the Earth \citep{tarduno2020}, precession could help to overcome
inconsistencies related to the available energy budget
\citep{olson2013}.  The question of precession-driven dynamos in
liquid planetary cores is intimately connected with the energy budget
which would be available for (self-)sustaining of the magnetic field
\citep{landeau2022}.  A rough estimate of the order of the flow
amplitude that is expected to be directly driven from precessional
forcing shows that the corresponding flow probably is too small to
sustain the Earth's magnetic field \citep{loper1975,rochester1975}.
However, in the turbulent case, it might be possible to maintain the
geomagnetic field, if the parameters involved (shear, viscosity) adopt
extreme values that lie at the edge of their assumed value range
\citep{landeau2022}.  Indeed, from experimental investigations of
precession-driven flows with a small precession angle, it is known
that the directly driven flow is unstable, giving rise to chaotic
small-scale flows as well as large-scale flow patterns different from
the structure of the volume force due to the precession
\citep{manasseh1992,manasseh1994,manasseh1996}, a phenomenon that has
been called resonant collapse \citep{mcewan1970}.  In this
case, precession acts as a kind of catalyst that allows a transfer of
kinetic energy from the rotational movement of the liquid into a
different flow pattern that may be capable of generating a magnetic
field via electromagnetic induction.

In order to examine the ability of a precession-driven flow to excite
and sustain a magnetohydrodynamic dynamo, an experimental facility is
under construction at Helmholtz-Zentrum Dresden-Rossendorf (HZDR),
which will be used to drive a flow of liquid sodium solely by
precession \citep{stefani2015,giesecke2015a,giesecke2018}.  In this
experiment, named DRESDYN, the forcing of a fluid flow by precession
will be realized in a cylindrical container, similar to the smaller
experiment conducted by \citet{gans1970,gans1971} in the 1970s, where
a threefold amplification of an external magnetic field was found,
indicating that dynamo action can be expected in the vicinity of the
transition from a laminar flow state to vigorous turbulence if the
system is sufficiently large.  This was recently supported by applying
a combination of hydrodynamic experiments and numerical simulations
with a kinematic dynamo model \citep{giesecke2018,giesecke2019}. In
these studies, it was shown that, in the planned precession experiment,
dynamo action is best possible in a parameter regime where the flow
structure is determined by a combination of the non-axisymmetric
directly forced flow and an axisymmetric double roll, which emerges in
the transitional regime of the global flow state mentioned above.
This transition goes along with a significant generation of a mean
azimuthal circulation (zonal flow), which is known from earlier
experiments \citep{kobine1995,kobine1996}, and recent simulations
indicate that this is a feature specific for the large-angle
precession \citep{lopez2016}.

The present study takes up this preliminary work and aims at a
comprehensive classification of the flow state of a precession-driven
flow in a cylindrical cavity.  We are conservative regarding a
classification as ``laminar'' or ``turbulent'', because in all cases
large-scale components dominate the flow and in particular the
Reynolds number that is achievable in the simulations is too low
to associate the flow state with developed turbulence.
Therefore we label the individual regimes as subcritical and
supercritical states with a transitional regime separating the two
states (comparable to the classification of \citet{pizzi2022} applying
a low-state and a high-state). We focus on the series of changes that
take place at the transition regime between subcritical and
supercritical state, which represents the most striking phenomenon in
the case of a large precession angle.

The outline of the study is as follows: In
Section~\ref{sec::02_equations_and_setup} we list the basic equations
and explain the set-up of our model.  Our analyses are continued in
Section~\ref{sec::03_evolution_forcing}, where we characterize the
spatial structure of the flow and the evolution of the kinetic energy
when the forcing due to precession is increased.  In
Section~\ref{sec::nonlin_amp} we compare the amplitude of the directly
forced flow with the result of the nonlinear model recently presented
by \citet{gao2021}.  The extension of the expected flow behaviour to
more extreme parameters that cannot be achieved in numerical
simulations is carried out in
Section~\ref{sec::06_scaling_with_rotation} by means of experimental
observations at a down-scaled water experiment hosted at HZDR.
Finally, we summarize and conclude our results in
section~\ref{sec::conclusions}.

%%%%%%%%%%%%%%%%%%%%%%%%%%%%%%%%%%%%%%%%%%%%%%%%%%%%%%%%%%%%%%%%%%%%%%%%%%%%%%%

\section{Problem set-up and base state}
\label{sec::02_equations_and_setup}

\begin{figure}
   \centerline{\includegraphics[width=0.66\textwidth]{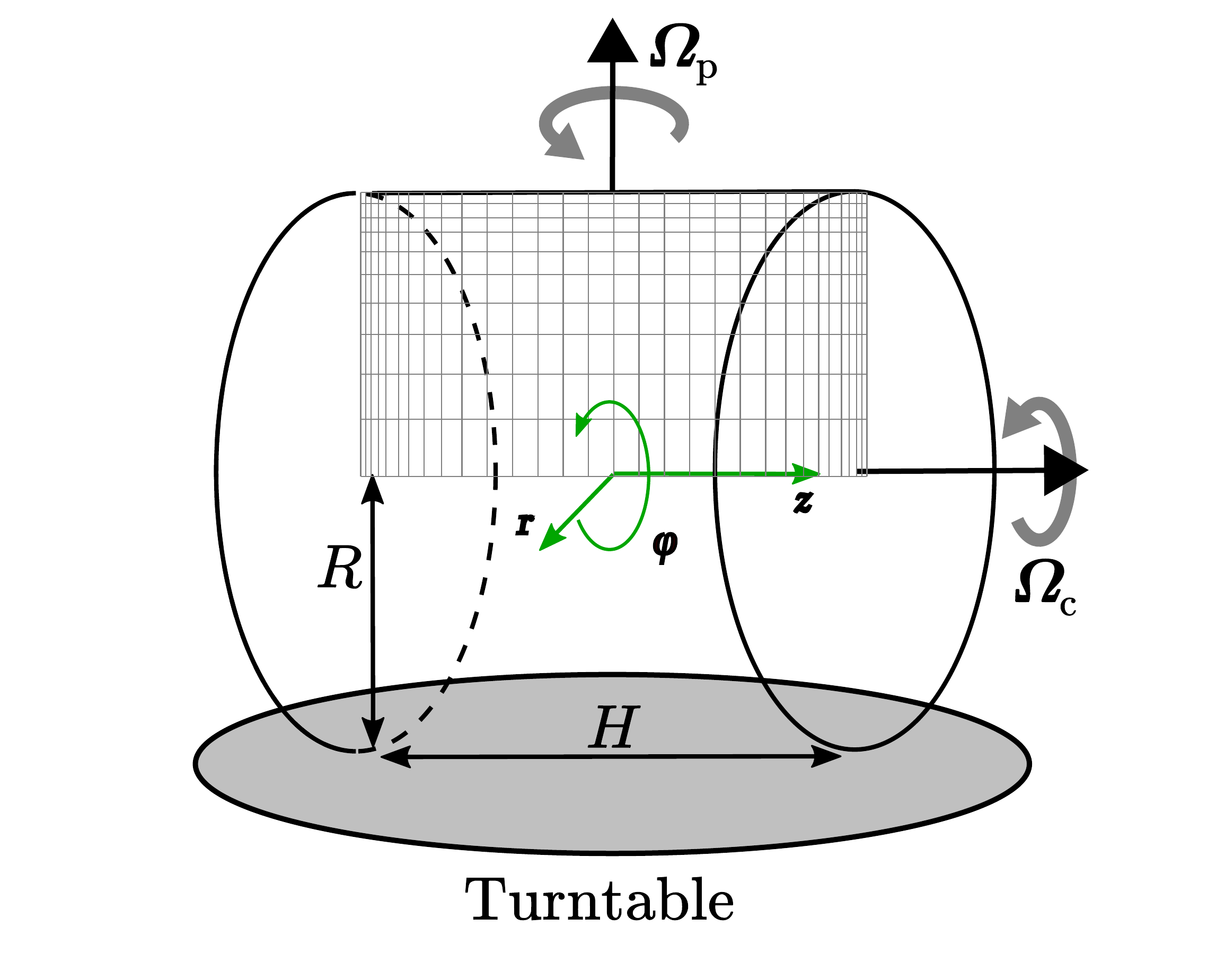}}
   \caption{
     \label{fig::01_sketch_numgrid}
     \raggedright 
     Sketch of the cylindrical domain subject to precessional
     forcing with the illustration of the direction of rotation and
     precession axes.
     The grey grid in the meridional plane shows the
     distribution of the spectral elements utilized in the numerical simulations
     with the code {\tt{SEMTEX}}.}
\end{figure}

The general set-up used for the description of a precessing flow in a
cylindrical cavity is sketched in
figure~\ref{fig::01_sketch_numgrid}. The system is determined by the
height $H$ and the radius $R$ of the cylinder. The direction of the
motion of the container is described by the orientation of two
rotation axes, which are given by the unit vectors
${\vec{k}}_{\rm{p}}$ for the precession and ${\hat{\vec{z}}}$
for the rotation.  The corresponding angular velocities are then given by
$\vec{\varOmega}_{\rm{c}}=\varOmega_{\rm{c}}\hat{\vec{z}}$
due to the rotation of the container and
$\vec{\varOmega}_{\rm{p}}=\varOmega_{\rm{p}}{\vec{k}}_{\rm{p}}$ due to
the precession of the rotation axis.  The nutation angle $\alpha$
describes the relative orientation of $\vec{\varOmega}_{c}$ and
$\vec{\varOmega}_{\rm{p}}$ and is defined by
$\cos{\alpha}=\vec{\varOmega}_{\rm{c}} \cdot
\vec{\varOmega}_{\rm{p}}/(\varOmega_{\rm{c}}\varOmega_{\rm{p}})$.  In
the present study we always assume $\alpha$ to be fixed at
$90^{\circ}$.  We further consider a fixed geometry with the aspect
ratio $\Gamma=H/R=2$.
In the following, we use the time scale $\varOmega_{\rm{c}}^{-1}$ and the length
scale $R$ so that ${\vec{u}}$ denotes the velocity in units of the
rotation velocity at the outer boundary of the cylinder
($\varOmega_{\rm{c}} R$) and ${P}$ denotes the scaled pressure in
units of $\varOmega_{\rm{c}}^2R^2$.

In our analysis we refer to two different frames of reference.  In the
precession frame of reference the observer follows the rotational
motion due to the precession ($\vec{\varOmega}_{\rm{p}}$) and looks at
the spinning cylinder ($\vec{\varOmega}_{\rm{c}}$).  Thus the observer
rotates around ${\vec{k}}_{\rm{p}}$ at $\varOmega_{\rm{p}}$ whereas
the cylinder axis (rotating along $\hat{\vec{z}}$) is fixed in this
frame of reference.
The fluid flow $\vec{u}$ is described by a Navier-Stokes equation that
simply reads
\begin{equation}
  \frac{\partial\vec{u}}{\partial t} + \vec{u} \cdot \vec{\nabla}
  \vec{u}  +  2{\rm{Po}}\vec{k}_{\rm{p}} \times \vec{u}
  =-\vec{\nabla} P + \frac{1}{\rm{Re}} \nabla^2\vec{u},\label{eq::navsto_precframe} 
\end{equation}
\noindent
where we introduce the Reynolds number,
${\rm{Re}}=\varOmega_{\rm{c}}R^2/\nu$, defined with the rotation
velocity of the outer wall, the kinematic viscosity $\nu$ and the
Poincar{\'e} number ${\rm{Po}}=\varOmega_{\rm{p}}/\varOmega_{\rm{c}}$.
The boundary conditions applied in our models reflect the rotation of
the container,
i.e. $\vec{u}_{\rm{bc}}={\vec{\varOmega}}_{\rm{c}}\times\vec{r}$.
Explicitly, this gives $u_{r,z}=0$ at $R=1$ and $z=\pm H/2=\pm 1$ and
$u_{\varphi}=\Omega_{\rm{c}}r$ at $z=\pm 1$.  Since it allows a simple
numerical implementation which only requires appropriate boundary
conditions for the description of the rotation of the cylinder, all
numerical simulations are carried out in this reference system.  This
has the additional advantage that the flow structure, which in this
reference frame is essentially dominated by a standing wave that goes
along with a stationary geometric structure of the flow, allows a
simple calculation of a mean flow.

In order to compute the time evolution of a precession-driven flow in
a cylindrical container we use the code {\tt{SEMTEX}}, which provides
numerical solutions of the incompressible Navier-Stokes equations
applying a coupled continuous Galerkin nodal spectral element Fourier
spatial discretization with semi-implicit temporal integration via a
time-splitting scheme in Cartesian as well as in cylindrical
coordinates.  The algorithm is described in detail in
\citet{blackburn2004} and \citet{blackburn2019}, including various test
problems that demonstrate the robustness and accuracy of the
scheme. Precessional forcing has been analysed with {\tt{SEMTEX}} and
shows a good agreement between flow data from simulations and
experiments
\citep{albrecht2015b,albrecht2015a,albrecht2016,giesecke2018,giesecke2019}.
The code uses a standard Fourier decomposition in the azimuthal
direction ($\varphi$) and quadrilateral spectral elements with
standard nodal Gauss-Lobatto-Legendre basis functions in the
meridional plane $r,z$.  Within a spectral element, the solution is
approximated by a polynomial of degree eight.  The maximum Reynolds
number applied in our simulations is ${\rm{Re}} = 10^4$.  For larger
values it is not possible to resolve the boundary layers while still
maintaining a reasonable computation time on standard HPC systems.  We
use a non-uniform grid with increasing resolution in the vicinity of
the boundaries so that Ekman and Stewartson layers are properly
resolved (see figure~\ref{fig::01_sketch_numgrid}).  All simulations
start with a solid body rotation profile
$\vec{u}(t=0)=\varOmega_{\rm{c}} r {\vec{e}}_{\varphi}$ with the
impact of precession being abruptly switched on at $t=0$.

\begin{figure}
\centerline{\includegraphics[width=0.4\textwidth]{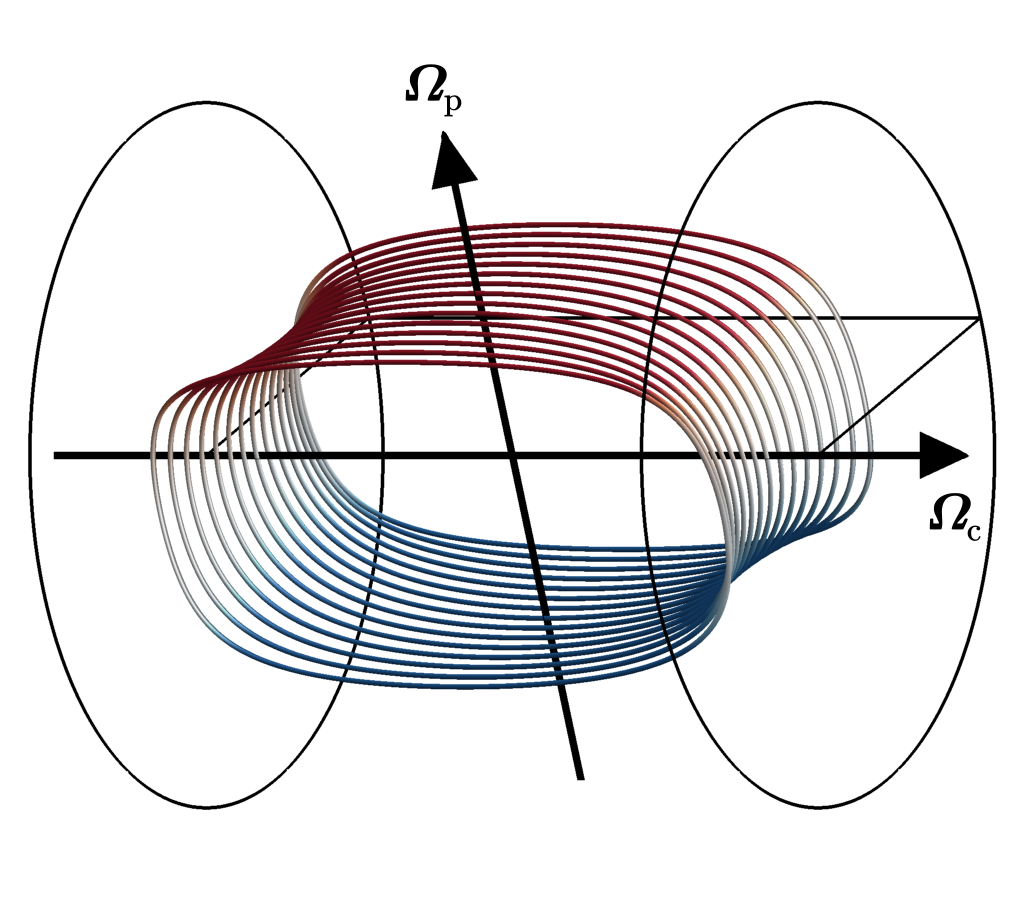}}
\caption{\label{fig::directforcedmode}
\raggedright 
Streamlines of the directly
forced flow with the simplest possible axial structure. This mode
is an inertial wave that is standing in the precession frame of
reference.}
\end{figure}

For small precession ratios in the precession frame, the flow is
predominantly co-rotating with the wall so that $u \sim
\mathcal{O}(1)$.  However, the usual disadvantage of the higher
velocities in this reference frame, which is associated with smaller
time steps in the numerical solution of the equations, turns out to be
much less serious, since the typical flow velocities reach the same
order of magnitude in both the precession frame of reference and the
reference frame that is co-rotating with the container wall.  In this
reference system the precession axis is no longer stationary and the
evolution of the flow is described by the incompressible Navier-Stokes
equation, including the Coriolis force and a time-dependent volume
force, the Poincar{\'e} force caused by the perpetual acceleration due
to the periodic change of the orientation of the rotation axis. The
equation reads \citep{tilgner1998}
\begin{equation}
  \frac{\partial}{\partial t}\,{\vec{u}}
  +{\vec{u}}\cdot{\nabla}\vec{{u}}=  
  -{\nabla}{P}
  \underbrace{-2({\rm{Po}}\vec{k}_{\rm{p}}+\hat{\vec{z}})\times{\vec{u}}}_{\mbox{Coriolis force}} 
  \underbrace{-{\rm{Po}}(\vec{k}_{\rm{p}}\times\hat{\vec{z}}) 
    \times\vec{r}}_{\mbox{Poincar{\'e} force } \vec{F}_{\!\!\rm{p}}}
  +\displaystyle\frac{1}{\rm{Re}}{\nabla}^2{\vec{u}},
  \label{eq::navier_scaled}
\end{equation}
subject to no-slip boundary conditions $\vec{u}_{\rm{bc}}=0$ at all
boundaries.  The precession vector $\vec{k}_{\rm{p}}(t)$ is now
time dependent and executes a retrograde gyroscopic motion which
causes the additional forcing term that describes the Poincar{\'e}
force on the right-hand side of Eq.~(\ref{eq::navier_scaled}).  The
dynamically relevant part of the Poincar{\'e} force is
$\vec{F}_{\rm{p}}=
-{\rm{Po}}\left(\vec{k}_{\rm{p}}(t)\times\hat{\vec{z}}\right)\times\vec{r}=
-r{\rm{Po}} \sin\alpha \cos(t+\varphi)\hat{{\vec{z}}}$, which is
responsible for the primary flow with an amplitude characterized by
the Poincar{\'e} number
${\rm{Po}}=\varOmega_{\rm{p}}/\varOmega_{\rm{c}}$.  The structure of
the forcing is antisymmetric with respect to the equatorial plane and
proportional to $\cos\varphi$ in the azimuthal direction, which is
immediately transferred to the geometric structure of the directly
driven flow so that only a direct forcing of inertial modes with
azimuthal wavenumber $m=1$ and with an odd axial wavenumber $k$ is
possible.  With an aspect ratio $\Gamma=H/R=2$ the simplest case
(i.e. with the smallest possible wavenumbers) results in a
characteristic flow pattern given by an axial recirculation, as shown
in Figure~\ref{fig::directforcedmode}.

In the co-rotating frame of reference the directly forced flow
constitutes an inertial wave that rotates opposite to the rotation of
the container with the frequency of the cylinder, whereas in the
precession frame of reference these modes are standing waves since the
forcing has the frequency of the rotation of the cylinder.  Higher
inertial modes arise due to instabilities \citep{kerswell1999} or
nonlinear (self-)interactions of the primary flow in the bulk and in
the boundary layers \citep{meunier2008}.  However, according to
\citet{greenspan1969}, first-order nonlinear interactions are
forbidden on the long time scale with an order given by the strength
of the nonlinear terms.  Therefore, it is usually assumed that all
higher modes are predominantly generated due to interactions (linear
and nonlinear) in the boundary layers.  The breaking of the axial
symmetry, i.e. the emergence of inertial modes with even axial
wavenumber was emphasized by \citet{tilgner2005}, who found that the 
dynamo effect was more efficient (i.e. started at smaller magnetic
Reynolds number or led to larger magnetic energies) if the axial
symmetry of the flow was broken.
%
%%%%%%%%%%%%%%%%%%%%%%%%%%%%%%%%%%%%%%%%%%%%%%%%%%%%%%%%%%%%%%%%%%%%%%%%

\section{Dependence on the forcing by precession}
\label{sec::03_evolution_forcing}

\subsection{Flow structure}

Previously, we have described the base flow in a precessing container
as a forced standing inertial wave superimposed on the solid body
rotation caused by the spinning container.  In a complementary
approach, \citet{busse1968} assumed that the fluid flow in a precessing
body obeys a uniform vorticity solution (as originally proposed by
\citet{poincare1910}) as the zeroth-order solution, and a balance of
torques is responsible for maintaining a steady flow.  The
corresponding flow is a rotational motion around an axis which 
coincides neither with the rotation axis of the container nor with the
precession axis.  Physically, the existence of the distinguished
rotation axis of the fluid is the result of a perpetual spin-up
process that reflects the effort of the fluid to align the rotational
motion due to the rotation of the container with the motion due to the
precession, whose axis direction is permanently changing.  The
explicit calculation in a precessing spheroidal cavity includes the
saturation due to the formation of boundary layers and yields an
implicit equation for the fluid rotation axis $\vec{\omega}_{\rm{f}}$
(see also \citet{noir2003} for a calculation based on the balance of
torques and the recent results from \citet{kida2020,kida2021} that
include higher-order terms).  The solution of Busse has the
interesting property that, for fixed forcing and sufficiently large
ellipticity, bistability exists in terms of two stable solutions with
a possible abrupt transition between them. The transition features a
hysteresis, which means that, when increasing the forcing, the
transition to the supercritical state occurs at a critical
Poincar{\'e} number ${\rm{Po}}^{\rm{c}}_{1}$ which is larger than the
value ${\rm{Po}}^{\rm{c}}_{2}$ marking the transition from the
supercritical state to the subcritical state when reducing the
forcing.  The theory has been confirmed in numerics
\citep{tilgner1999,tilgner2001} and experiments \citep{noir2003}, and
more recently it was adopted to the case of a precessing ellipsoid
\citep{buhrmann2022}.  However, Busse's theory cannot be
straightforwardly applied to the case of the cylinder, because of the
presence of corners at the end caps of the cylinder, which perform a
rotational motion due to the precession and prevent the realization of
a simple torque balance so that a uniform vorticity solution is not a
good representation for a base state in a cylindrical geometry.  This is
also reflected, for example, in the emergence of wave beams
originating in the corners or the emergence of turbulent injections
from the sharp shear layer close to the sidewalls
\citep{marques2015}.

\begin{figure}
\subfloat[][]{\includegraphics[width=0.3\textwidth]{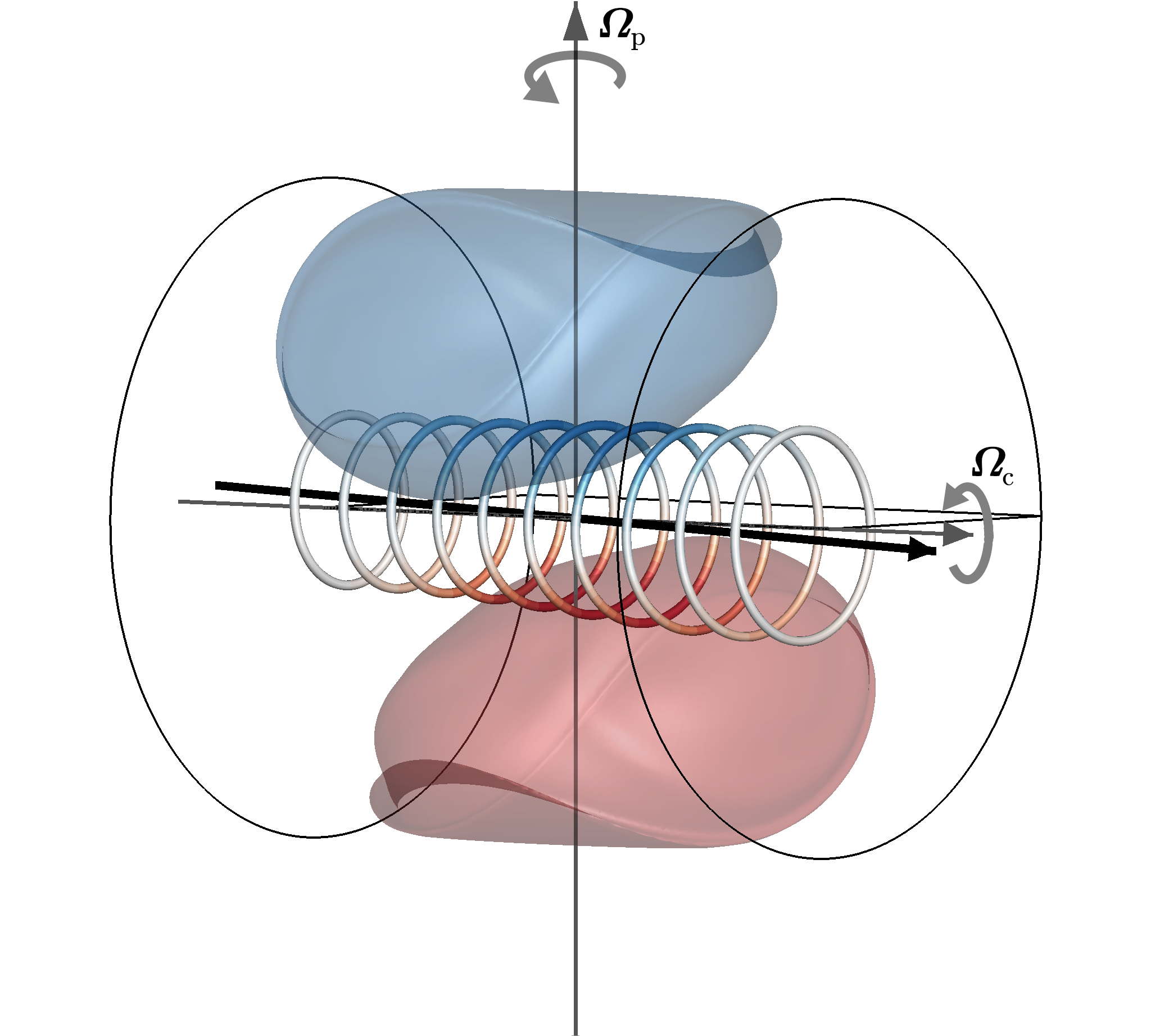}\label{fig::fluid_rotaxis_a}}
\subfloat[][]{\includegraphics[width=0.3\textwidth]{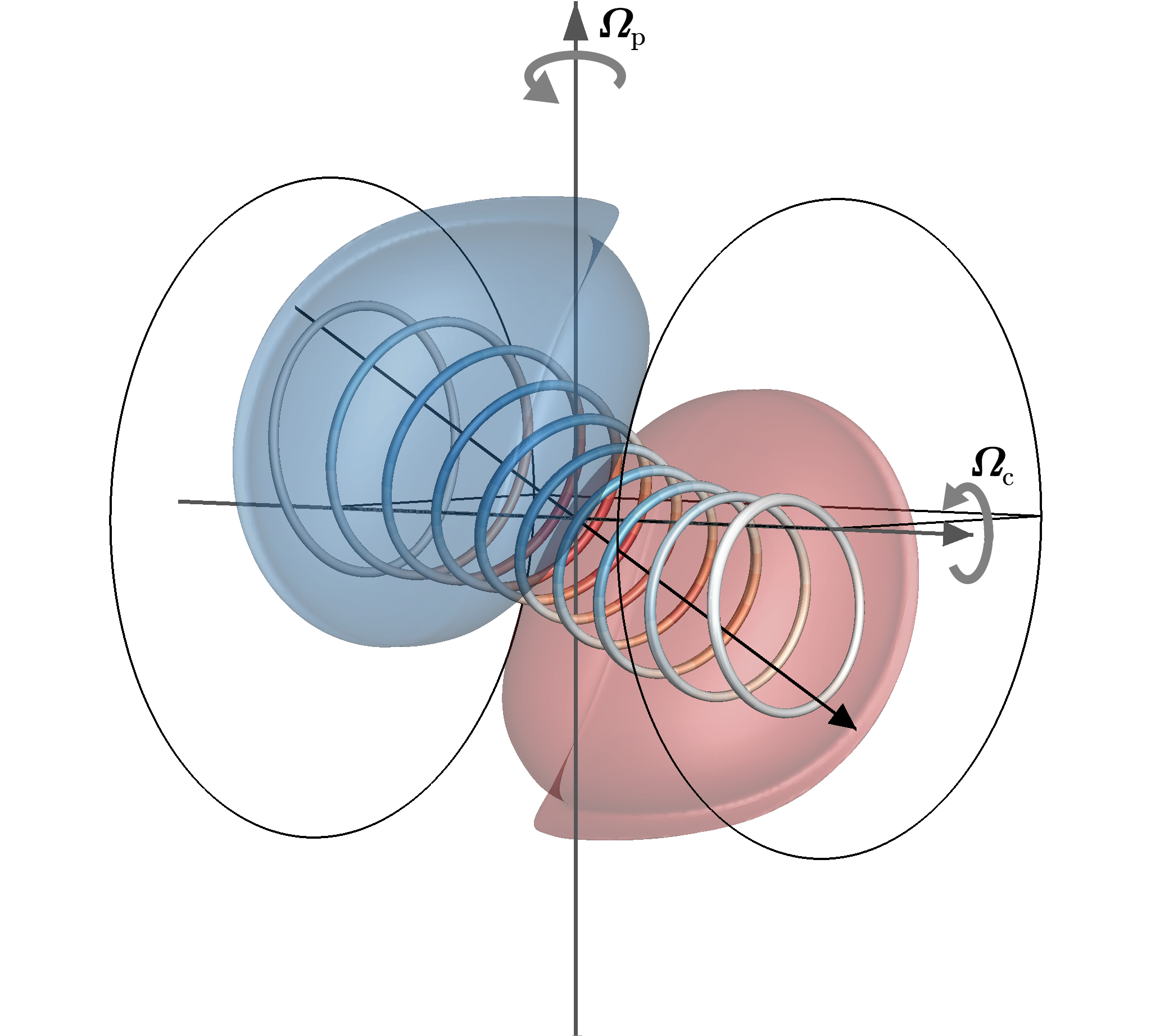}\label{fig::fluid_rotaxis_b}}
\subfloat[][]{\includegraphics[width=0.3\textwidth]{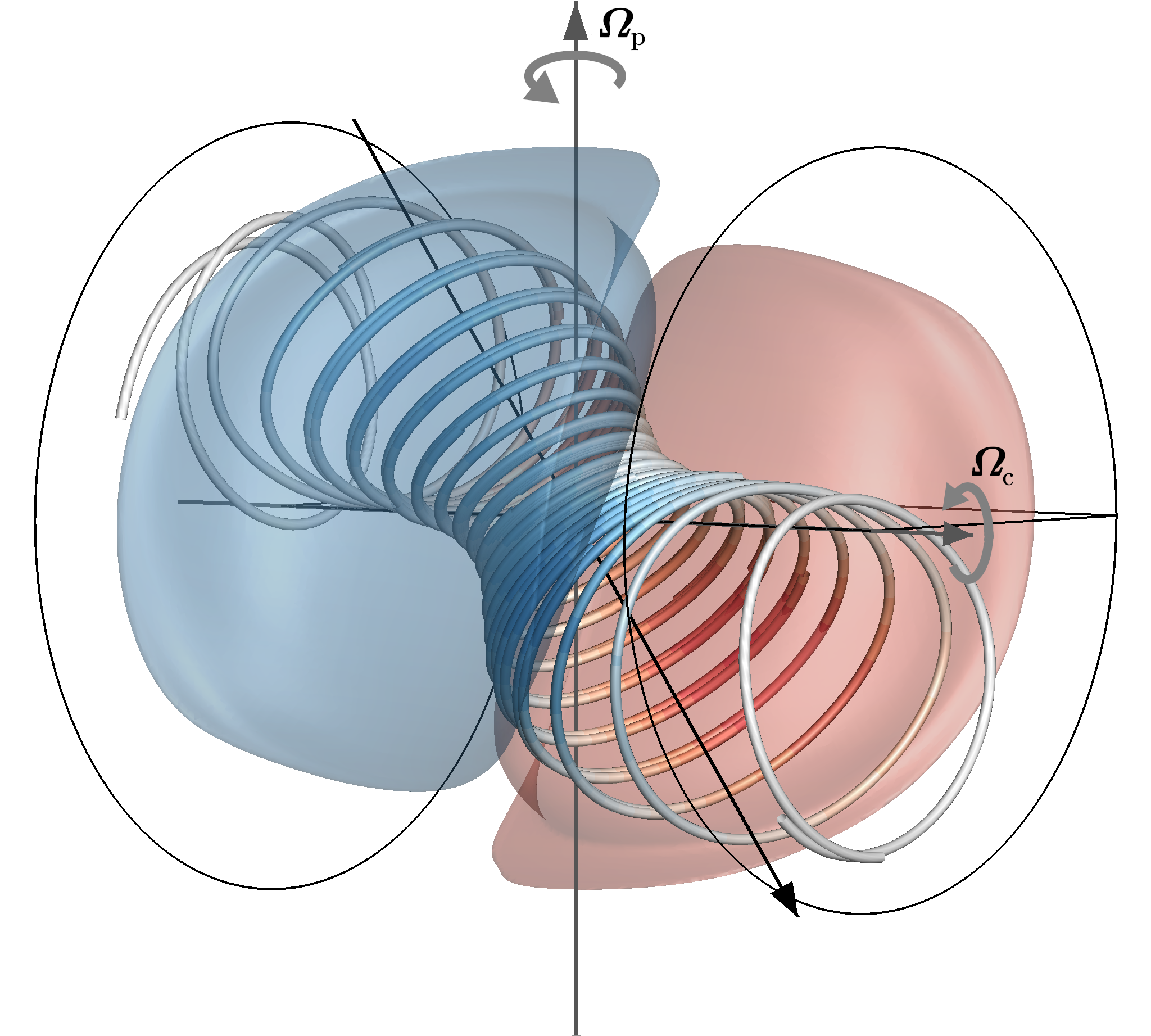}\label{fig::fluid_rotaxis_c}}
\\
\subfloat[][]{\includegraphics[width=0.3\textwidth]{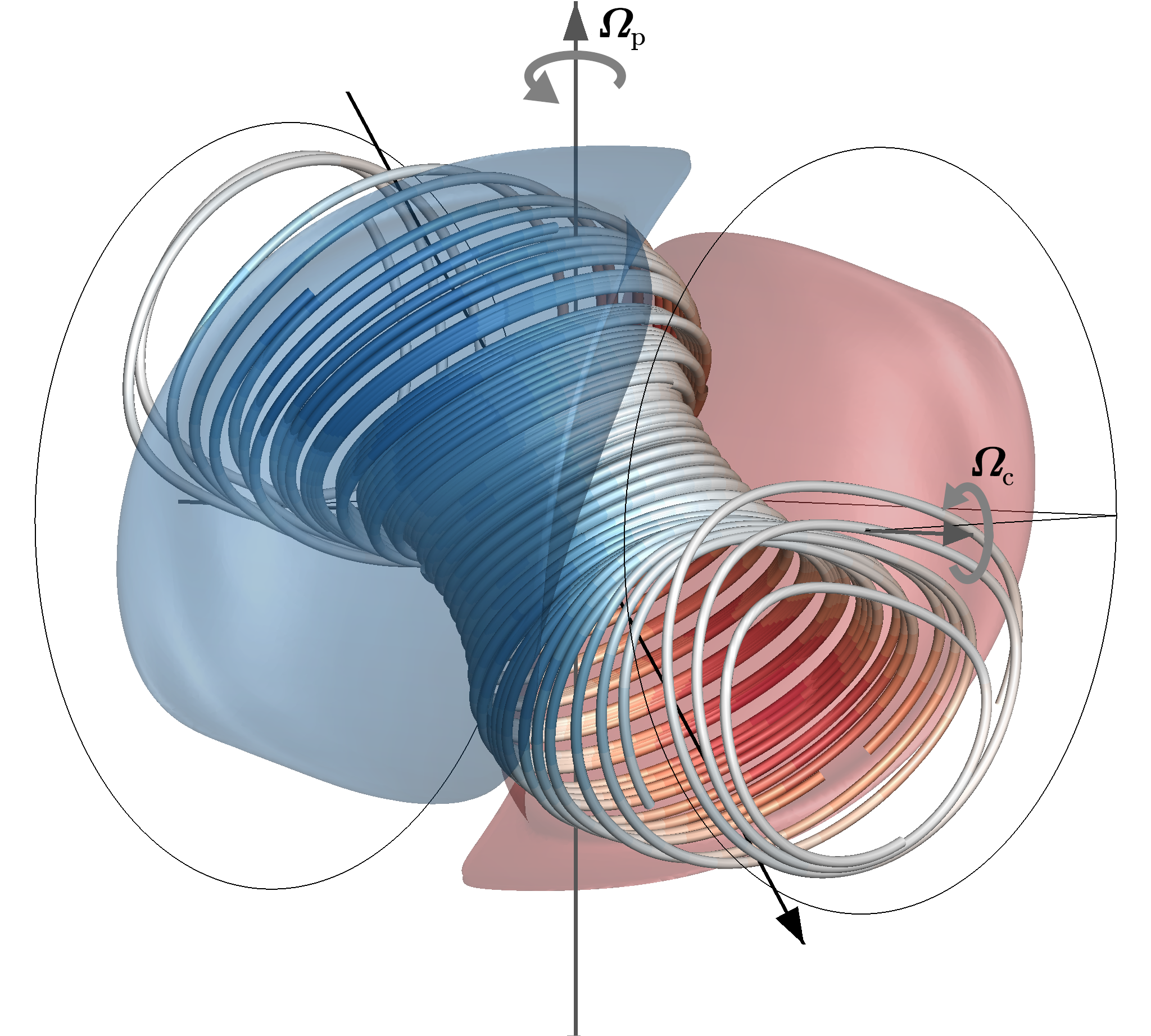}\label{fig::fluid_rotaxis_d}}
\subfloat[][]{\includegraphics[width=0.3\textwidth]{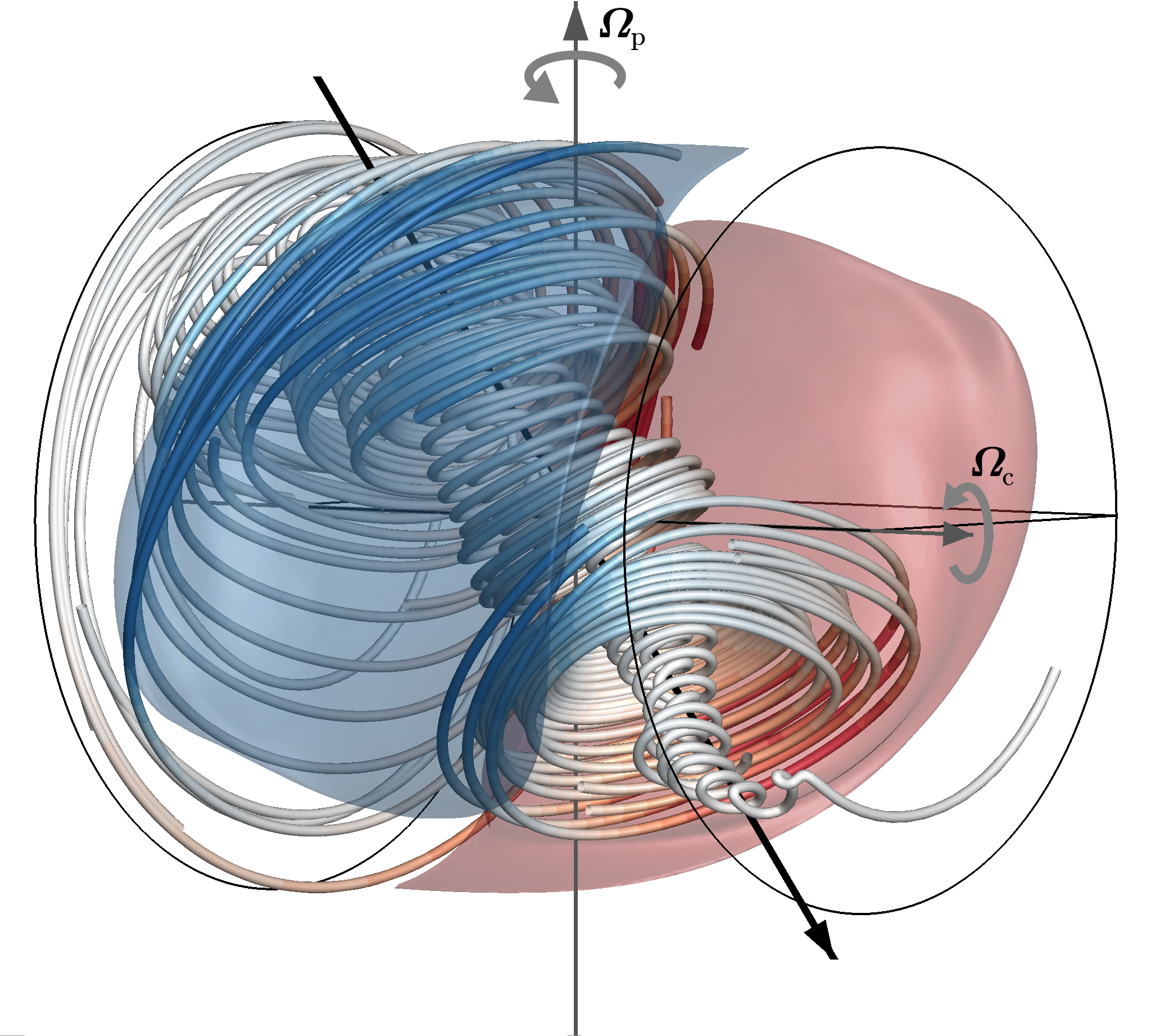}\label{fig::fluid_rotaxis_e}}
\subfloat[][]{\includegraphics[width=0.3\textwidth]{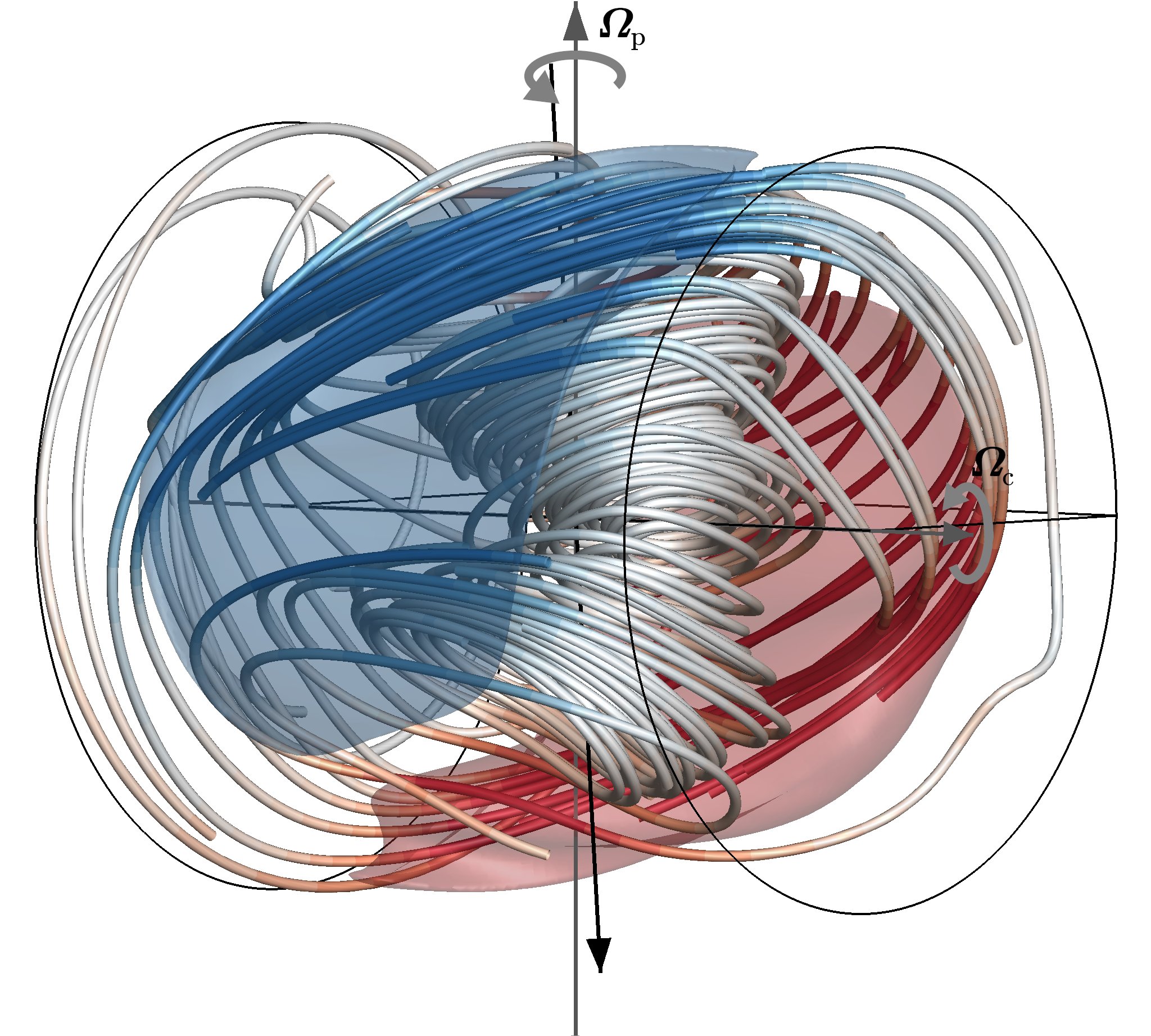}\label{fig::fluid_rotaxis_f}}
\\
\subfloat[][]{\includegraphics[width=0.3\textwidth]{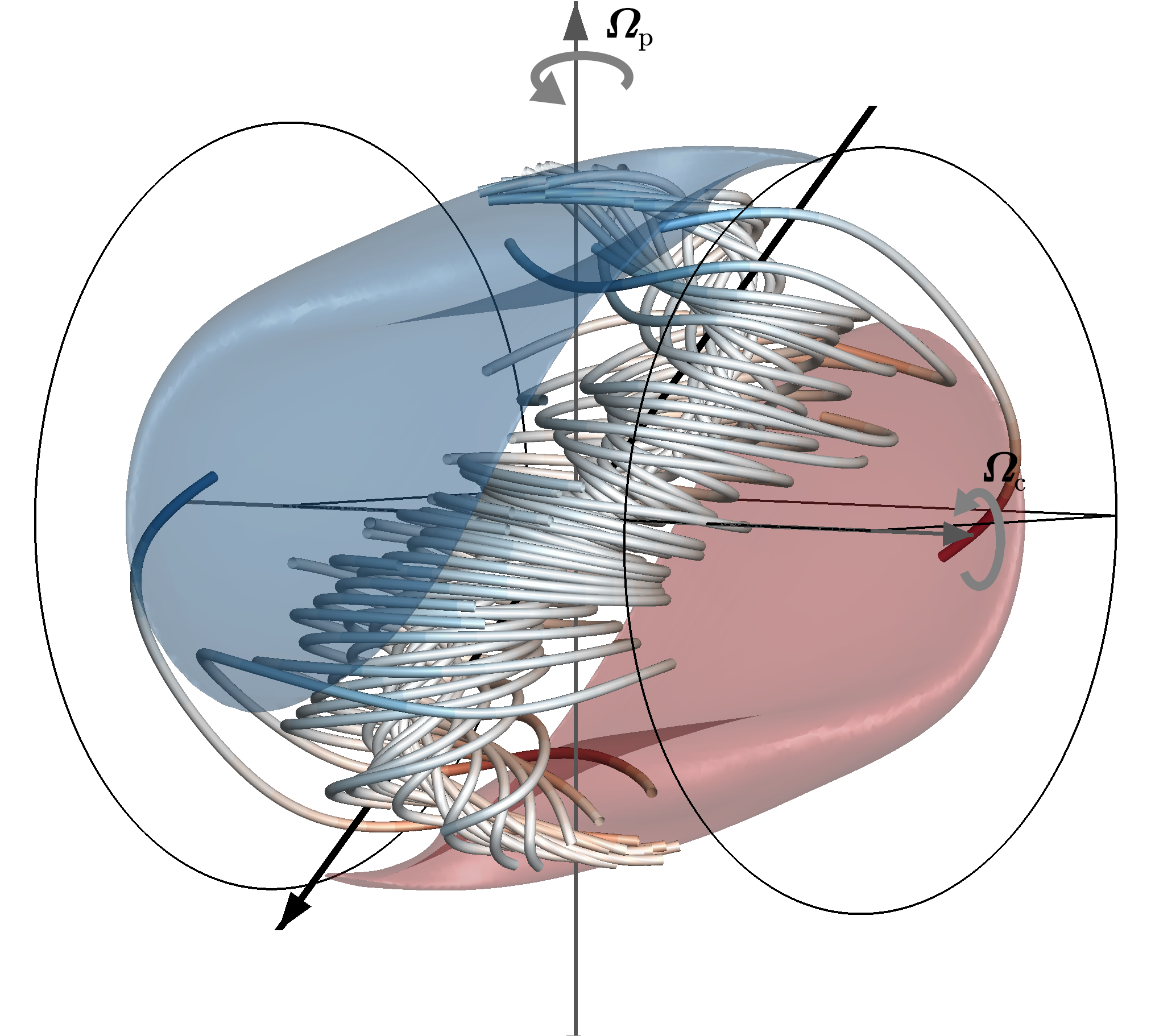}\label{fig::fluid_rotaxis_g}}
\subfloat[][]{\includegraphics[width=0.3\textwidth]{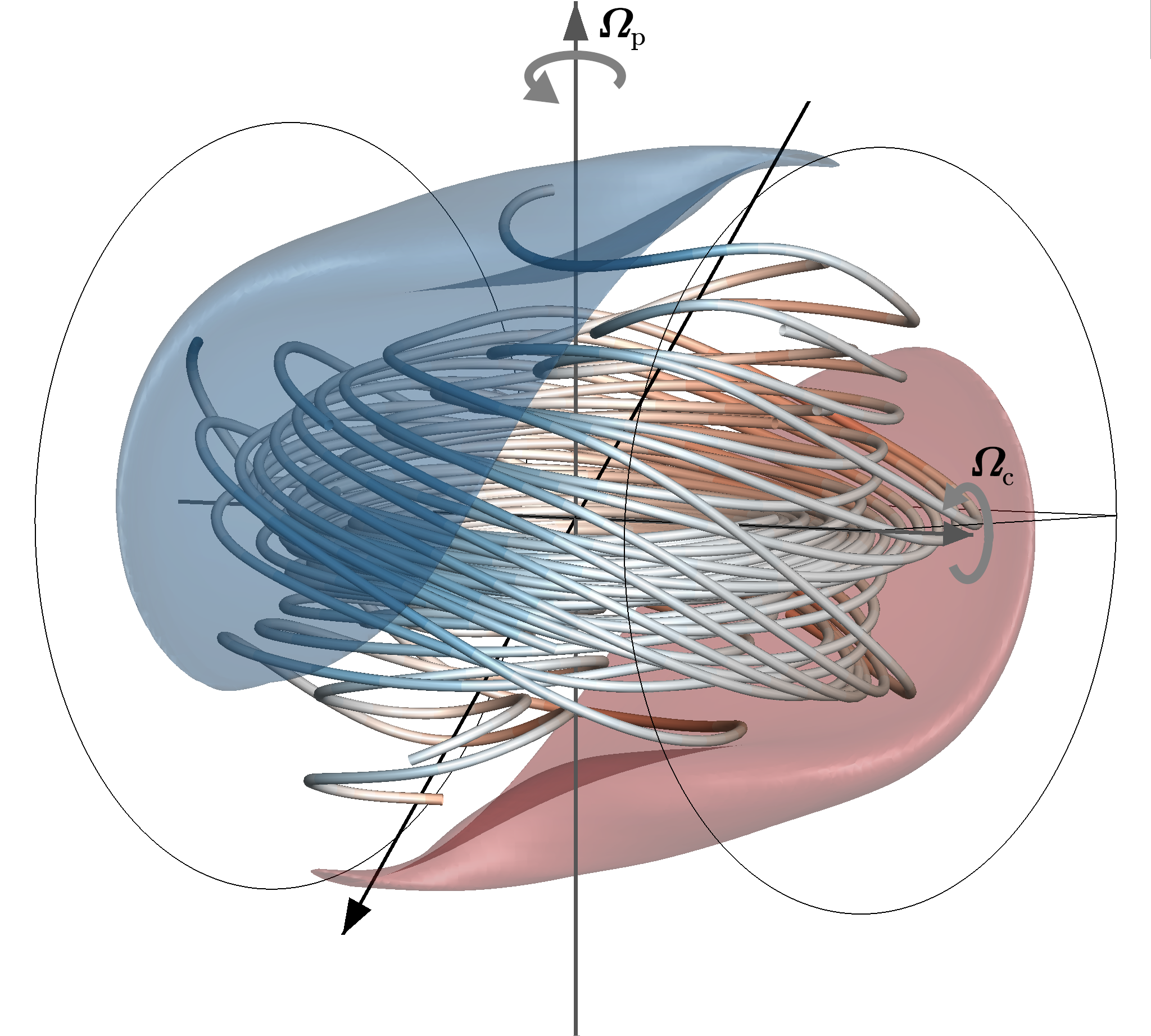}\label{fig::fluid_rotaxis_h}}
\subfloat[][]{\includegraphics[width=0.3\textwidth]{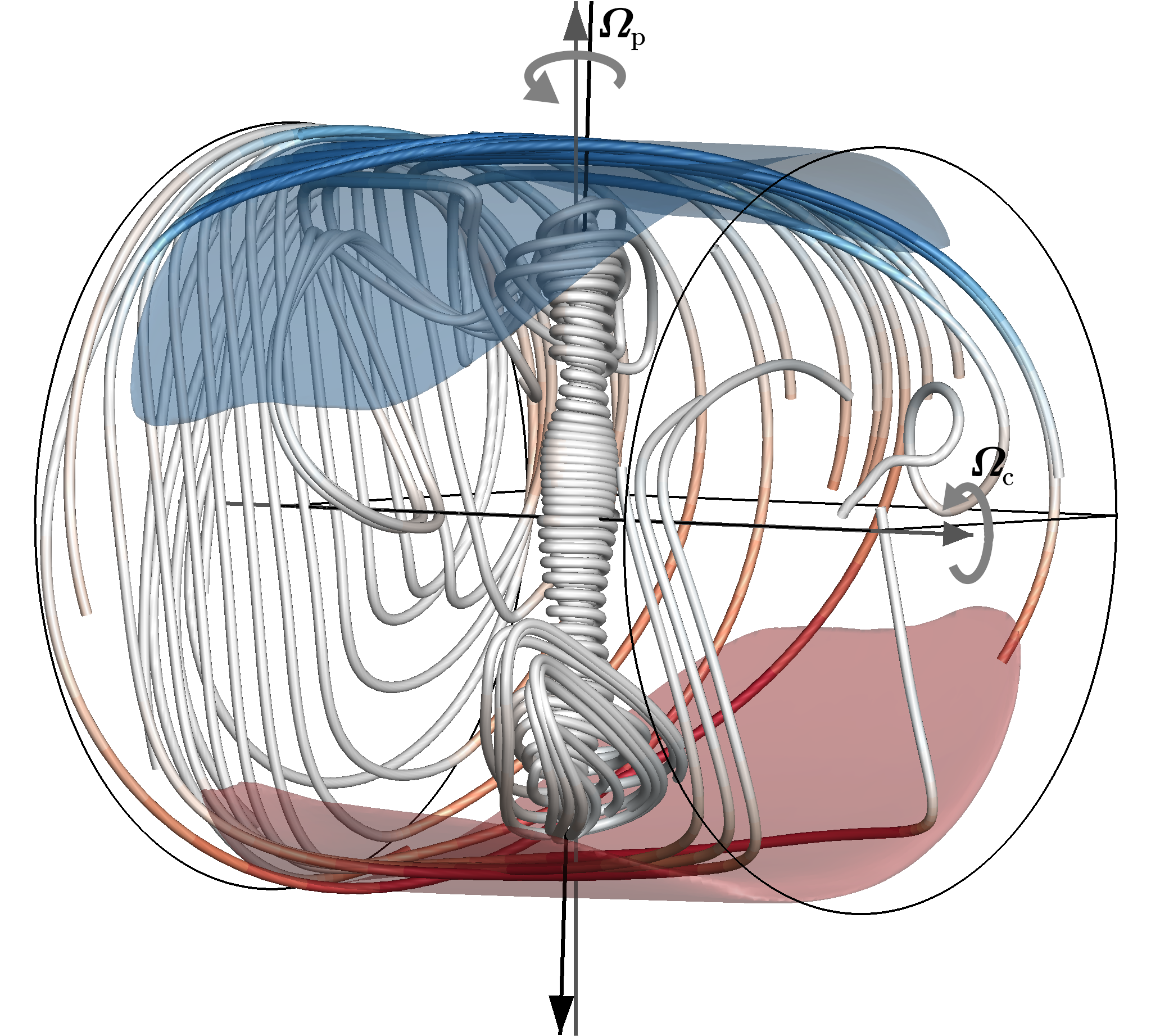}\label{fig::fluid_rotaxis_i}}
\caption{\label{fig::fluid_rotaxis}
\raggedright 
Structure of the time-averaged flow. From upper left
to lower right:
${\rm{Po}}=0.001,0.01,0.05,0.075, 0.0875,0.1,0.125,0.1375,0.2$.  
The streamlines represent the rotational fluid motion in the bulk. The
coloured isosurfaces show the axial velocity $u_z$. The colour of the
streamlines represents the axial flow as well. However, since this
component is small in the region where the streamlines are calculated,
the streamlines are mostly white showing that the axial flow in this
location is quite small.}
\end{figure}

In the following, we examine the change in the geometry of the flow in
dependence on the strength of the precession. For this we consider the
time-averaged flow field defined as $\left<\vec{u}\right>=(\Delta
t)^{-1}\int\vec{u}dt$, where $\Delta t$ represents a time period in the
statistical stationary regime. This representation reflects the flow
behaviour rather well, which can be recognized by comparison with the
time-resolved behaviour of the flow fields which is available as a
movie in the supplemental material at
{\tt{https://doi.org/10.1017/jfm.2024.60}}. 
This visualization also shows that the scale of the spatial
fluctuations decreases with increasing ${\rm{Po}}$ and for large
${\rm{Po}}$ these fluctuations form a (weak) wave pattern that rotates
around the precession axis.  Averaging in time, the fluctuations and
the time-space periodic variations cancel out so that in all cases the
overall structure is dominated by the non-axisymmetric mode with $m=1$,
which is standing in the precessing system. This is illustrated in
Figure~\ref{fig::fluid_rotaxis}, which shows the structure of the core
flow in the precessing cylinder for different precession ratios.
These plots also show that, with increasing ${\rm{Po}}$, the flow
changes its azimuthal position and becomes more and more concentrated
adjacent to the sidewall.  The streamlines in the bulk visualize the
presence and orientation of a particular fluid rotation axis whereas
the iso-surfaces in the same figure denote the axial velocity $u_z$ at
50\% of the rotation velocity of the sidewall of the cylinder.  For
small forcing (${\rm{Po}}=0.001$, figure~\ref{fig::fluid_rotaxis_a}) it
is justified to classify the internal bulk flow as a rotational motion
around an own particular rotation axis, which is (slightly) different
from the rotation axis of the cylinder.  However, already the case
${\rm{Po}}=0.01$ (figure~\ref{fig::fluid_rotaxis_b}) demonstrates the
impact of the end caps, which enforces the fluid rotation motion to be
parallel to the end cap and thus cause a bending of the fluid rotation
axis, leading to an S-shaped pattern.  When the forcing is
sufficiently large so that the angle between the fluid rotation axis
and the rotation axis of the cylinder is large enough, the involvement
of the end caps is reduced and the ``bending'' of the fluid rotation
axis decreases again.

We thus can identify two quasi-stationary states, one characterized by
the fluid rotation axis being roughly directed parallel the rotation axis of
the cylinder, and the second with the fluid rotation axis directed
perpendicular to the direction of the rotation axis of the cylinder,
with an abrupt transition between these states. 
 
We emphasize that the flow in the bulk, which determines the
rotational motion and hence the direction of the fluid rotation axis,
is weak in comparison with the flow in the vicinity of the sidewalls,
which -- especially for large values of the Poincar{\'e} number in the
supercritical region -- is largely independent of the fluid rotation
axis.

%%%%%%%%%%%%%%%%%%%%%%%%%%%%%%%%%%%%%%%%%%%%%%%%%%%%%%%%%%%%%%%%%%%%%%%%%%%%%%%%%%%

\subsection{Kinetic energy}
\label{subsec::3_2_kinetic_energy}

The change in the flow structure, as described in the previous
section, is accompanied by a significant change in the amplitudes of
the individual components of the flow.  We will discuss this in the
following in terms of the kinetic energy.  For this discussion we
switch to the co-rotating frame of reference, which better allows to
analyse the decomposition of the energy into axisymmetric and
non-axisymmetric parts.  This affects the axisymmetric part of the
azimuthal component, since the associated transformation consists of
only a rotation, leaving the other contributions unchanged.

\subsubsection{Onset of the triadic instability}

\begin{figure}
  \subfloat[][]{\includegraphics[width=0.475\textwidth]{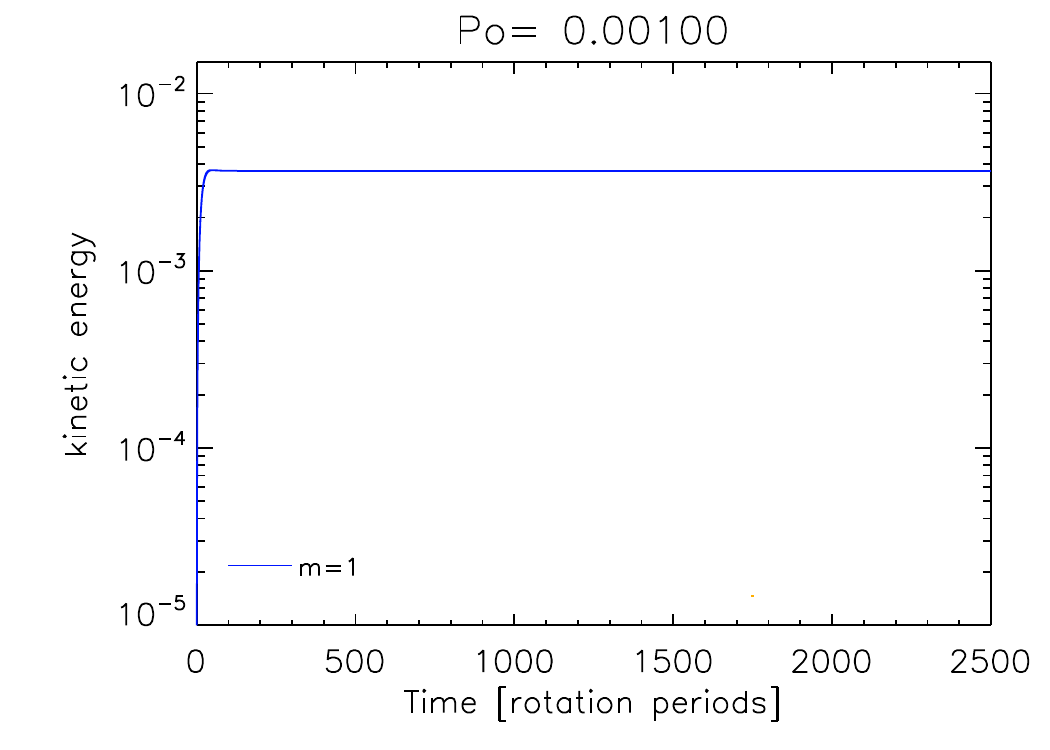}\label{fig::ekin_vs_time_onset_a}}
  \subfloat[][]{\includegraphics[width=0.475\textwidth]{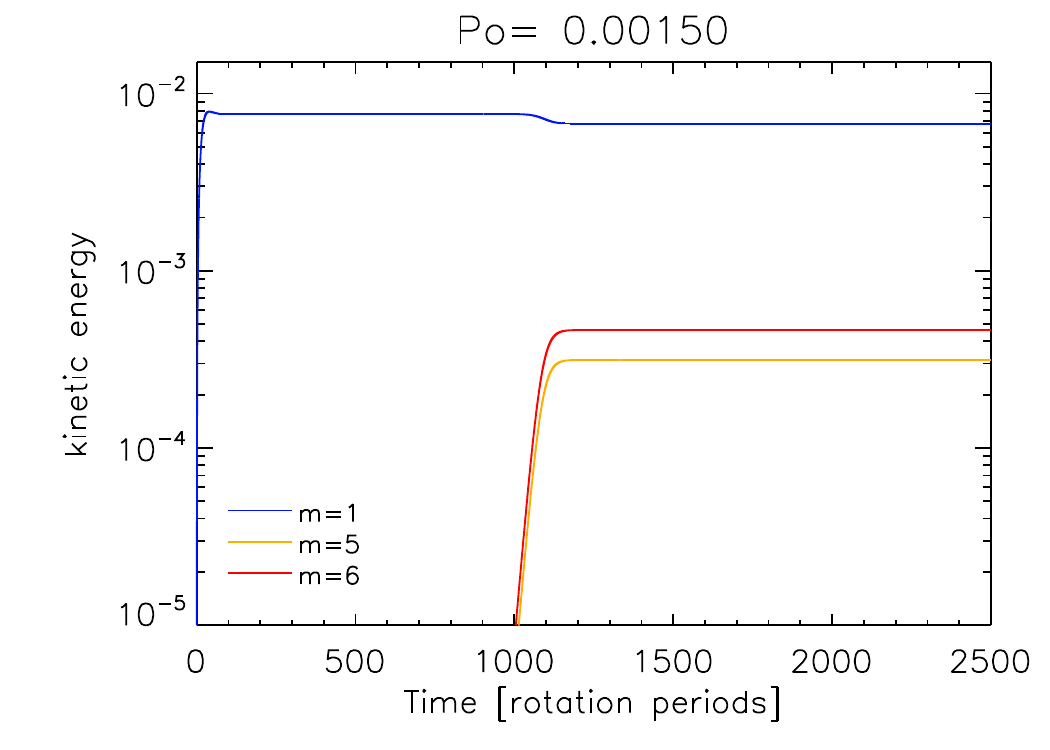}\label{fig::ekin_vs_time_onset_b}}
  \\
  \subfloat[][]{\includegraphics[width=0.475\textwidth]{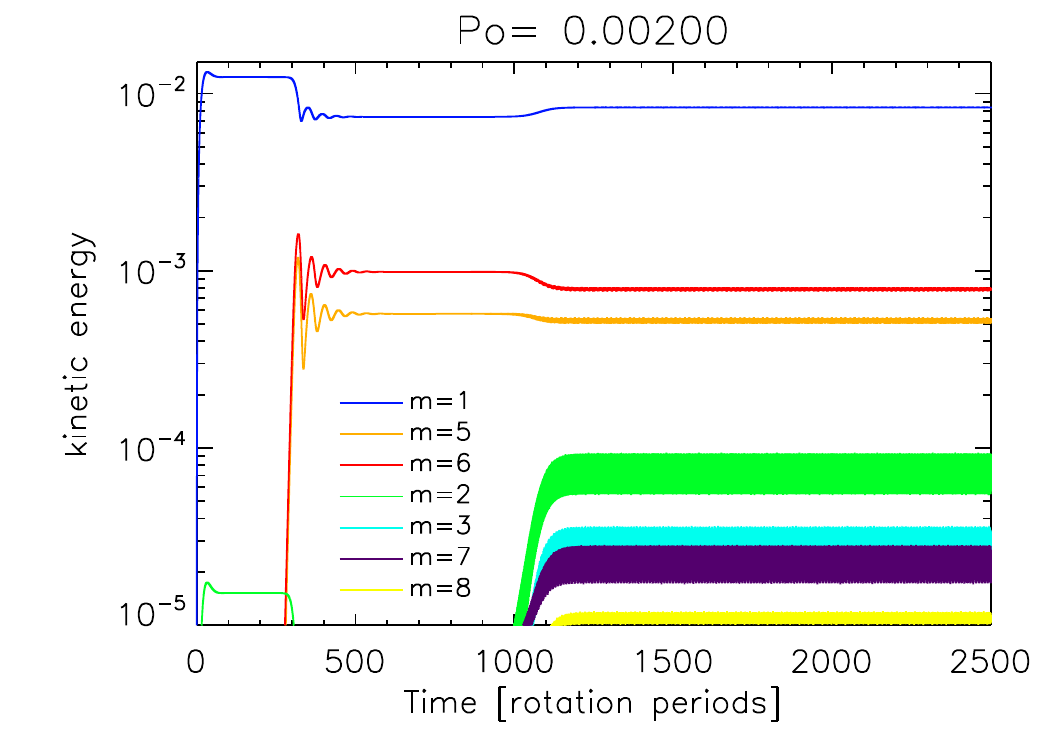}\label{fig::ekin_vs_time_onset_c}}
  \subfloat[][]{\includegraphics[width=0.475\textwidth]{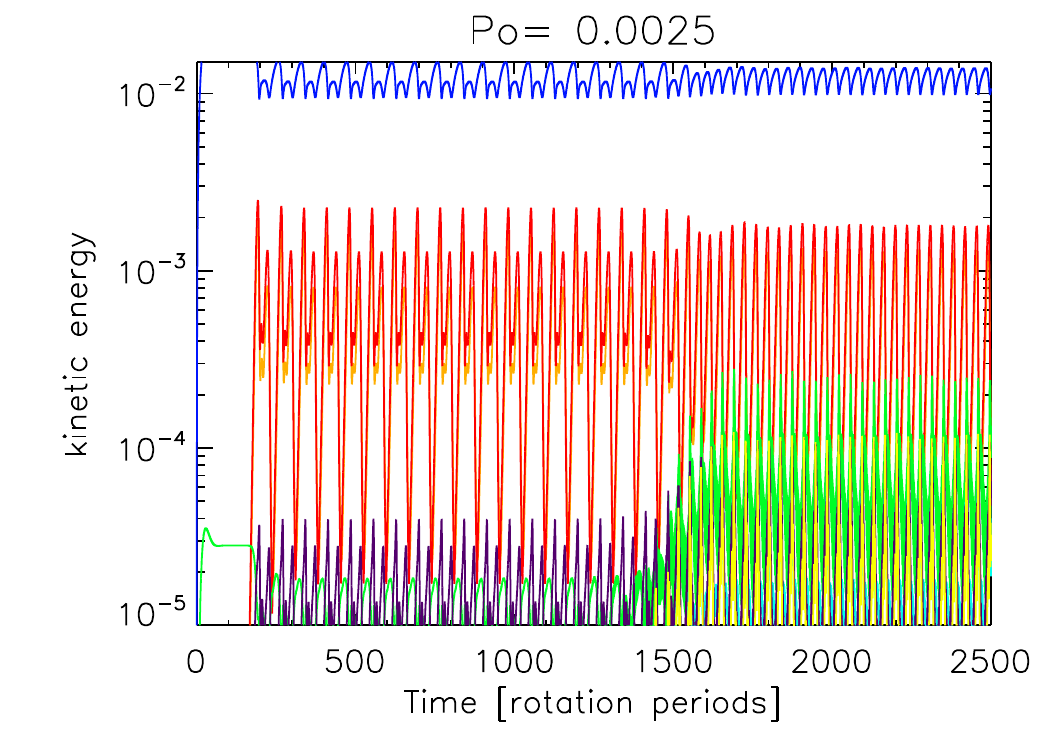}\label{fig::ekin_vs_time_onset_d}}
  \caption{\label{fig::ekin_vs_time_onset}
    \raggedright (a-d) Kinetic energy of the flow contributions with
    the lowest azimuthal wavenumbers. From upper left to lower right
    ${\rm{Po}}=0.001,0.0015,0.002,0.0025$.  Note that, for case (d), the
    final state is reached only after a second transition around
    $t=1000$ and $t=1500$, respectively. In these cases the periodic
    variation of the energies indicates the transition from rotating
    wave to a modulated rotating wave.  In all cases we have
    ${\rm{Re}}=10^4$, $H/R=2$, and $\alpha=90^{\circ}$.
  } 
\end{figure}

We start with weak forcing in Figure~\ref{fig::ekin_vs_time_onset},
which shows the time evolution of the kinetic energy for the leading
azimuthal wavenumbers $m$ for four exemplary ${\rm{Po}}$ with
${\rm{Po}} \leq 0.0025$.  Panel ~\ref{fig::ekin_vs_time_onset_a}
represents the case ${\rm{Po}}=0.001$ which is stable and
time independent and only consists of the forced (resonant) mode
corresponding to figure~\ref{fig::fluid_rotaxis_a}.  Increasing
${\rm{Po}}$, the $m=1$ contributions remain highly dominant in all
cases (blue curves in figure~\ref{fig::ekin_vs_time_onset}), however,
the directly driven flow is prone to a triadic resonance which adds
two free inertial modes.  Theoretical
\citep{kerswell1999,lagrange2011}, experimental \citep{herault2019}
and numerical \citep{giesecke2015b,albrecht2015b} investigations have
proven the occurrence of triadic resonances in precession-driven flows
and have demonstrated their characteristic properties in the form of
selection rules for the wavenumbers and frequencies.  These relations
read $m_{\rm{f}} = 1 = m_2-m_1, k_{\rm{f}} = \left|k_2-k_1\right|$ and
$\omega_{\rm{f}} =1 = \omega_2-\omega_1$, with the triplet
$\{m_{\rm{f}},k_{\rm{f}},\omega_{\rm{f}}\}$ comprising the azimuthal
wavenumber, axial wavenumber and the frequency of the directly
forced mode and the triplets with the indices $1$ and $2$ describe the
corresponding characteristics of two resonantly interacting free
inertial modes.  At aspect ratio $\Gamma=2$ the simplest possible
triadic resonance that is involved with the forced mode at $m=1,k=1$
and $\omega_{\rm{f}}=1$ emerges in terms of two free inertial modes
with wavenumbers $m=5, k=1$ and $m=6, k=2$
\citep{lagrange2011,marques2015,lopez2018}.  This instability sets in
at ${\rm{Po}}\approx 0.00125$ and can be identified in the temporal
evolution of the kinetic energy by means of the flow contributions
with $m=5$ and $m=6$, as shown in
figure~\ref{fig::ekin_vs_time_onset_b}.
The growth of these modes is accompanied by a corresponding drop in
energy of the directly forced flow.  The onset of the triadic
instability complies with a first Hopf bifurcation from a standing
wave to a rotating wave.  However, the energies of the $m=5$ and $m=6$
contributions do not show any time dependency, because these modes
exhibit a stationary geometric structure, which rotates with constant
frequency around the axis of the cylinder.

If one increases the Poincar{\'e} number further, additional
contributions with larger $m$ appear
(figure~\ref{fig::ekin_vs_time_onset_c}) caused by interactions of the
three initial modes among themselves.  The emergence of the new modes
is accompanied by a secondary transition to an oscillating energy that
corresponds to the transition from a rotating wave to a modulated
rotating wave.  This secondary transitions occurs earlier when further
increasing the forcing (see
figure~\ref{fig::ekin_vs_time_onset_d}). Furthermore, the amplitude of
the oscillations increases as well as the number of involved
frequencies.  In figures~\ref{fig::ekin_vs_time_onset_c}
and~\ref{fig::ekin_vs_time_onset_d} four different branches of
solutions can be identified.  In
figure~\ref{fig::ekin_vs_time_onset_c} we recognize two different
behaviours: for $t \in [500,1000]$ and for $t>1000$. In the first
interval, there are no oscillations so the solution looks like the
solution of figure~\ref{fig::ekin_vs_time_onset_b}. This behaviour is
unstable, since the final state ($t>1000$) displays
oscillations. Dynamically speaking, this means that the solution is
approaching, for $t \in [500,1000]$, an unstable configuration with
the same dynamical properties (non-oscillatory) as the solution of
figure~\ref{fig::ekin_vs_time_onset_b}. This configuration is unstable
since the final state is oscillatory, indicating a Hopf-like
bifurcation, which basically consists in adding a new temporal
frequency to the flow.  This final state is a modulated rotating wave,
so branch 2 appears to be due to a second Hopf bifurcation. However, the
first part of figure~\ref{fig::ekin_vs_time_onset_d} (unstable up to
$t=1500$) seems to be not of the type of any final state of previous
panels, so it should belong to another branch number 3, which may
arise from further bifurcations on branch 2. The branch number 4
emerges for $t > 1500$ in figure~\ref{fig::ekin_vs_time_onset_d} and seems
to be related with branch 3 by means of period halving.  All involved
modes and their frequencies are still resulting from combinations of
the originally occurring waves of the first instability and we do not
observe the occurrence of a linearly independent (incommensurable)
further triadic resonance.  This is in contrast to the spherical
Couette system, where the base flow is axially symmetric so that a
first bifurcation is required in order to break this axisymmetry and
therefore another incommensurable triad is observed
\citep{garcia2020,garcia2021}.

\begin{figure}
\centerline{\includegraphics[width=0.66\textwidth]{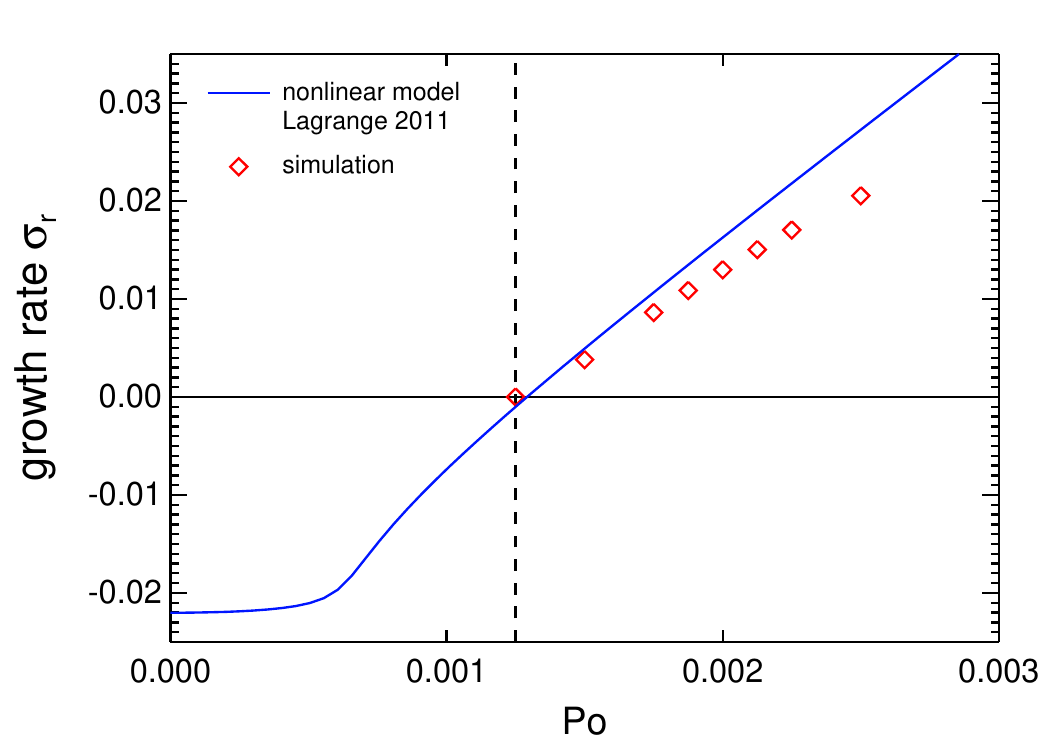}}
\caption{\label{fig::triades}
\raggedright   Growth rates calculated from numerical
  simulations in comparison with the weakly nonlinear theory
  developed in \citet{lagrange2011}.
  The dashed vertical line at ${\rm{Po}}=0.00125$ denotes
  the threshold for the onset of the triadic instability.
}
\end{figure}

Figure~\ref{fig::triades} compares the growth rates calculated from
the exponential growth of the kinetic energy of the unstable modes, as
shown in figure~\ref{fig::ekin_vs_time_onset}, with the results
calculated from the theory presented in \citet{lagrange2011} (dashed
curve, calculated with their Eq.~(4.17) adapted to the aspect ratio
$\Gamma=2.0$; see page 120 of \citet{lagrange2011}).  For all
${\rm{Po}}$ values the modes $m=5$ and $m=6$ exhibit a similar growth rate.
We see a good agreement between theory and simulations, in particular
the critical value of ${\rm{Po}}$ that is required for the onset of
the instability agrees nearly perfectly. However, for increasing
forcing the growth rates from the simulations are systematically
smaller than the theoretical predictions. This can be explained by the
reduced theoretical model of \citet{lagrange2011} that only considers
four different modes (namely the axisymmetric mode, the directly
forced mode and the two free inertial waves with ${m=5,k=1}$ and
${m=6,k=2}$), whereas the simulations include further modes even at
relatively low supercriticality (see
figure~\ref{fig::ekin_vs_time_onset_c}
and~\ref{fig::ekin_vs_time_onset_d}), which extract energy from the
forced mode so that its amplitude decreases as soon as the free Kelvin
modes appear (this can be seen explicitly in
figure~\ref{fig::nonlin_evolution_amplitudes_a} later on).  Another
possible cause for the deviation might be that the theoretical model
is only valid for the exact resonance so that detuning effects are
neglected, and saturation is only caused by surface and volume viscous
damping.

%%%%%%%%%%%%%%%%%%%%%%%%%%%%%%%%%%%%%%%%%%%%%%%%%%%%%%%%%%%%%%%%%%%%%%%%%%%%

\subsubsection{Chaotic regime and transition}

\begin{figure*}%[t!]
  \subfloat[][]{\includegraphics[width=0.4875\textwidth]{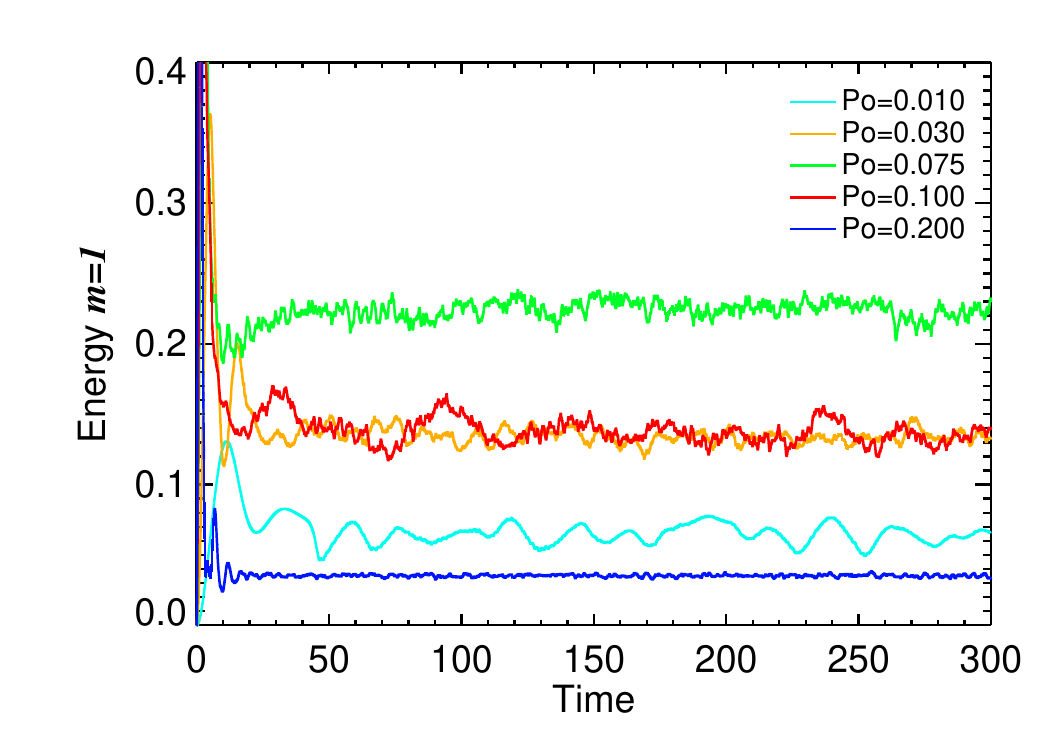}\label{fig::ener_vs_tim_a}}
  \quad
  \subfloat[][]{\includegraphics[width=0.4875\textwidth]{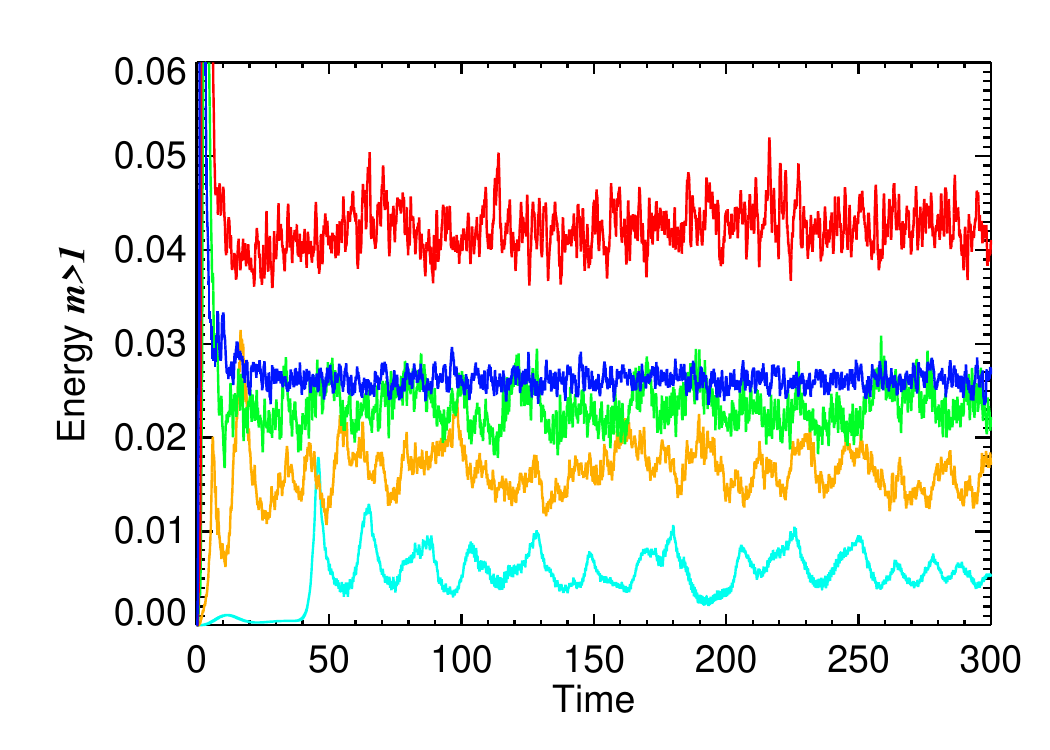}\label{fig::ener_vs_tim_b}}
  \\
    \centering{
    \subfloat[][]{\includegraphics[width=0.4875\textwidth]{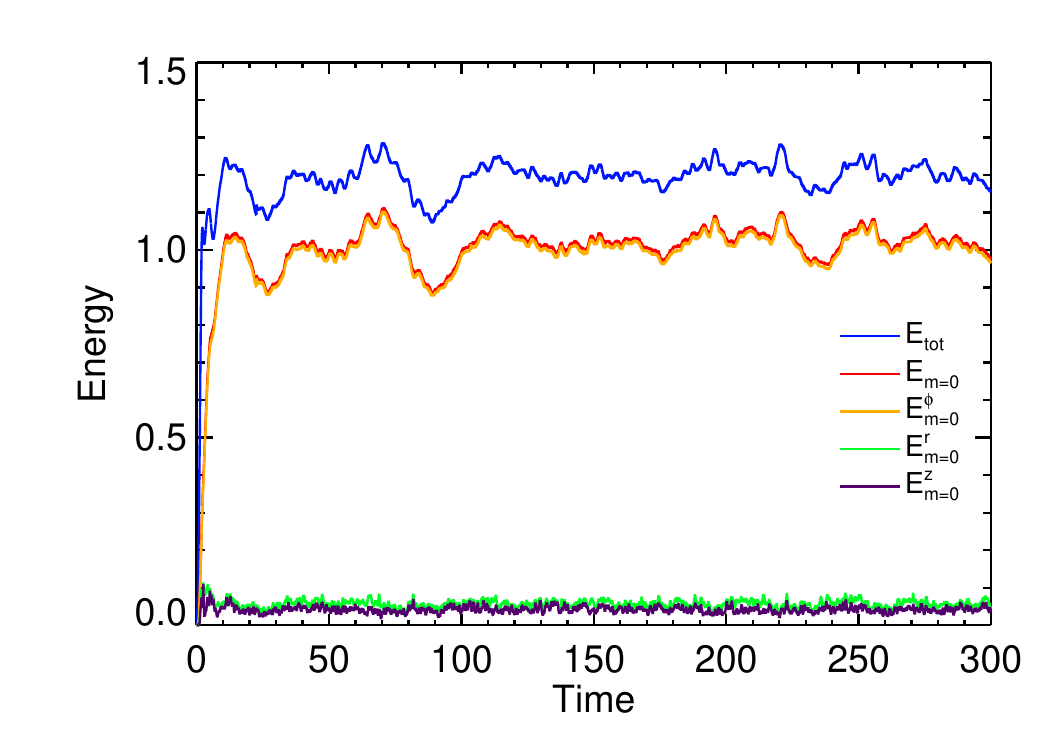}\label{fig::enerm0_vs_tim}}
    }
  \caption{\label{fig::ener_vs_tim}
    \raggedright
    (a) Time evolution of the kinetic energy of the flow, directly driven by
    precessional forcing for various values of ${\rm{Po}}$.
(b) Higher non-axissymmetric contributions with $m>1$.
(c) Time evolution of the kinetic energy 
    for the particular case ${\rm{Po}}=0.1$. The
    axisymmetric energy (red curve) reaches roughly 90\% of the total kinetic
    energy, and furthermore, is almost contained in the $\phi$
    component (orange curve). Note that the curves for the energy from
    the $r$ and the $z$ components are enhanced by a factor of $10$.
    In all cases we have ${\rm{Re}}=10^4$, $H/R=2$,
    $\alpha=90^{\circ}$
    and time is denoted in terms of rotation periods.
  }
\end{figure*}

When further increasing the forcing parameter ${\rm{Po}}$, the
behaviour of the flow becomes chaotic, but still results in a
stationary state in the statistical sense.  The time evolution of the
kinetic energy for various cases with ${\rm{Po}}\geq 0.01$ is shown
in figure~\ref{fig::ener_vs_tim}.
Because of the coupling of the individual inertial modes, it makes
less sense to consider individual $m$ so that we show the kinetic
energy of the directly forced flow ($m=1$,
figure~\ref{fig::ener_vs_tim_a}) in comparison with 
the integrated energy of all higher non-axisymmetric modes with $m>1$
(figure~\ref{fig::ener_vs_tim_b}). 
\begin{figure*}
   \subfloat[][]{\includegraphics[width=0.475\textwidth]{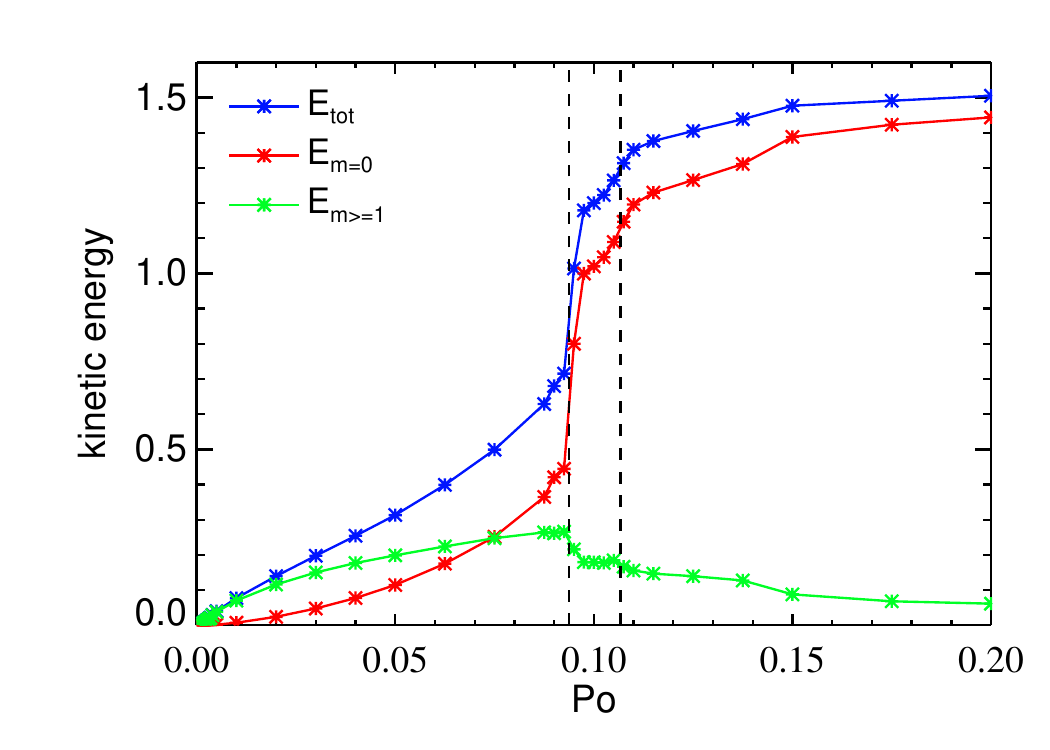}\label{fig::ener_vs_po_nonaxisym_a}}  
   \quad
   \subfloat[][]{\includegraphics[width=0.475\textwidth]{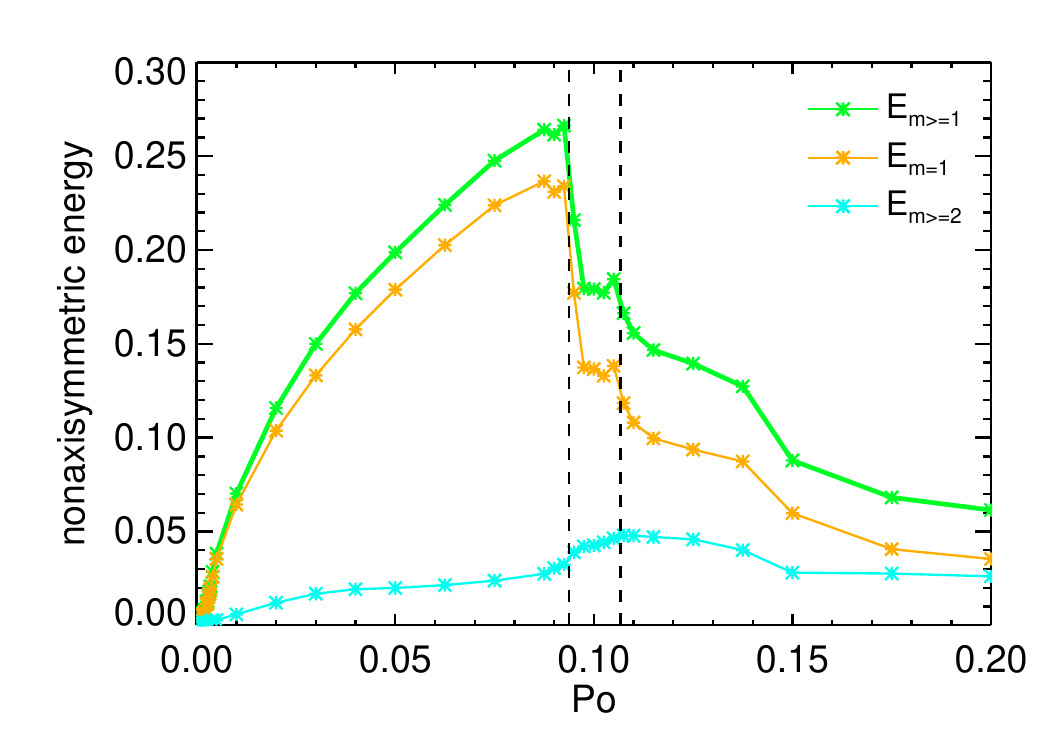}\label{fig::ener_vs_po_nonaxisym_b}}
\\[-0.6cm]
   \subfloat[][]{\includegraphics[width=0.475\textwidth]{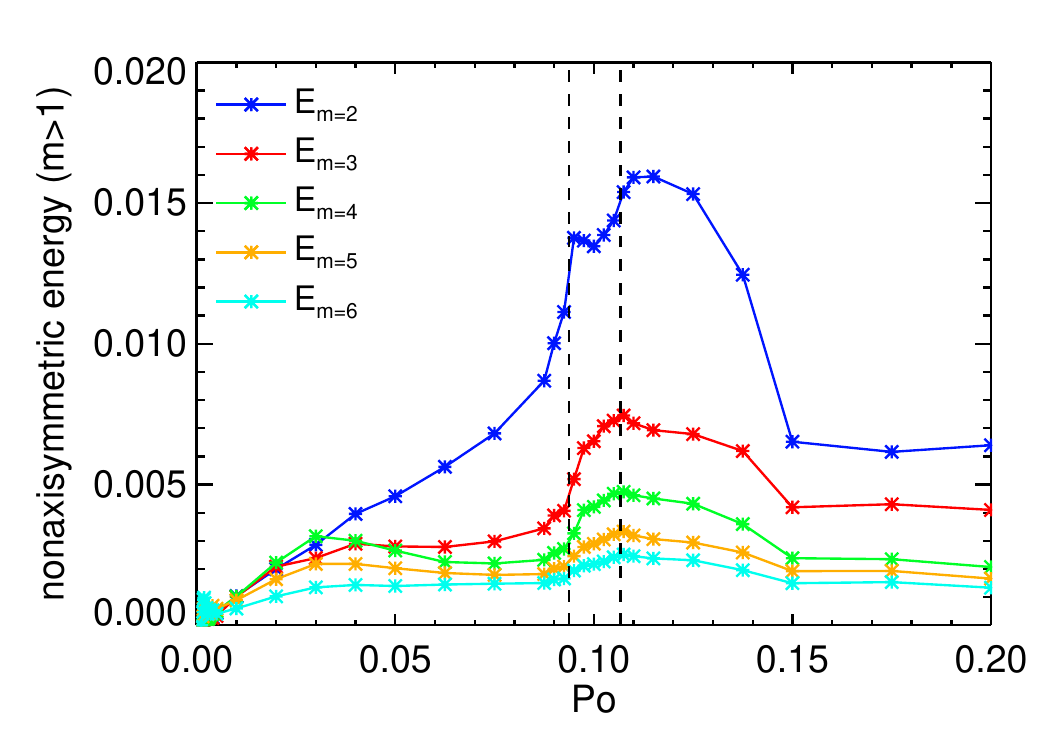}\label{fig::ener_vs_po_nonaxisym_c}}  
   \quad
   \subfloat[][]{\includegraphics[width=0.475\textwidth]{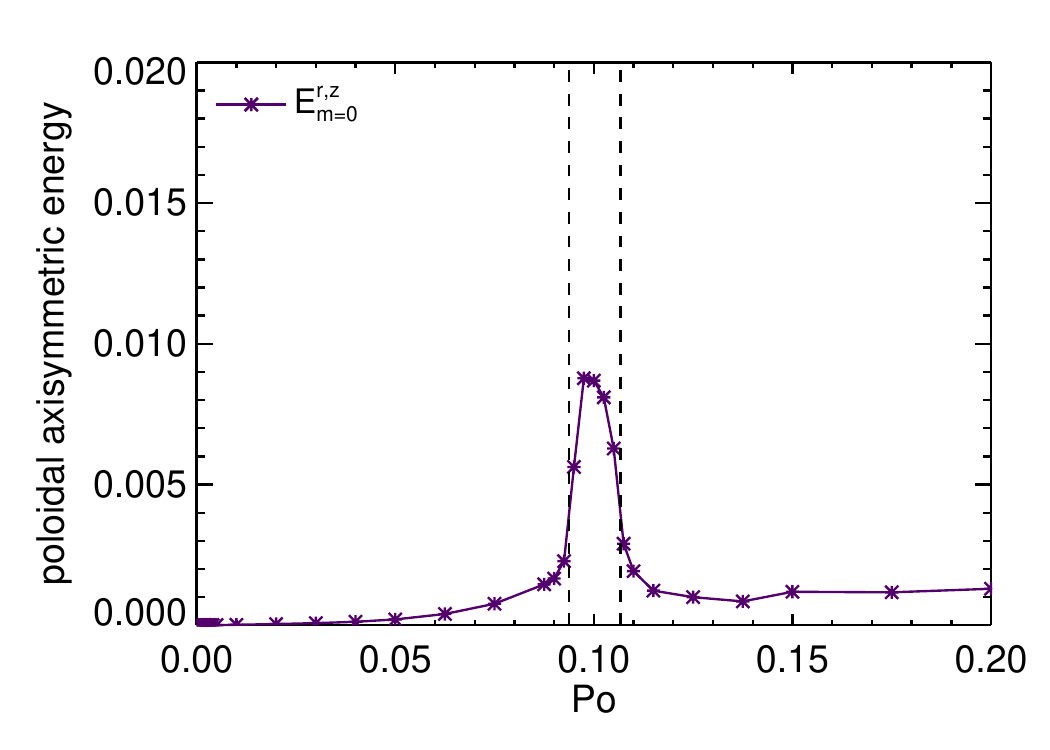}\label{fig::ener_vs_po_nonaxisym_d}}
\caption{\label{fig::ener_vs_po_nonaxisym}
  \raggedright 
  Time-averaged kinetic energy calculated in the co-rotating frame of
  reference.  (a) Decomposition of the total kinetic energy (blue)
  into axisymmetric contributions (red) and non-axisymmetric
  contributions (green).  (b) Decomposition of the non-axisymmetric
  contributions (green) into directly forced flow ($m=1$, orange) and
  the remainder ($m\geq 2$, light blue).  (c) Kinetic energy of
  individual Fourier modes beyond $m=1$.  (d) Kinetic energy of the
  poloidal axisymmetric flow ($u_r, u_z$, double-roll mode). In all
  cases we have ${\rm{Re}}=10^4$, $H/R=2$, $\alpha=90^{\circ}$.  The
  vertical dashed lines denote the transition regime as discussed in
  the text.
}
\end{figure*}
As in the previous cases, we see a strong peak at the beginning,
followed by a transient phase, which, depending on the precession
ratio, lasts approximately ten to thirty rotation periods. Afterwards, the
system is in a quasi-stationary state without regular periodic
behaviour.  We also see that the kinetic energy does not increase
monotonically with the strength of the force.  For instance, in
figure~\ref{fig::ener_vs_tim_a}, the energy at ${\rm{Po}}=0.075$ is
almost one order of magnitude larger than at the strongest forcing
with ${\rm{Po}}=0.2$.  A particular case is ${\rm{Po}}=0.1$ for which
it takes a longer time period to reach a quasi-stable equilibrium
which may indicate a bistable state, where the system needs some time
to reach the final state.  We also see that, for this ${\rm{Po}}$, the
proportion of contributions with $m>1$ is higher than for other
parameters (see red curve in figure~\ref{fig::ener_vs_tim_b}), which
indicates that the flow is more complex at ${\rm{Po}}=0.1$.  Finally,
figure~\ref{fig::enerm0_vs_tim} shows the proportion of the
axisymmetric contributions (red curve) in relation to the total
kinetic energy (blue curve) for this case, and it is obvious that
almost the entire kinetic energy of the axisymmetric contributions is
contained in the azimuthal component (orange curve).  Note that the
energetic contributions of the poloidal component (green and violet
curves) are enhanced by a factor of $10$ in order to make these
contributions a little clearer.

Due to the moderate fluctuations of the kinetic energy around a mean
value, it makes sense to consider the time-integrated values as a
function of the forcing.  Figure ~\ref{fig::ener_vs_po_nonaxisym_a}
shows the total kinetic energy (blue curve), and its decomposition
into an axisymmetric part (red curve) and a non-axisymmetric part
(green curve).  For small and intermediate ${\rm{Po}}$, the
contribution of the axisymmetric flow remains small, and the course of
action is determined by the directly driven non-axisymmetric flow.
This only changes around ${\rm{Po}}\approx 0.075$.  From this value
onwards, the dominant component is the axisymmetric component.  The
abrupt increase around ${\rm{Po}}\approx 0.094$ occurs parallel to a
drop in the kinetic energy of the non-axisymmetric components whereby
this drop is significantly smaller than the increase in the axially
symmetric energy.  The sudden increase with a maximum of roughly
$E_{\rm{kin}}\approx 1.5$ at ${\rm{Po}}=0.2$ means that nearly all
kinetic energy available from the initial rotational motion has been
transferred into a fluid flow with a fairly simple structure, namely a
zonal flow (in our scaled units $\varOmega_{\rm{c}}=1, R=1, H=2$, the
kinetic energy of the initial solid body rotation is
$E_{\rm{kin}}=\int\vec{u}^2/2dV=\pi
H\int\varOmega_{\rm{c}}r^3dr=\pi/2$).  This strong zonal flow is
oriented opposite to the direction of rotation of the cylinder and
therefore corresponds to a deceleration of the purely azimuthal flow
at the beginning.  This zonal flow is the result of nonlinear
interactions, for which nonlinear models list three different sources,
namely the nonlinear interaction of the forced mode with itself and
with its viscous modification and the nonlinear interactions in the
end cap boundary layers \citep{gao2021}.

\begin{figure}
%%      \includegraphics[width=0.55\textwidth]{./plots_final/contour_flow_struc/vec2d_sim_uzur_axisim_totalplane_po0p0100}
%%      \hspace*{-1.4cm}
%%      \includegraphics[width=0.55\textwidth]{./plots_final/contour_flow_struc/vec2d_sim_uzur_axisim_totalplane_po0p0750}
%%      \\
%%      \includegraphics[width=0.55\textwidth]{./plots_final/contour_flow_struc/vec2d_sim_uzur_axisim_totalplane_po0p1000}
%%      \hspace*{-1.4cm}
%%      \includegraphics[width=0.55\textwidth]{./plots_final/contour_flow_struc/vec2d_sim_uzur_axisim_totalplane_po0p2000}
%      \hspace*{-0.2cm}
%      \includegraphics[width=0.575\textwidth]{giesecke_global_state_cylinder_precession_fig_14a.pdf}
%      \hspace*{-1.6cm}
%      \includegraphics[width=0.575\textwidth]{giesecke_global_state_cylinder_precession_fig_14b.pdf}
%      \\[-0.3cm]
%      \hspace*{-0.2cm}
%      \hspace*{-1.6cm}
%      \includegraphics[width=0.575\textwidth]{giesecke_global_state_cylinder_precession_fig_14d.pdf}
  \subfloat[][]{\begin{minipage}{0.49\textwidth}
      {
        \includegraphics[width=0.99\textwidth]{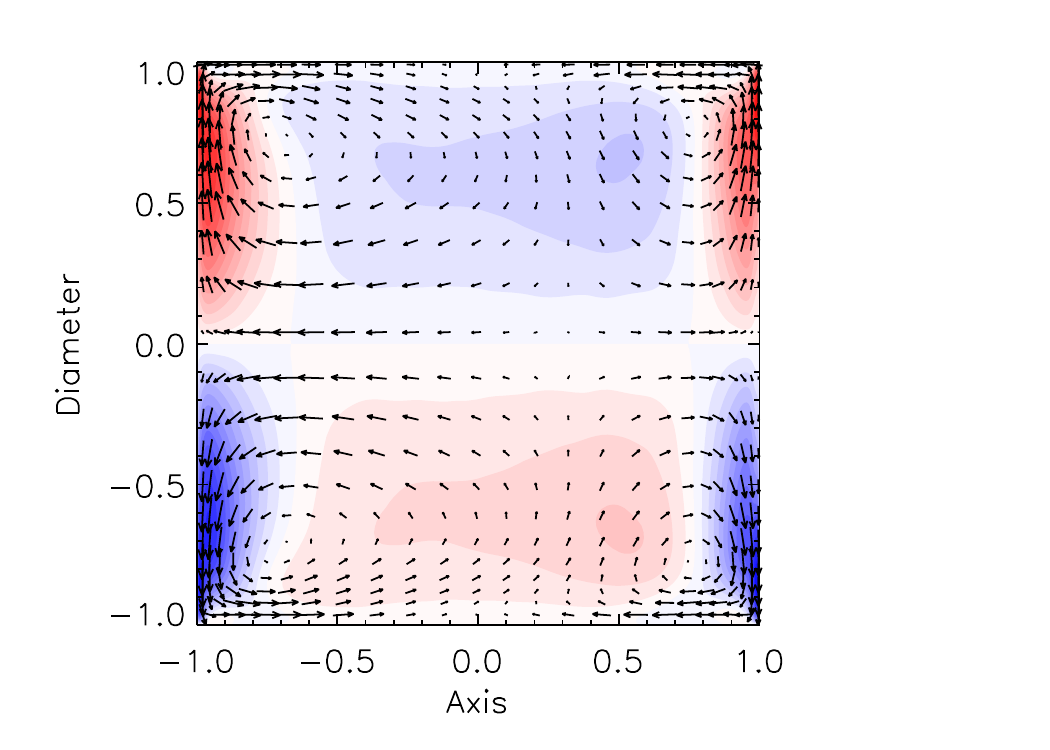}
      }
    \end{minipage}\label{fig::drm_contour_uphi_axisym_a}}
  \hspace*{-1.7cm}
  \subfloat[][]{\begin{minipage}{0.5\textwidth}
      {
        \includegraphics[width=1.0\textwidth]{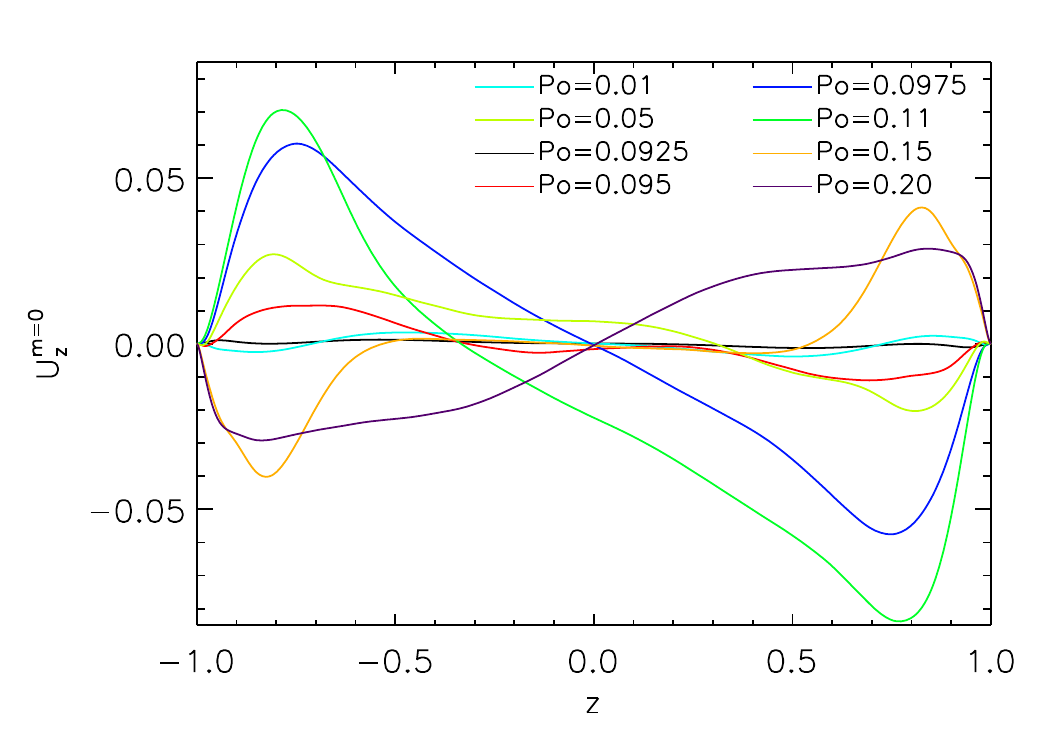}
      }
    \end{minipage}\label{fig::drm_contour_uphi_axisym_b}}
  \caption{\label{fig::drm_contour_uphi_axisym}
    \raggedright 
    (a) Time-averaged pattern of the poloidal axisymmetric flow components
    $u_r\hat{\vec{r}}$ and $u_z\hat{\vec{z}}$ (vector arrows). The
    coloured contour plots show 
    the radial flow component $u_r$. From upper left to lower right:
    ${\rm{Po}}=0.01, 0.075, 0.1, 0.2$.
    (b) Axial profile of the axisymmetric part of $u_z$
    for various ${\rm{Po}}$ taken at $r=0.9$.
    %% daten liegen auf
    %% /bigdata/dresdyn/precession/hydrosim/analysis/flow_structure_compare_exp_sim/radial%_profiles/ (inklusive preprocessed data and script) 
  }
\end{figure}

The composition of the non-axisymmetric flow is shown in more detail
in figure~\ref{fig::ener_vs_po_nonaxisym_b}, which presents the
division of the non-axisymmetric energy (green curve) into the
components with the Fourier component $m=1$ (orange curve) and the
remaining part (with $m>1$, light blue curve).  After reaching the
maximum shortly before ${\rm{Po}}\approx 0.1$ with an amount of up to
15\% of the kinetic energy of the initially purely rotating flow, we
see a sharp drop due to the collapse of the Fourier component with
$m=1$.  The most interesting region is the regime $0.094 \leq
{\rm{Po}} \leq 0.105$ that is indicated by the dashed vertical lines
in all plots of figure~\ref{fig::ener_vs_po_nonaxisym}. Within this
interval of ${\rm{Po}}$ we find a plateau-like behaviour for the $m=1$
contributions and, at the same time, a secondary axisymmetric flow,
which is characterized by a regular large-scale pattern in the
poloidal components $u_r$ and $u_z$
(figure~\ref{fig::ener_vs_po_nonaxisym_d}).  In contrast to the
previous results at smaller ${\rm{Re}}$ presented in
\citet{pizzi2021b}, here the simulations are sufficiently detailed in
${\rm{Po}}$ to illustrate the plateau-like behaviour of the forced mode
and the exact correspondence of plateau and occurrence of the
double-roll mode.  The energy contribution of this double-roll
structure is small compared with the directly driven flow or with the
total axisymmetric energy which confirms the initially made statement
about the dominance of a pure rotational motion ($E^{m=0}_{\rm{tot}}
\approx E^{m=0}_{\rm{\varphi}} \gg E^{m=0}_{\rm{r,z}}$).  The shape of
the double-roll mode with $m=0$ and $k=2$ is shown in
figure~\ref{fig::drm_contour_uphi_axisym_a} together with the
corresponding axial profiles of $u_z$ in
figure~\ref{fig::drm_contour_uphi_axisym_b}. These curves show, on the
one hand, the variation of the strength of this double roll and a
reversal of the direction of the flow for ${\rm{Po}} >0.125$.  This
particular flow structure is interesting because of the appearance and
disappearance for a small range of ${\rm{Po}}$, and because it bears a
remarkable resemblance to a large-scale flow structure that is known
to be beneficial for dynamo action \citep{dudley1989}.

%*****************************************************************
%

The energy contributions of the higher non-axisymmetric modes are
shown in figure~\ref{fig::ener_vs_po_nonaxisym_c}. It is particularly
noticeable here that the $m=2$ 
contributions increase more strongly in comparison with even higher
$m$-modes, and this increase also begins before the transition described
above.  Increasing ${\rm{Po}}$, in the further course, the $m=1$
component continues to fall in several stages, until finally, at
${\rm{Po}} =0.2$, an almost equal distribution between the energy of
the $m=1$ component and the contributions with $m>1$ is achieved.
Note that the energies in figure~\ref{fig::ener_vs_po_nonaxisym} are
only decomposed according to the azimuthal symmetry, and therefore all
turbulent components (each with the corresponding azimuthal symmetry)
are also included.

\begin{figure*}%[t!]
  \subfloat[][]{\includegraphics[width=0.4875\textwidth]{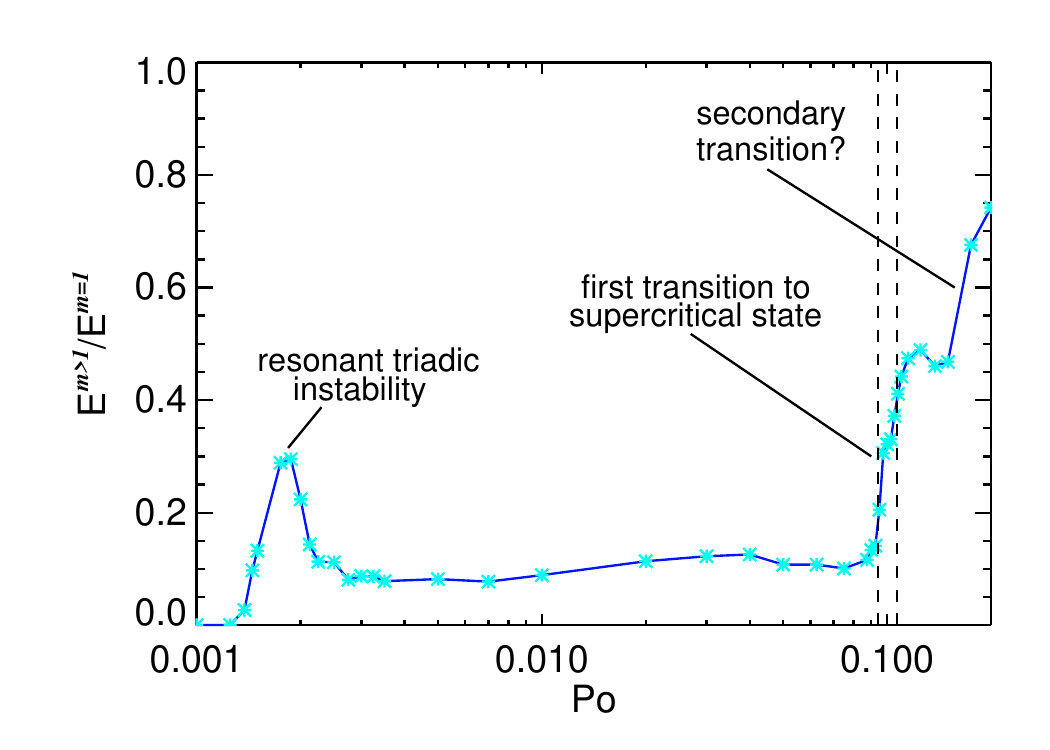}\label{fig::ener_tur_a}} 
  \subfloat[][]{\includegraphics[width=0.4875\textwidth]{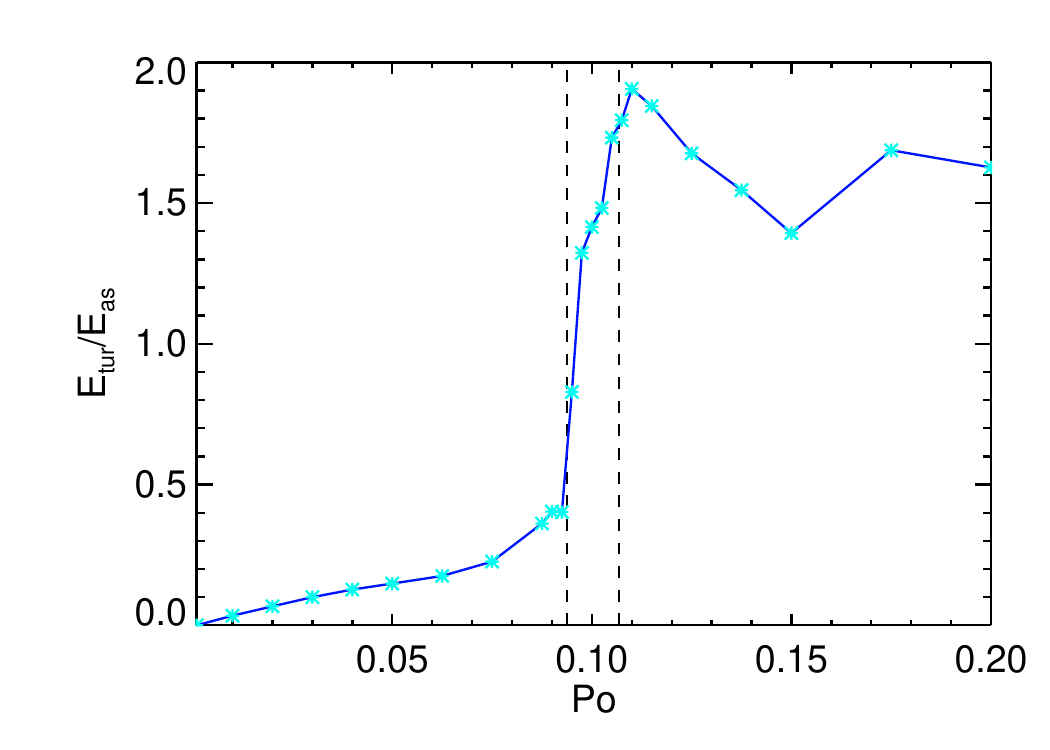}\label{fig::ener_tur_b}} 
 \caption{\label{fig::ener_tur}
\raggedright 
(a) Time-averaged values of the relation of energy of the
non-axisymmetric contributions with $m>1$ to the directly
driven contributions with $m=1$.
(b) Time-averaged values of the turbulent energy in relation to the
energy of the axisymmetric flow showing the transition from
a regime dominated by rotation to a regime dominated by
non-rotating turbulence (in the precession frame of reference).
}
\end{figure*}

Figure~\ref{fig::ener_tur_a} summarizes the changes in the flow state
on the basis of the ratio of the energy of higher Fourier modes
(i.e. with $m>1$) to the energy of the directly driven component
($m=1$) in a logarithmic representation of the dependence on
${\rm{Po}}$. In that representation we can clearly see the resonant
triadic instability on the left side briefly above ${\rm{Po\approx
    0.001}}$ and the multi-step transition to the supercritical state
on the right-hand side characterized by an increasing contribution of
the higher $m$ modes, which occurs in at least two different stages.

We have determined the degree of turbulence by integrating the
quadratic deviations of the flow from its (temporal) average
\begin{equation}
E_{\rm{tur}}=\frac{1}{2}\int U_{\rm{rms}}^2 {\rm{d}}V = \frac{1}{2\pi H R^2 (t_1-t_0)}
  \int\limits_0^{2\pi}\!{\rm{d}}\varphi\!\int\limits_0^1\!{\rm{d}}r\!\!\int\limits_0^H\!{\rm{d}}z
  \!\!\int\limits_{t_0}^{t_1}\!\left(u(\vec{r},t)-\left<u(\vec{r})\right>\right)^2{\rm{d}}t.
  \label{eq:def_urms}
\end{equation}
The result in relation to the energy of the axially symmetric flow in
figure~\ref{fig::ener_tur_b}  shows no, or only a moderate
degree of, turbulence 
before the transition, i.e. in this region the flow is dominated by
rotation.  Within the transition region, the energy of the
turbulent fluctuations increases sharply and the system transitions
from a state dominated by rotation into a state dominated by
turbulence (note that the
steep increase is also caused by the strong deceleration of the
rotational motion, as discussed in \citet{pizzi2021a}).

%%%%%%%%%%%%%%%%%%%%%%%%%%%%%%%%%%%%%%%%%%%%%%%%%%%%%%%%%%%%%%%%%%%%%%%%%%%%%%

\section{Nonlinear evolution of the directly forced flow}
\label{sec::nonlin_amp}

A weakly nonlinear theory for the amplitude of the directly driven
mode was developed in \citet{meunier2008} and later extended to the
case of the triadic instability \citep{lagrange2011}.  Recently,
\citet{gao2021} presented an improved nonlinear model that describes
the generation of the axisymmetric circulation (zonal flow) and
includes nonlinear interactions of the inviscid mode with itself and
its viscous correction as well as the nonlinear interactions in the
boundary layers.  The results show a reasonable agreement with flow
structures at small-angle precession obtained from simulations
\citep{albrecht2021} and experiments \citep{meunier2008}.
Interestingly, in this model the terms that are specific to the
cylinder geometry cancel out, leaving only contributions that are
independent of the geometry.  This may possibly be an explanation for
the similarity to the behaviour in the precessing spheroid in large
parts of the subcritical state, as described by \citet{horimoto2018}
or \citet{komoda2019}. For more complex geometries, e.g. spherical
shells or for parameter ranges in which the approximation of
\citet{gao2021} is no longer relevant, this should not be the case.

\begin{figure*}%[t!]
  \subfloat[][]{\includegraphics[width=0.495\textwidth]{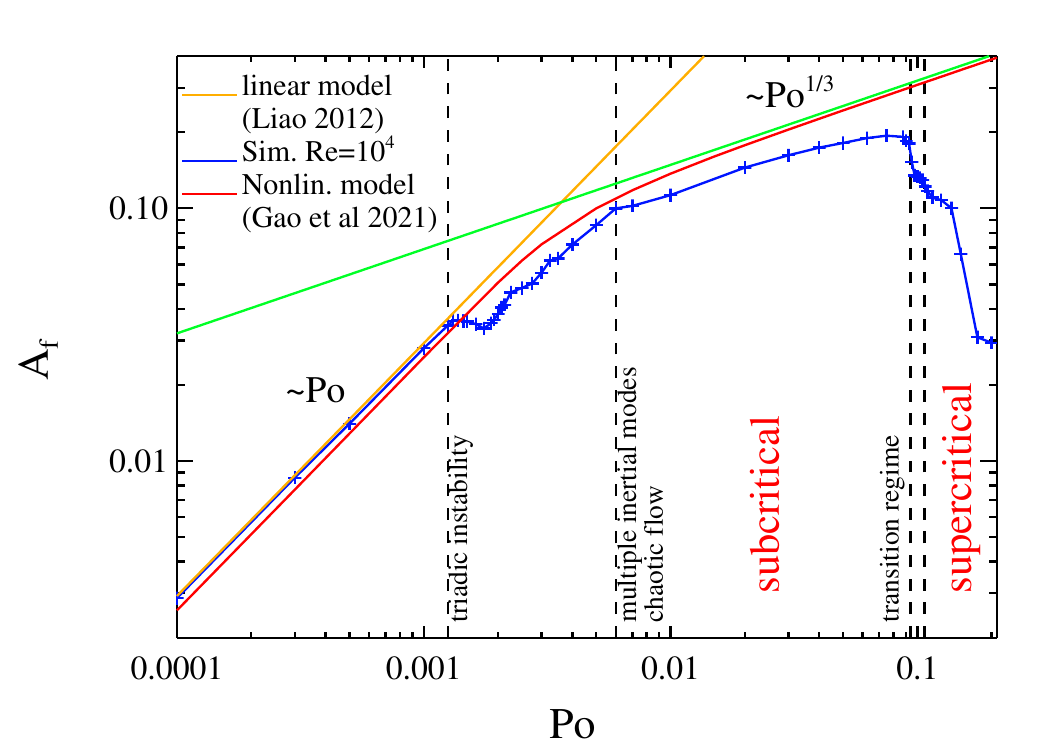}\label{fig::nonlin_evolution_amplitudes_a}}
  \quad
  \subfloat[][]{\includegraphics[width=0.495\textwidth]{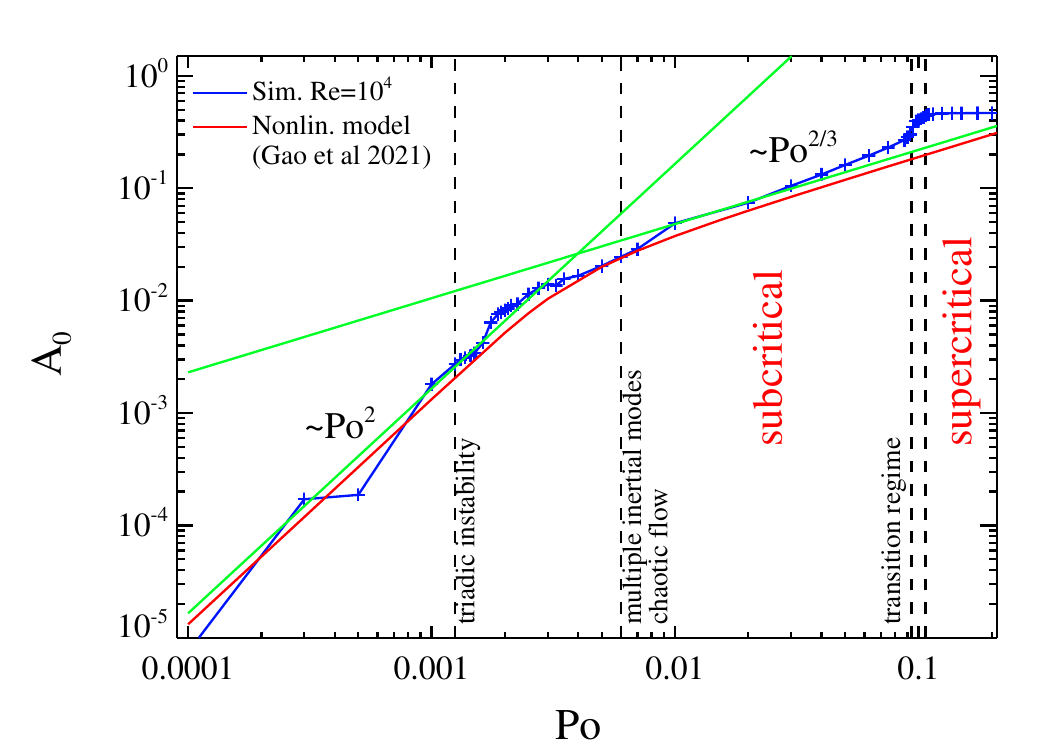}\label{fig::nonlin_evolution_amplitudes_b}}
\caption{\label{fig::nonlin_evolution_amplitudes}
\raggedright 
  Comparison of amplitudes from simulations (blue curves) and the
  nonlinear model of \citet{gao2021} (red curves) for fixed
  ${\rm{Re}}=10^4$. (a) Shows the 
  directly forced mode with $m=1, k=1$. The orange curve denotes the linear
  viscous solution of \citet{liao2012}, which behaves $\propto {\rm{Po}}$.
  The green lines illustrate the scaling
  $\propto {\rm{Po}}^{\nicefrac{1}{3}}$ in the subcritical regime. 
  (b) Shows the axisymmetric geostrophic mode with $m=0$ and $k=0$
  (i.e. the zonal flow in the co-rotating frame of reference).
  The green lines illustrate the scaling in ${\rm{Po}}$ below the
  transition to the supercritical regime.
  The vertical dashed lines in both plots indicate the onset of the
  triadic instability, the transition to 
  chaotic behaviour where no individual peaks for free inertial waves
  are discernible in the spectrum anymore, and the transition region
  before the supercritical regime as already marked in
  Figs.~\ref{fig::ener_vs_po_nonaxisym} and~\ref{fig::ener_tur} (from left to
  right). 
}
\end{figure*}

In the following, we use our simulation results to explore the
validity range of the nonlinear model.  The nonlinear model simplifies
the problem by considering only the axisymmetric geostrophic flow
($m=0, k=0$) and the directly forced mode ($m=1,k=1$), which finally
leads to a coupled system of two ordinary differential equations for
the amplitudes, in the following denoted by $A_{0}$ and $A_{{\rm{f}}}$.
The system is solved using the coefficients for the aspect ratio
$H/R=2$ given in \citet{gao2021} (see their Eqs.~(4.23) and
Tab.~1). In all cases, after a brief transient phase the solutions
converge to a steady state.  The corresponding amplitudes are shown in
figure~\ref{fig::nonlin_evolution_amplitudes} with the directly forced
mode in panel (a) and the axisymmetric geostrophic mode in panel
(b) in comparison with the amplitudes derived from the simulations
(blue curves) obtained by projection on the eigenfunctions of
inertial waves \citep{pizzi2021a}.

The numerical solutions basically exhibit three striking features,
allowing for the distinction of four different regimes, whereby not
all corresponding features are reflected in the nonlinear
model. Initially, we see a linear evolution, where the directly driven
mode scales $\propto {\rm{Po}}$.  A slight deviation between the
nonlinear model and numerical solutions occurs with the onset of the
triadic instability just above ${\rm{Po}}\approx 0.001$.  Due to the
coupling with the free Kelvin modes, a slight dip occurs here in the
simulations, which is not captured by the nonlinear model that
neglects higher-order contributions.  A second transition occurs in
the range of ${\rm{Po}}\approx 0.006$.  Here, the amplitude scaling
changes, transitioning into a behaviour $\propto
{\rm{Po}}^{\nicefrac{1}{3}}$.  The transition range is relatively
broad, and the behaviour is evident in both simulations and the
nonlinear model.  In the further course, the amplitude of the
simulated flow remains below the nonlinear model, which is easy to
understand, since the simulations also involve modes with higher $m$
and $k$ originating in the directly driven mode.  Subsequently, we
refer to these three regimes, where simulations and the nonlinear
model essentially coincide (except for minor deviations caused by the
triadic instability), as the subcritical state.  In these regimes we
have $A_0\propto A_{\rm{f}}^2$, for which, in principle, three different
interactions are responsible: the nonlinear self-interaction of the
directly driven mode, the nonlinear interaction of the inviscid forced
mode and the viscous correction and the nonlinear interactions in the
boundary layers at the end caps.  However, numerical simulations with
free-slip boundary conditions at $z=\pm H/2$ (so that no boundary
layers develop at the end caps) showed that there is nearly no change
regarding the transition and the occurrence of the double-roll mode
\citep{pizziphd}.  This suggests that the main contributions are
caused by the first two interaction terms.  Interestingly,
$A_{\rm{f}}$ exhibits a slightly lower slope in the simulation data,
while the behaviour for $A_0$ is reversed. We attribute this to the
contribution of higher modes, which draw energy from the directly
forced mode, and ultimately, through an inverse cascade, provide
additional contributions to the zonal flow similar to the behaviour
in the local box model by \citet{pizzi2022}.  Around a
Poincar{\'e} number of ${\rm{Po}}\approx 0.1$, a multi-stage
transition follows, culminating in the supercritical state.  However,
the collapse of the directly driven flow is not reflected by the
nonlinear model so that its validity ceases in this regime.
Furthermore, the double-roll mode that is merely perceptible in the
transition region is not included in the nonlinear model, which only
considers the axisymmetric azimuthal flow component, whereas the
double-roll mode is related to the poloidal components
$u_r\hat{\vec{r}}$ and $u_z\hat{\vec{z}}$.  Therefore, the transition
and the further evolution in the supercritical region must have a
cause that is beyond the effects considered in \citeauthor{gao2021}'s
model.  In our very first studies, we had proposed that the transition
and the emergence of the double-roll mode are associated with a
centrifugal instability \citep{giesecke2018,giesecke2019}.  However,
recent simulations indicate that this assessment is not entirely
correct, as the violation of the Rayleigh criterion necessary for the
occurrence of a centrifugal instability only occurs for ${\rm{Po}}
\gtrsim 0.1075$ \citep{pizzi2021b}, which is roughly $15\%$ beyond the
value of ${\rm{Po}}$ at which the collapse of the directly forced mode
and the appearance of the double-roll mode take place.  In any case, a
complete understanding would require the consideration of the
influence of the entire large-scale flow including the
non-axisymmetric, i.e. the directly forced, part.

Other possibilities for an instability mechanism could be the
elliptical instability \citep{kerswell1993}, enabled due to the tilt of
the fluid rotation axis with respect to the rotation axis of the
cylinder (see figure~\ref{fig::fluid_rotaxis}), or an instability
within turbulent boundary layers.  The occurrence of an elliptical
instability could be supported by the evolution of the $m=2$ mode,
which, unlike the other contributions, exhibits a relatively steep
increase in the pre-transition region (see
figure~\ref{fig::ener_vs_po_nonaxisym_c}).  The occurrence of a
boundary layer instability, on the other hand, might be supported by
the discovery of irregular turbulent bursts originating in the corners
of a precessing cylinder \citep{marques2015}.  Such an instability has
also been suspected in \cite{pizzi2021a}, but only at a significantly
larger Reynolds number than achievable in our simulations. This is
also in line with the results of \cite{buffett2021}, who found that an
instability for oscillating boundary layers (as generated by
precession) occurs at significantly higher Reynolds numbers compared
with the case of axisymmetric boundary layers in conventionally rotating
flows. This must be mentioned with some restriction, because at the
transition the nature of the boundary layers changes from oscillating
to axially symmetric \citep{pizzi2021a}, so that further
investigations are required in order to disentangle the mutual
relationship between boundary layers and internal flow.

%%%%%%%%%%%%%%%%%%%%%%%%%%%%%%%%%%%%%%%%%%%%%%%%%%%%%%%%%%%%%%%%%%%%%%%%%%%%%%%%%%%

\section{Experimental scaling for increasing rotation}
\label{sec::06_scaling_with_rotation}

Despite the rapid development of algorithms and advances in their
implementation on high performance computers in recent years and
decades, it is still not possible to carry out simulations of a
precession-driven flow in our set-up with a Reynolds number much larger
than ${\rm{Re}}\sim 10^4$.  In order to scale our models to more
realistic rotation rates and thus larger Reynolds numbers we resort to
experimental investigations on a water test experiment which has been
in operation at HZDR for several years
\citep{herault2015,herault2019,giesecke2018,kumar2023}.
\begin{figure*}
\subfloat[][]{\includegraphics[width=0.465\textwidth]{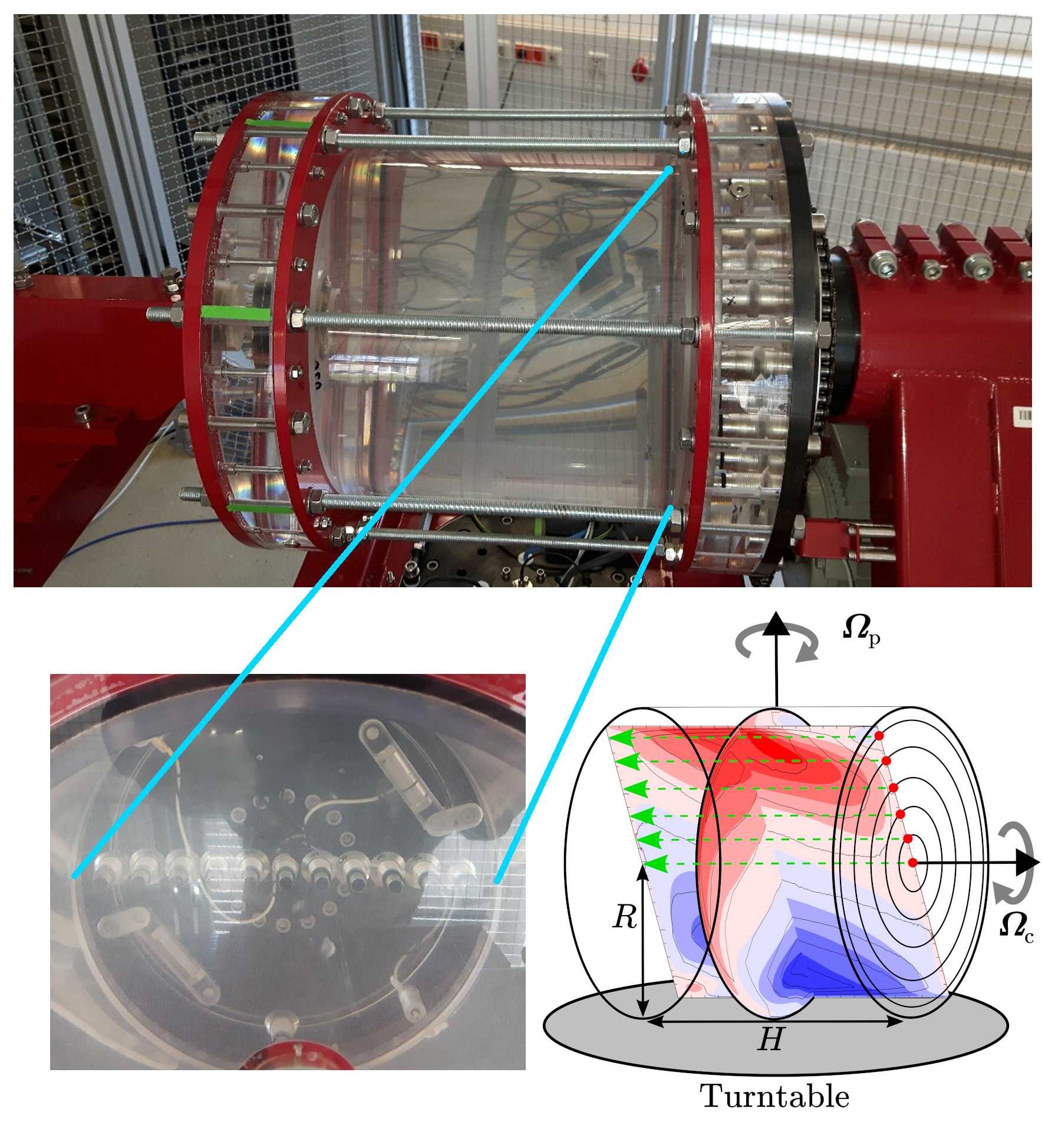}\label{fig::experiment_a}}
\subfloat[][]{\includegraphics[width=0.49\textwidth]{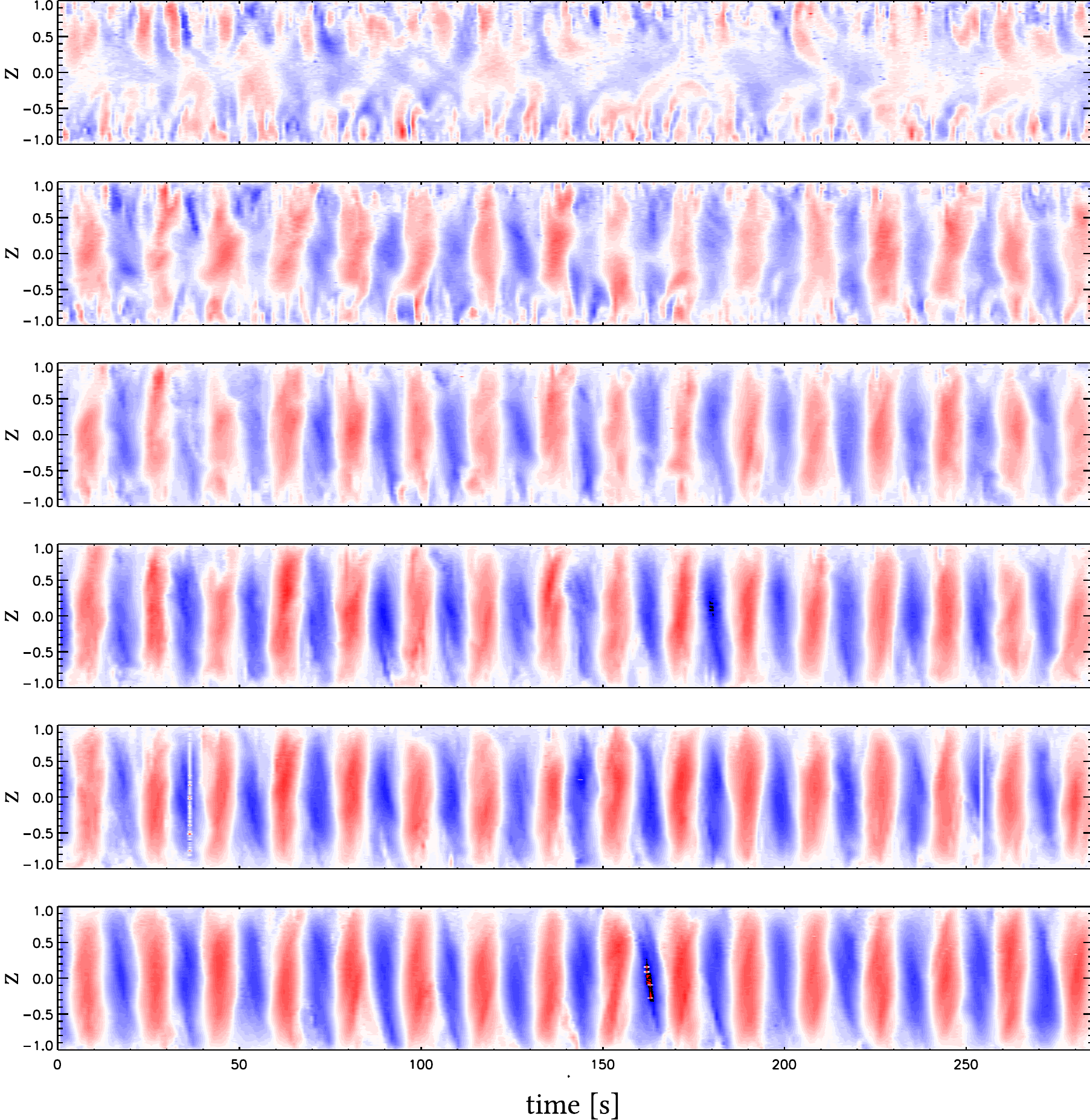}\label{fig::experiment_b}}
\caption{\label{fig::experiment}
\raggedright 
  (a) Picture of the water experiment consisting of a transparent
  cylinder filled with water, and fastening and safety brackets to fix the
  end caps.
  The bearing for the transmission of the drive for
  the rotation can be seen on the right side of the device.
  The lower panels of the left part show the UDV probes mounted on one endcap and
  a sketch that illustrates the propagation of the ultrasonic beams
  when performing a flow measurement with six radially mounted UDV
  probes. 
  (b) Temporal evolution of the axial flow component
  $u_z(z,t)$ as measured along the axis with six UDV probes mounted at
  $r=0, 30, 60, 90, 120$ and $150\mbox{ mm}$ (from top to bottom) for
  a run at ${\rm{Re}}=10^4$ and ${\rm{Po}}=0.05$.}
\end{figure*}
The experiment consists of a plexiglass cylinder with height
$H=0.326\mbox{ m}$ and radius $R=0.163\mbox{ m}$ so that the aspect
ratio is $\Gamma=H/R=2$ (figure~\ref{fig::experiment_a}).  The
cylinder is held by a frame structure that ensures the stability and
tightness of the system by means of solid screw clamps that run
parallel to the sidewalls.  The rotation of the container is driven
by a chain through a motor with a maximum power of \SI{3600}{\watt}
that is controlled by three power sources.  The turntable is driven by
a second motor with a maximum power of \SI{2200}{\watt}, which is
controlled individually so that the precession and the rotation
frequency can be adjusted independently with the rotation frequency in
the range of $0~\si{\hertz} < f_{\rm{c}} < 10~\si{\hertz}$ and the
precession frequency in the range of $0~\si{\hertz} < f_{\rm{p}} <
1~\si{\hertz}$.  Thus, Reynolds numbers
${\rm{Re}}=\varOmega_{\rm{c}}R^2/\nu$ with an order of magnitude up to
$2\times 10^6$ are achievable with distilled water with an assumed
viscosity of $\nu=10^{-6}\mbox{ m$^2$/s}$.  Similar to the numerical
models we fix the angle between precession axis and rotation axis to
$\alpha=90^{\circ}$ (other angles are discussed in
\citet{kumar2023}). We analyse the flow by means of velocity
measurements using ultrasonic Doppler velocimetry (UDV), which
provides spatially and temporally well-resolved profiles of the axial
velocity along an ultrasonic beam parallel to the rotation axis (see
lower right panel in figure~\ref{fig::experiment_a}).  Similar
measurements allowed us to identify axisymmetric inertial waves in a
rotating liquid metal driven by a Lorentz force that results from a
rotating external magnetic field \citep{vogt2014}.  In the precession
experiment, spatially resolved velocity profiles can be obtained up to
a Reynolds number of ${\rm{Re=10^5}}$.  A more detailed description of
the measurement system can be found in \citet{giesecke2019} and
\citet{kumar2023}.  Here, we show results from new measurement
campaigns that resolve the radial structure of the axial velocity
component as well as the behaviour of the critical Poincar{\'e} number for
high rotational frequencies.

\begin{figure*}
  \includegraphics[width=0.98\textwidth]{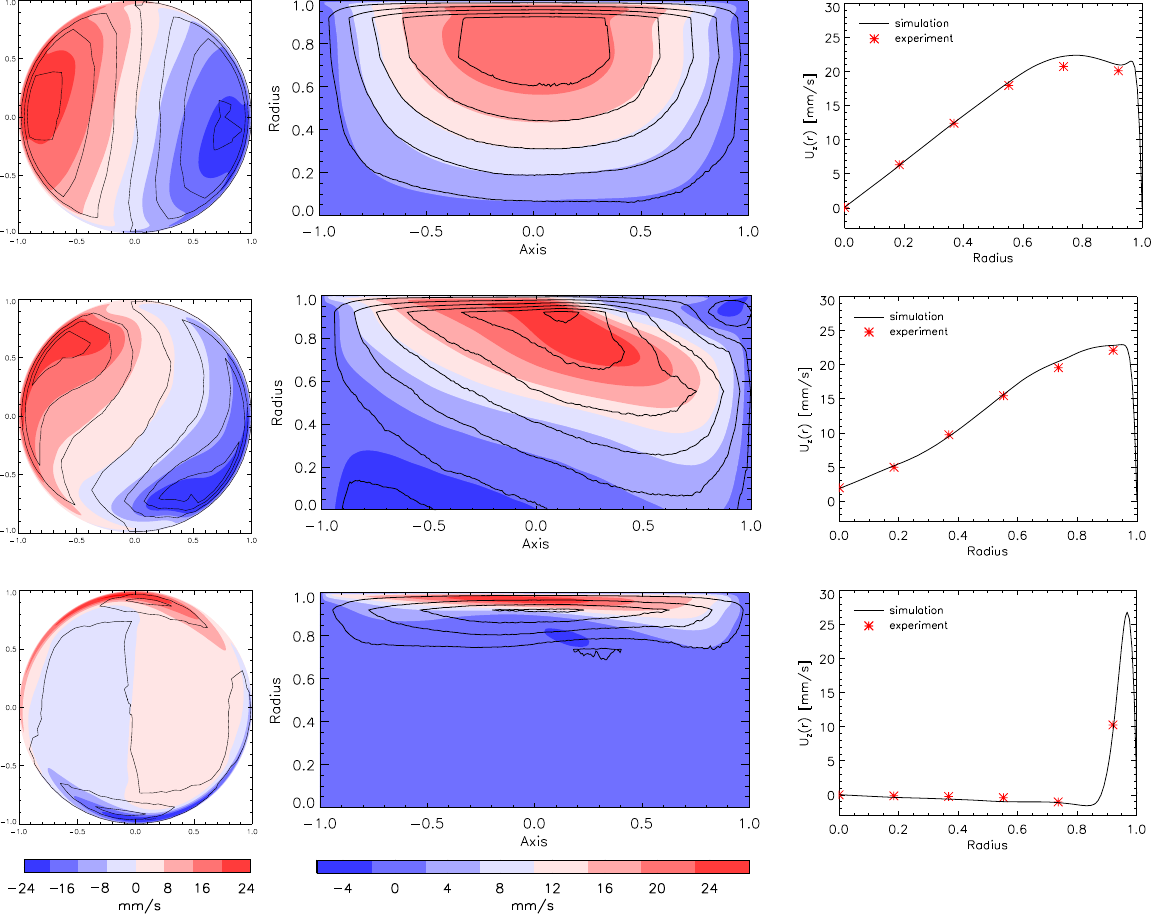}
\caption{\label{fig::mean_flow_structure_experiment}
\raggedright 
  Left and central column: Structure of the time-averaged axial flow
  $u_z$ in the equatorial
  plane (left) and in a meridional plane where the flow amplitude is
  maximum (center). The colour coded structures present
  the results of the simulations and the solid black contour lines show the
  results from the experiment. The black contour lines
  follow the same level scheme as the coloured structures.
  Right column:
  radial profile of the axial flow component $u_z$ in the equatorial
  plane at the angle where the axial flow is maximum. The solid line
  shows the results from the simulations and the stars show the
  results obtained from phase averaging of the UDV measurements. From
  top to bottom: ${\rm{Po}}=0.05, 0.1, 0.2$, ${\rm{Re}}=10^4$.}
\end{figure*}
At lower Reynolds number, such as ${\rm{Re}}=10^4$, it is possible to
perform (nearly) simultaneous measurements with six UDV probes with
the help of a multiplexer circuit, so that a coarsely resolved radial
profile of the axial velocity can be measured.  A typical time series
is shown in figure~\ref{fig::experiment_b}, which presents the axial
velocity measured simultaneously at the six probes mounted at the same
azimuthal angle but at different radii $r=0,30,60,90,120,150~\si{mm}$
(from top to bottom).  For the post-processing we divide the measured
time series into individual chunks, each covering exactly one rotation
period and the time average of the axial component is calculated by
superimposing these sections.
Figure~\ref{fig::mean_flow_structure_experiment} shows the results in
comparison with the time-averaged data from the simulations.  The
colour-coded contours in the left and the central column of
figure~\ref{fig::mean_flow_structure_experiment} denote the results
from the simulations while the superimposed black contour lines show
the data from the measurements, with the individual colour gradations
corresponding to the same values of the contour lines.  For all three
Poincar{\'e} numbers we see a good agreement between simulation and
experiment both in structure and amplitude. This is additionally
confirmed in detail by the radial profiles of $u_z$ shown in the
right-hand column of figure~\ref{fig::mean_flow_structure_experiment},
which illustrate the increasing concentration of the flow close to the
outer wall for large ${\rm{Po}}$.

To determine the amplitudes of the double-roll mode (and thus to
evaluate its occurrence or disappearance), we use the fact that the
measurement with UDV is carried out in the co-rotating system, and
thus the non-axisymmetric inertial modes do not make any contribution
due to their periodicity in $\varphi$ if a sufficiently long time
averaging is applied.  The time average of $u_z$ therefore corresponds
directly to the axially symmetrical component provided that there is
no pulsation of this contribution (which, as we know from simulations,
does not exist).
\begin{figure}
   \subfloat[][]{\includegraphics[width=0.485\textwidth]{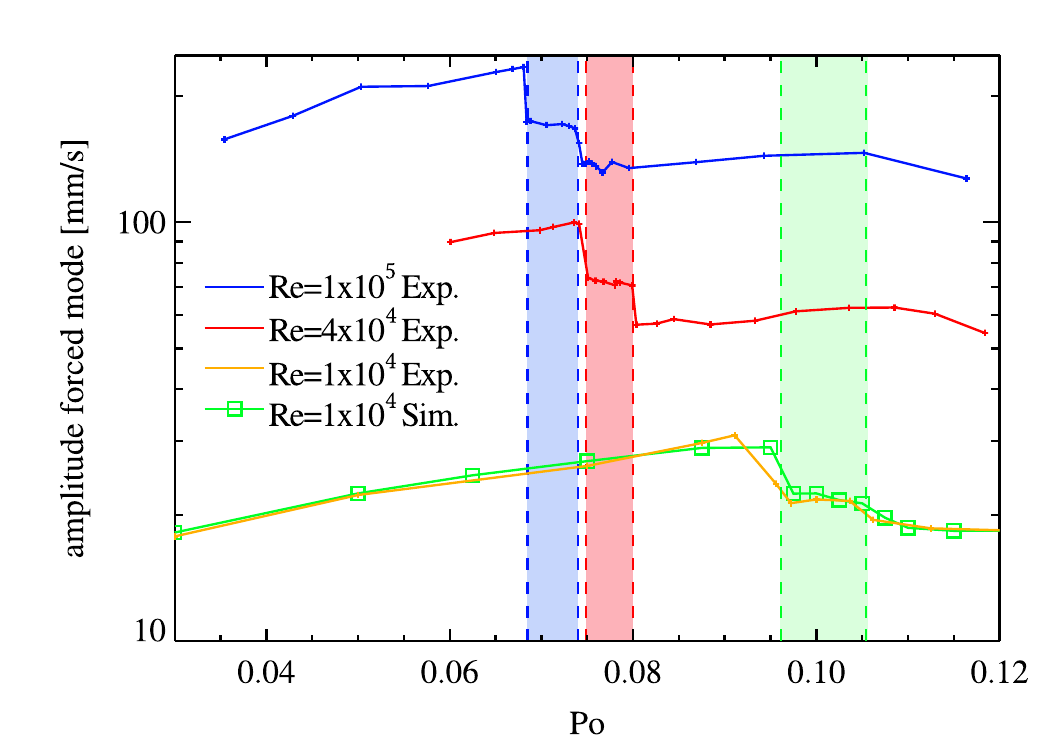}\label{fig::amplitudes_exp_vs_rey_a}}
   \subfloat[][]{\includegraphics[width=0.485\textwidth]{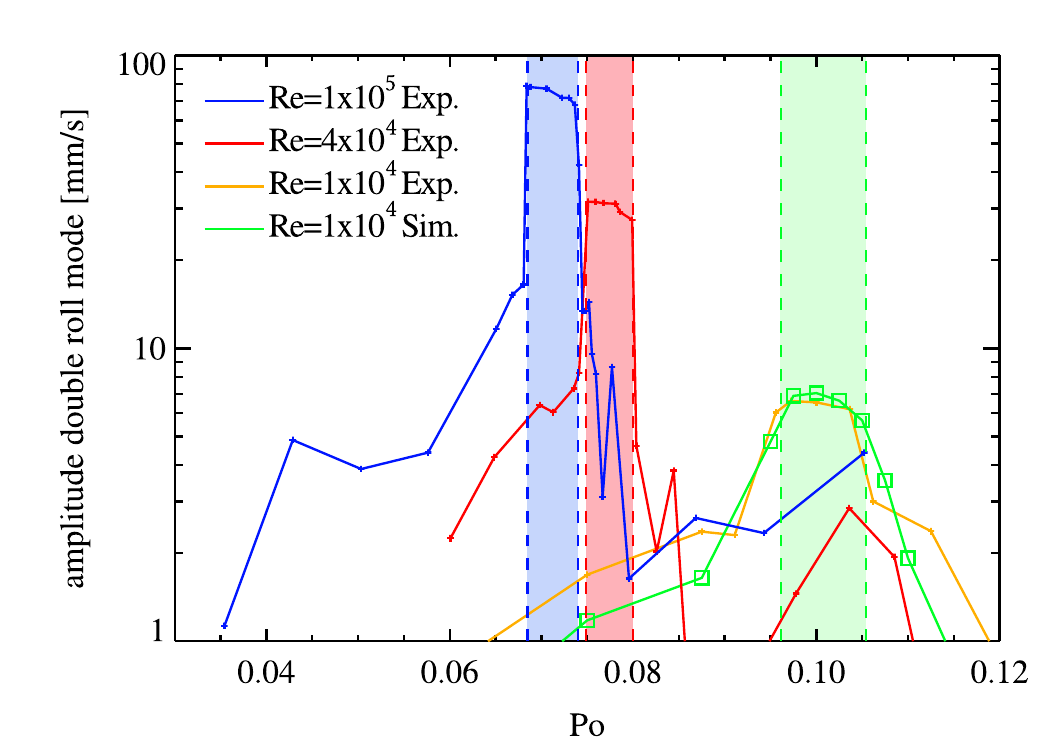}\label{fig::amplitudes_exp_vs_rey_b}}
  \caption{\label{fig::amplitudes_exp_vs_rey}
\raggedright 
(a) Amplitudes of the directly forced mode with $m=1$
and $k=1$ versus forcing
parameter ${\rm{Po}}$ for various Reynolds numbers. The
transitional regime is marked by the vertical dashed lines.
(b) Same as left, but for the double-roll mode with
$m=0$ and $k=2$ (data taken from \citet{giesecke2018,giesecke2019}).
  }
\end{figure}
The corresponding amplitudes, calculated as described in
Appendix~\ref{appendix::experimental_amplitudes}, are similar to
those presented in \cite{giesecke2018} and \cite{giesecke2019},
which are reproduced in figure~\ref{fig::amplitudes_exp_vs_rey}
for the cases ${\rm{Re}}=10^4,4\times 10^4$ and $10^5$ for the two
most interesting modes $\tilde{\tilde{u}}_{km}$ with $(m,k)=(1,1)$ and
$(m,k)=(0,2)$.  Qualitatively, the curves show a similar shape as the
kinetic energy in figure~\ref{fig::ener_vs_po_nonaxisym_b}
and~\ref{fig::ener_vs_po_nonaxisym_d}.  We see an increase of the
forced mode up to a critical point at which the amplitude collapses
and then remains on an intermediate plateau for a small range before
another, smaller collapse occurs.  At the same time, we see an interim
appearance of the double-roll mode in the plateau regime. Especially
for the larger Reynolds numbers, the appearance of the double-roll
structure is sharply delineated from the remaining part of the
parameter range and when comparing the critical thresholds as
indicated by the vertical dashed lines, we see that the regime of the
plateau between the first and second drops of the directly driven mode
correspond exactly to the appearance and disappearance of the
double-roll mode.

\begin{figure}%[t!]
  \subfloat[][]{\includegraphics[width=0.485\textwidth]{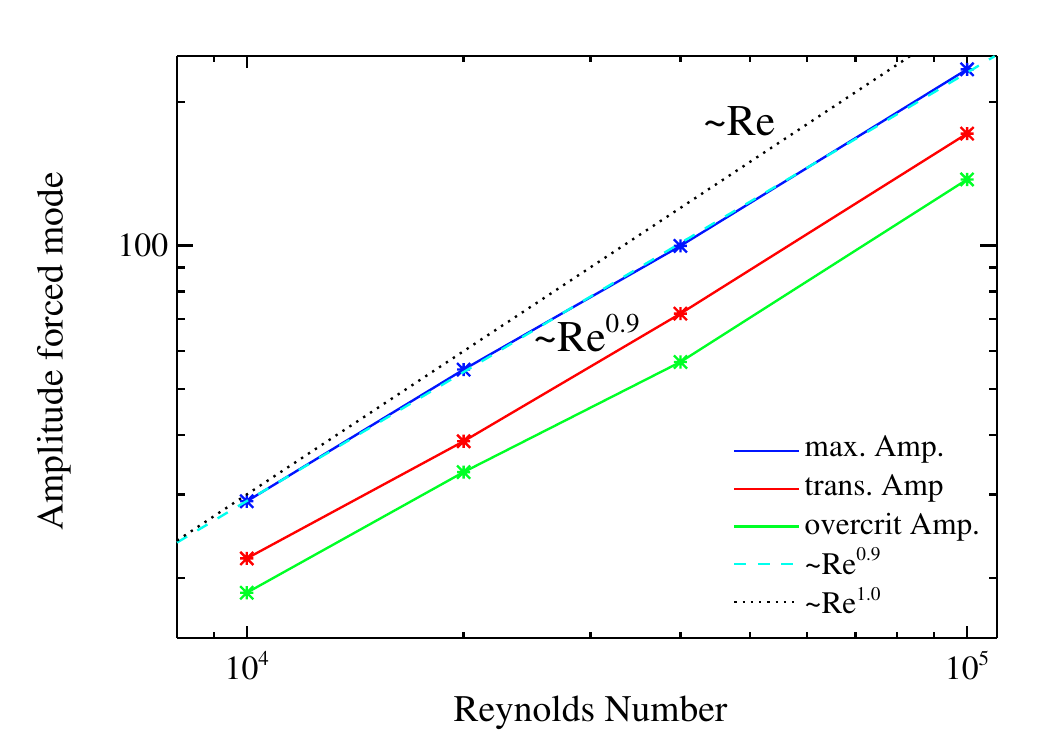}\label{fig::scaling_amplitudes_a}}
  \quad
  \subfloat[][]{\includegraphics[width=0.485\textwidth]{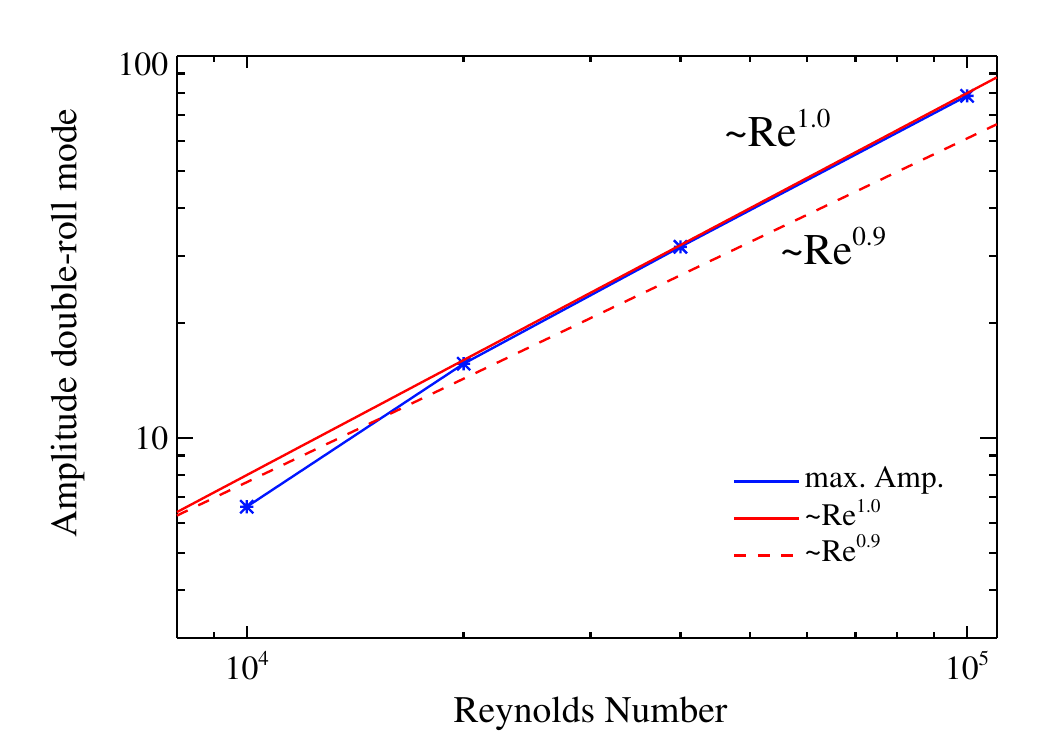}\label{fig::scaling_amplitudes_b}}
  \caption{\label{fig::scaling_amplitudes}
    \raggedright 
    (a) Scaling of
    the (measured) amplitudes (maximum, transition region and
    overcritical regime) for the directly forced mode in dependence
    of ${\rm{Re}}$.
    (b) Same as left, but for the double-role mode.
  }
\end{figure}

Figure~\ref{fig::scaling_amplitudes} supplements these results and
shows the scaling of the amplitudes in the
different regimes as a function of ${\rm{Re}}$.
The amplitudes of the directly forced mode, taken within the three
different regimes, are shown in
Figure~\ref{fig::scaling_amplitudes_a}, which compares the amplitude
of the directly forced mode taken at the maximum (i.e. at the critical
threshold), an average of the amplitudes obtained in the plateau
regime and the average of the amplitudes in the overcritical regime
(i.e. after the second collapse). We see that all three amplitudes
share the same scaling behaviour which is $\propto {\rm{Re}}^{0.9}$,
which can clearly be distinguished from a scaling linear in
${\rm{Re}}$ indicated by the black dotted curve.  Saturation by
viscous damping in the bulk would be expected to result in a purely
linear scaling $\propto {\rm{Re}}$. To confirm the deviation observed
by us in a robust manner, measurements at significantly larger
${\rm{Re}}$ are therefore still required.

Figure~\ref{fig::scaling_amplitudes_b} shows the related 
scaling for the double-roll mode, where we restrict the analysis to
the respective maximum amplitude.
It is striking that this amplitude follows a linear scaling
$\propto {\rm{Re}}$, so that the
relative weighting of the double-roll mode should become more important
as ${\rm{Re}}$ increases.
Unfortunately, reliable measurements with UDV are not possible above
${\rm{Re}}\sim 10^5$
due to the excessive flow velocity so that it cannot be ruled out that a change
in the scaling behaviour occurs for ${\rm{Re}}>10^5$. This applies in
particular to the double-roll mode, the occurrence of which is
associated with exceeding the critical ${\rm{Po}}$. This threshold
decreases with increasing Reynolds number, and we use the occurrence
and disappearance of the double-roll mode as a proxy to identify the
transition of the flow state.  This feature can still be identified
with the help of the UDV 
probes for Reynolds numbers up to ${\rm{Re}}=2\times 10^6$
without relying on a spatially resolved measurement of the axial profile.
\begin{figure}%[h!]
  \centerline{\includegraphics[width=0.66\textwidth]{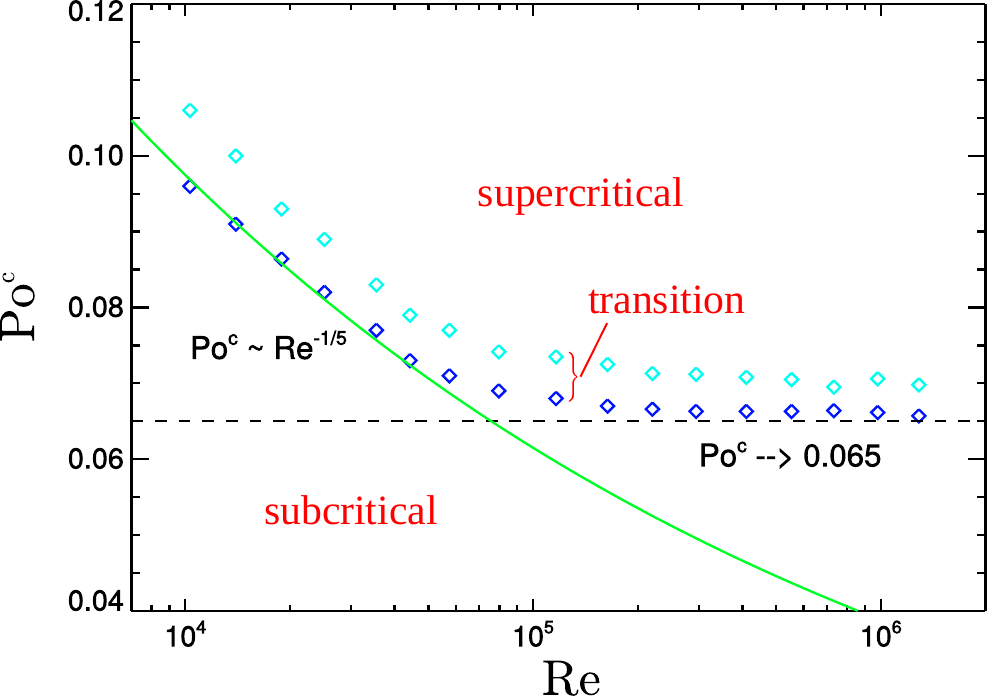}}
  \caption{\label{fig::scaling_transition}
    \raggedright 
    State diagram showing the
    critical thresholds for appearance and disappearance of the double-roll mode in dependence on the Reynolds number.}
\end{figure}
Figure~\ref{fig::scaling_transition} shows that the critical threshold
for the occurrence of the double-roll mode follows a scaling $\propto
{\rm{Re}}^{-\nicefrac{1}{5}}$ for ${\rm{Re}}\lesssim 10^5$. A similar
scaling arises for the second threshold that characterizes the
disappearance of the double roll mode.  Above a Reynolds number of
approximately ${\rm{Re}}\approx 10^5$ we see a transition of the
behaviour and the critical thresholds follow an asymptotic regime with
${\rm{Po}}^{\rm{c}}\rightarrow 0.065$.  There are two further
observations that are worth emphasizing.  Firstly, the width of
the transition area, i.e. the regime between the emergence and
disappearance of the double-roll mode, is only slightly dependent on
${\rm{Re}}$.  The second interesting fact is the similarity of the
regime with ${\rm{Re}}\gtrsim 10^5$ to the measurements in a
spheroid with small ellipticity carried out by
\citet{horimoto2018}. These measurements reveal a bistable system in a
small range of ${\rm{Po}}$ with the possibility of a quiescent
turbulent state and a state of developed turbulence.  The authors
also observed that the range for the bistable state and thus the
transition to a developed turbulent state only weakly depends on
${\rm{Re}}$ but did not observe a particular scaling $\propto
{\rm{Re}}^{\beta}$ for small ${\rm{Re}}$ as we found for the cylinder.
This suggests that, for the cylinder, the scaling in the range
${\rm{Re}} \lesssim 10^5$ may be caused by the Ekman layers at the end
caps, for which the extrapolation of simulation data indicates the
possibility of the transition to turbulent boundary layers at approximately
${\rm{Re}}\approx 10^5$ \citep{pizzi2021a}.

%%%%%%%%%%%%%%%%%%%%%%%%%%%%%%%%%%%%%%%%%%%%%%%%%%%%%%%%%%%%%%%%%%%%%%%%%%%%%%%

\section{Summary and Conclusion}
\label{sec::conclusions}

We have examined the fluid flow in a precessing cylinder using
numerical simulations and UDV measurements in a water experiment.  We
have limited our investigations to the case with maximum precession
angle, since in this case the volume force caused by precession is
also maximum.  In contrast to earlier investigations performed at
small nutation angles that show multiple dynamical features (e.g. flow
breakdowns or quasi-periodic bursts,
\citet{manasseh1992,manasseh1994,manasseh1996}), in our study a unique
characterization of the flow state is possible using the Poincar{\'e}
number ${\rm{Po}}$.

The volume force caused by the precession directly drives an inertial
mode whose amplitude can be calculated from linear theory. In the
present case, the configuration is resonant, and the calculation of
the amplitude must take into account viscosity in the boundary layers
\citep{gans1970,meunier2008,liao2012}.  This directly forced flow is
unstable \citep{kerswell1993} and becomes time dependent in terms of a
triadic resonance when a critical forcing is exceeded
\citep{kerswell1999}.  The emergence of a triadic resonance and the
transition from a laminar flow into a chaotic flow can be understood
in terms of a bifurcation of a rotating wave into a modulated rotating
wave \citep{garcia2021}.  In the case of precession the symmetry of the
system is already broken right from the beginning due to the
non-axisymmetric forcing, which differs for example from the
axisymmetric base state in the spherical Couette set-up that allows the
occurrence of an additional frequency because of the need for the
breaking of the axisymmetric symmetry before becoming chaotic
\citep{garcia2019, garcia2021}.  Despite the onset of the triadic
instability we still see a rather regular behaviour when further
increasing the forcing with a monotonically increasing flow amplitude
up to a first threshold, which initiates the transition to the
supercritical state.  Before the transition the directly forced flow
amplitude scales proportional to the cube root of the forcing
($\propto {\rm{Po}}^{\nicefrac{1}{3}}$) and the kinetic energy reaches
more than one third of the energy of the initial flow prescribed by a
rotation around the axis of the cylinder with
$\vec{u}=r\varOmega_{\rm{c}}\hat{\vec{\varphi}}$.

There is no unified nonlinear model that describes both the triadic
instability and the evolution of the two main modes (forced mode and
zonal flow).  However, since the energetic contribution of the free
Kelvin modes and the resulting chaotic contribution remain small over
a large range of parameters, this is not relevant, at least until the
transition to the supercritical regime.  Before this transition, a
nonlinear model based on viscous effects and two modes, the directly
driven Kelvin mode and the geostrophic axisymmetric circulation, is
capable of describing the behaviour of the precession-driven flow
\citep{gao2021}.  We conclude that the modifications of the flow
structure in the subcritical regime are the results of the nonlinear
interaction of the forced (inviscid) mode with itself and its viscous
correction and/or the corresponding interactions in the boundary
layers.  However, since simulations with free-slip boundary conditions
at the end caps show a rather similar behaviour, we believe that the
contributions within the Ekman layers are less important
\citep{pizziphd}.  Interestingly, Gao's system of equations does not
seem to allow two solutions, as they occur in the Busse model for
sufficiently large ellipticity.  The possibility of two solutions is
also observed in simulations and experiments, and it is known that the
abrupt change between the two states is associated with hysteresis
\citep{herault2015}.  This transition takes place in two consecutive
steps above a critical amplitude of the directly forced flow.  We show
that the remarkable change seen in the kinetic energy around
${\rm{Po}}\approx 0.1$ goes along with a significant modification of
the structure of the flow as well as with a decrease in the forced
mode's amplitude, the emergence of the double-roll mode, a sudden jump
of the orientation of the fluid rotation axis, an increased level of
turbulence, and a sudden stop of the rotational motion of the bulk fluid,
none of which are represented in the known nonlinear models.
Therefore, Gao's nonlinear model still seems incomplete, either
because certain interactions have not been taken into account, or
because higher-order contributions are necessary.  With regard to this
issue, we emphasize that the emergence of the double-roll mode is not
affected by Greenspan's theorem \citep{greenspan1969}, which forbids
nonlinear first-order interactions for axially symmetric, geostrophic
contributions on the long time scale with an order given by the
strength of the nonlinear terms.  Apart from the triadic instability
and the axisymmetric geostrophic mode, the double-roll mode is the
only clearly noticeable regular large-scale flow pattern in the
simulations with nutation angle $\alpha=90^{\circ}$ and at aspect
ratio $H/R=2$.

In other configurations, however, the spontaneous formation of
large-scale vortices has been observed in experiments of a
precession-driven flow at a different aspect ratio
\citep{mouhali2012}. In 
a spherical geometry, such vortices were associated with the conical
shear layer instability (CSI, \citet{lin2015}), which is known to be
beneficial for dynamo action \citep{lin2016}.  In a sphere, the conical
shear layers are spawned in the boundary layers at critical latitudes,
which do not exist in a cylindrical geometry.  In our simulations,
there is a slight indication of the emergence of such vortices around
${\rm{Po}}\approx 0.03$.  However, these vortices remain weak and
short lived (less than half a rotation period).  One could assume that
the formation of the vortices is triggered by wave beams originating
in the corners of the cylinder, as described by \citet{marques2015}.
The corners at the end caps of the cylinder would thus play a similar
role as the critical latitudes for the CSI. In fact, we can identify
such wave beams in the form of vectored fluctuations in the vicinity
of the corners.  The wave beams are clearly visible only in the
supercritical regime, while we see the formation of weak vortices in
the laminar region so that there is probably no connection between the
two phenomena.  The same applies to the scaling of the critical value
of ${\rm{Po}}$ as we observe it in the experiment.  Interestingly,
this particular scaling is not found in the scaling of the amplitudes
(in all regimes), whose behaviour follows a rather simple law with
respect to the rotation, although the observation of the deviation of
the forced mode from a scaling linear in ${\rm{Re}}$ might be
surprising.  However, in our measurements the data range for
determining the dependence on ${\rm{Re}}$ is limited, as it only
covers one order of magnitude, partly in a regime where the critical
Poincar{\'e} number for the transition is still decreasing. For more
robust statements, therefore, further measurements with larger
${\rm{Re}}$ are required, which, cannot be carried out with the
currently installed sensors.

Beyond the transition regime the energy of the axisymmetric flow
induced by precession reaches values up to $80 -- 90\%$ of a rigidly
rotating fluid. This large increase is essentially due to a
circulation that is opposite to the rotation of the container.  When
transferring to the precession frame of reference, this azimuthal flow
corresponds to a braking of the initial solid body rotation, giving
the impression that one observes a stagnant fluid with the container
rapidly rotating around it.  A significant fluid flow is observed only in
the vicinity of the sidewall, while being greatly weakened in the
interior, which results in a strong shear effect close to the sidewall
that should be helpful for dynamo action according to Landeau's 
criterion \citep{landeau2022}.

Finally, we emphasize that the axially symmetric component in the form
of a twofold roll as observed within this transition regime is
comparable to the mean poloidal flow in the VKS dynamo experiment
\citep{monchaux2007,giesecke2012a}.  It is well known that this type
of flow drives a dynamo at comparatively low values of the magnetic
Reynolds number \citep{dudley1989,ravelet2005,giesecke2012b}, which
might be of high interest with respect to the precession dynamo
experiment currently planned at Helmholtz-Zentrum Dresden-Rossendorf.
Indeed, kinematic simulations show that, in the range of the transition,
the possibility of achieving magnetic field self-excitation in the
planned dynamo experiment has an optimum
\citep{giesecke2018,giesecke2019, kumar2023}.  So far, the kinematic
models have been based on the assumption of dynamo action in the planned
experiment being driven by large-scale flow. However, this does not
need to be the only possibility. Recent self-consistent simulations
using the full set of magnetohydrodynamic equations and a geometric
set-up similar to the planned experiment showed small-scale dynamo
action with the magnetic energy saturating a rather low level
\citep{giesecke2024a}.  In these models, the regular dynamo state is
aperiodically interrupted by strong magnetic bursts (increasing the
magnetic energy by a factor of 3 to 5). The related velocity field is
also essentially small scale, and it might be that the randomly
excited inertial waves are responsible for the generation of magnetic
energy similar to the mechanism discussed by \citet{moffatt1970}
and \citet{soward1975}.

\backsection[Supplementary data]{\label{SupMat}Supplementary material
  and movies are available at \\https://doi.org/XX.XXXX/jfm.2024...} 

\backsection[Acknowledgements]{
AG would like to thank the Isaac Newton Institute for
Mathematical Sciences, Cambridge, for support and hospitality 
during the programme “Frontiers in dynamo theory: from the Earth to
the stars” where work on this paper was undertaken.}

\backsection[Funding]{This work benefited from support
through Project Nos. GR 967/7-1 and 
GI 1405/1-1 of the Deutsche Forschungsgesellschaft (DFG).
This work was supported by EPSRC grant no EP/K032208/1.
F. Garcia gratefully acknowledge the Serra H{\'u}nter and SGR
(2021-SGR-01045) programs of the Generalitat de Catalunya (Spain).} 

\backsection[Declaration of interests]{The authors report no conflict
  of interest.} 

\backsection[Data availability statement]{The data that support the
  findings of this study are available upon reasonable request.}
%  openly available in [repository name] at http://doi.org/[doi],
%  reference number [reference number].
%See JFM's
%\href{https://www.cambridge.org/core/journals/journal-of-fluid-mechanics/information/journal-policies/research-transparency}
%{research transparency policy} for more information}

\backsection[Author ORCIDs]{
  A. Giesecke, https://orcid.org/0000-0002-2009-3166;
  T. Vogt, https://orcid.org/0000-0002-0022-5758;
  T. Gundrum, https://orcid.org/0000-0002-5971-7431;
  V. Kumar, https://orcid.org/0000-0002-4662-4295;
  F. Stefani, https://orcid.org/0000-0002-8770-4080;
  F. Pizzi, https://orcid.org/0000-0003-2018-3185;
  F. Garcia, https://orcid.org/0000-0003-4507-0486
}

%\backsection[Author contributions]{Authors may include details of the contributions made by each author to the manuscript'}

%%%%%%%%%%%%%%%%%%%%%%%%%%%%%%%%%%%%%%%%%%%%%%%%%%%%%%%%%%%%%%%%%%%%%%%%%%%%%%%%%%%%%%%%%%%%%%%%%%%%%%%%%%%

\appendix

\section{Radial structure of inertial modes}
\label{appendix::00::poincare_equation}

We distinguish four different classes of inertial waves according to
the elementary geometric properties, i.e. axisymmetry ($m=0$) and
geostrophy ($k=0$).  These classes and the corresponding
representation ${\vec{u}}_{mkn}$ are listed in the following, whereby
our list additionally contains the corresponding dispersion relations,
which are required for the calculation of the radial wavenumbers
$\lambda_{mkn}$ and the frequencies $\omega_{mkn}$.  We use the
indices $m, k$ and $n$, which are the azimuthal wavenumber, the axial
wavenumber and the third index, $n$, that counts the solutions for the
parameter $\lambda$, which results from the dispersion relation that
provides a kind of a radial wavenumber.
\begin{enumerate}
\item
Axisymmetric geostrophic inertial modes ($m=0, k=0$):
\begin{eqnarray}
  {u}^r_{00n} & =  & 0\nonumber
  \\
  {u}^{\varphi}_{00n} & =  &
  J_{1}(\lambda_{00n}r)
  \label{eq::axi_geo}
  \\
  {u}^z_{00n} & =  & 0\nonumber
  \\
  J_1(\lambda_{00n}) & = & 0
  \nonumber\\
  \omega_{00n} & = & 0\nonumber
\end{eqnarray}
\item
Axisymmetric non-geostrophic mode ($m=0, k\neq 0$):
\begin{eqnarray}
  {u}^{r}_{0kn} & = &
  -i\frac{\omega_{0kn}\lambda_{0kn}\Gamma}{4-\omega_{0kn}^2}J_{1}(\lambda_{0kn}r)\sin\left(\frac{k\pi z}{H}\right)
  \nonumber
  \\
  {u}^{\varphi}_{0kn} & = &
  -\frac{\lambda_{0kn}\Gamma}{4-\omega_{0kn}^2} J_{1}(\lambda_{0kn}r)\sin\left(\frac{k\pi z}{H}\right)
  \label{eq::axi_nongeo}
  \\
  {u}^{z}_{0kn} & = & -i\frac{k\pi}{\omega_{0kn}}J_0(\lambda_{0kn}r)\cos\left(\frac{k\pi z}{H}\right)\nonumber
  \\
  J_1(\lambda_{0kn}) & = & 0 \nonumber \\
  \omega_{0kn} & = & \pm2\left[1+\left(\frac{\lambda_{0kn}H}{k\pi}\right)\right]^{-\nicefrac{1}{2}}\nonumber
\end{eqnarray}
\item
Non-axisymmetric geostrophic modes ($m\neq 0, k=0$):
\begin{eqnarray}
  {u}^r_{m0n} & = &
  -i\frac{mH}{2r} J_{m}(\lambda_{m0n}r)e^{im\varphi}\nonumber
  \\
  {u}^{\varphi}_{m0n} & = &
  \frac{H}{2}\left[\lambda_{m0n} J_{m-1}(\lambda_{m0n}r)
    -\frac{m}{r} J_{m}(\lambda_{m0n}r)\right]e^{im\varphi} 
  \label{eq::nonaxi_geo}
  \\
  {u}^z_{m0n} & = & 0\nonumber
  \\
  J_m(\lambda_{m0n}) & = & 0\nonumber
  \\
  \omega_{m0n} & = & 0\nonumber
\end{eqnarray}
\item
Non-axisymmetric non-geostrophic modes ($m\neq 0, k\neq 0$):
\begin{eqnarray}
{u}_{mkn}^r & = &\!\! 
\frac{-i}{4-\omega_{mkn}^2}\!\!
\left[\omega_{mkn}\lambda_{mkn} J_{m-1}(\lambda_{mkn} r)
+\frac{(2-\omega_{mkn})m}{r}J_m(\lambda_{mkn} r)\!\right]
\sin\left(\frac{(2k-1)\pi z}{H}\right)e^{im\varphi} 
\nonumber
\\ 
{u}^{\varphi}_{mkn} & = &\!\! 
\frac{1}{4-\omega_{mkn}^2}\!\!
\left[2\lambda_{mkn} 
J_{m-1}(\lambda_{mkn} r) 
-\frac{(2-\omega_{mkn})m}{r}J_m(\lambda_{mkn} r)\right]
\sin\left(\frac{(2k-1)\pi z}{H}\right)e^{im\varphi} 
\nonumber
\\
{u}^{z}_{mkn}  & = &\!\! \frac{i}{\omega_{mkn}}\frac{\left(2k-1\right)\pi}{H} 
J_m(\lambda_{mkn} r)
\cos\left(\frac{(2k-1)\pi z}{H}\right)e^{im\varphi} 
\label{eq::velfield2_app}
\\[0.2cm]
0 & = & \omega_{mkn}\lambda_{mkn}J_{m-1}(\lambda_{mkn}) +
(2-\omega_{mkn})mJ_m(\lambda_{mkn})
\nonumber\\[0.2cm] 
\omega_{mkn} & = &
\pm2\left[1+\frac{\lambda_{mkn}^2H^2}{\left(2k-1\right)^2\pi^2}\right]^{-\nicefrac{1}{2}}.
\nonumber
\end{eqnarray}
\end{enumerate}

%%%%%%%%%%%%%%%%%%%%%%%%%%%%%%%%%%%%%%%%%%%%%%%%%%%%%%%%%%%%%%%%%%%%%%%%%%%%%%%

\section{Calculation of inertial mode amplitudes from axial profiles
  measured in the experiment}\label{appendix::experimental_amplitudes}

The measured velocity profiles are analysed quantitatively by the
application of a reduced projection method similar to the approach
that is used for the numerical data in \citet{pizzi2021a}.
The basis for the procedure is
the time series of the axial velocity component $u_z(r=r_s,z,t)$,
where $r_s$ is one particular radius at which the UDV probes are
mounted. In a first step we apply a discrete sine transformation at
each time step $t_n$:
\begin{equation}
\tilde{u}_k(r_s,t_n)=\frac{1}{N_z}\sum\limits_{j=0}^{N_z-1}
  u_z(r_s,z_j,t_n)\cdot \sin\left(\frac{\pi z_j k}{H}\right),
\label{eq::axial_decomposition}
\end{equation}  
where $z_j=jH/(N_z-1)$ is the discrete and normalized axial coordinate.
\begin{figure}
  \subfloat[][]{\includegraphics[width=0.48\textwidth]{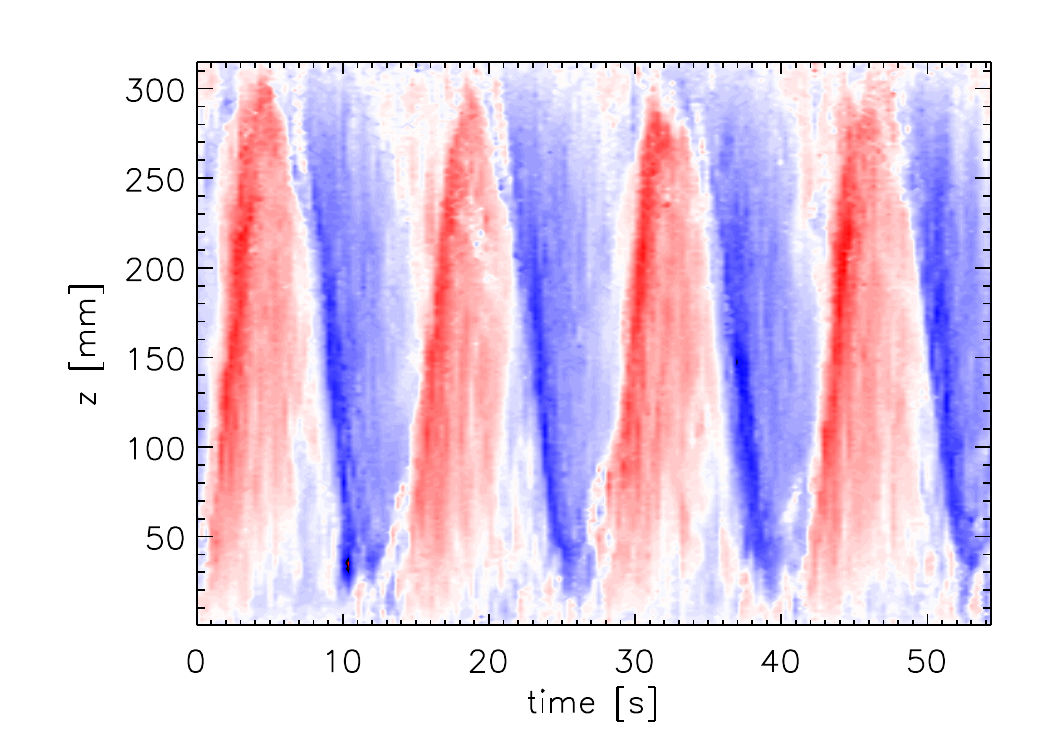}\label{fig::fft_dst_a}}
  \subfloat[][]{\includegraphics[width=0.48\textwidth]{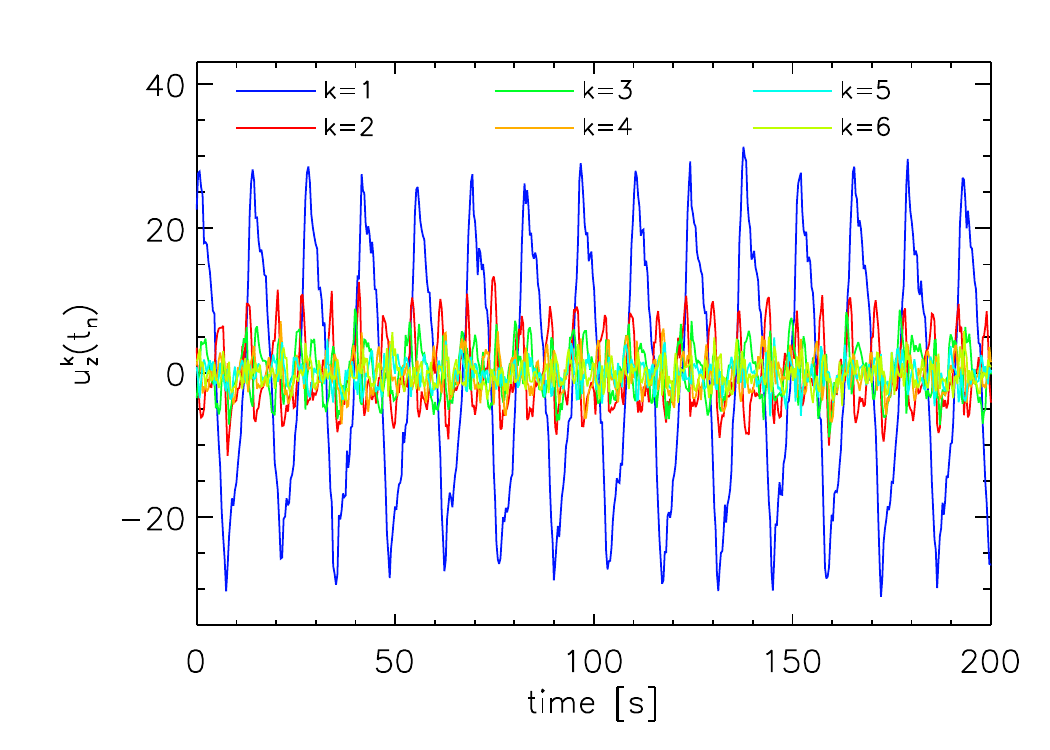}\label{fig::fft_dst_b}}
\\
  \subfloat[][]{\includegraphics[width=0.48\textwidth]{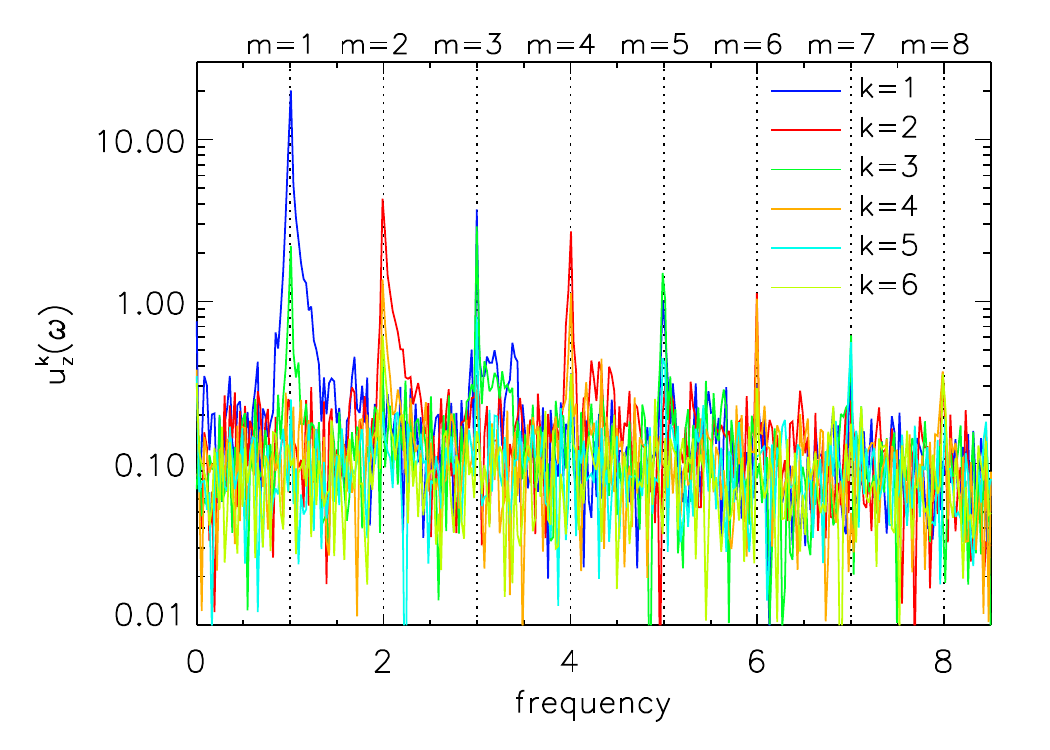}\label{fig::fft_dst_c}}
  \subfloat[][]{\includegraphics[width=0.48\textwidth]{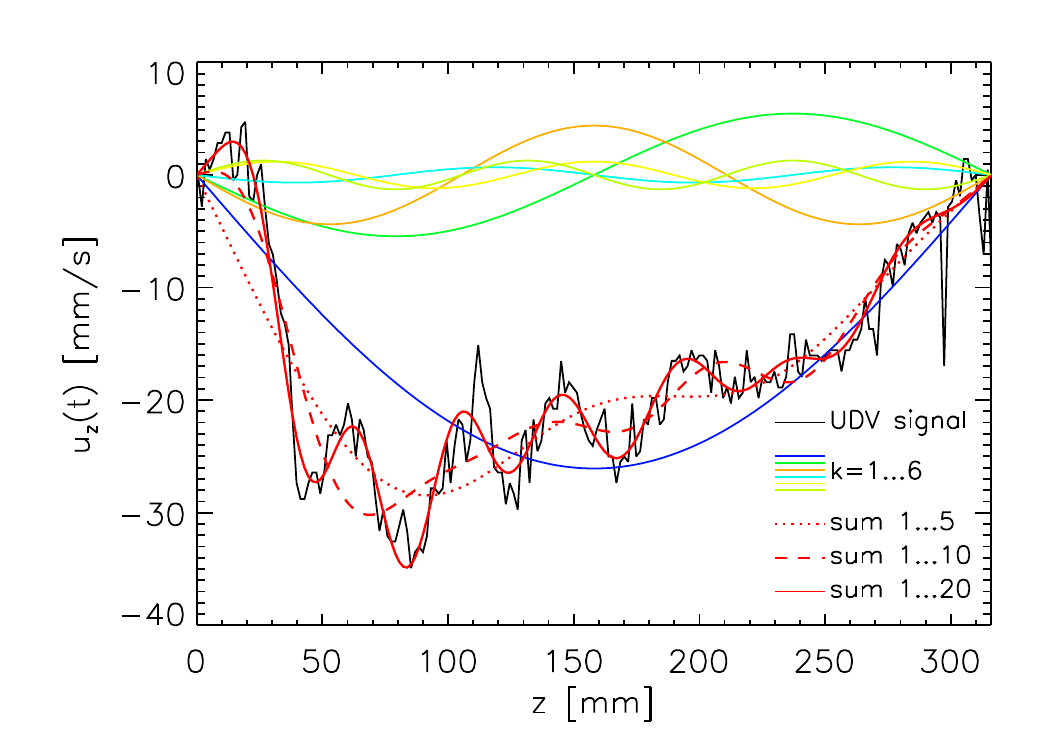}\label{fig::fft_dst_d}}
  \caption{\label{fig::fft_dst}
\raggedright 
(a) Original time series as measured by
an UDV probe mounted at $r=0.9$.
(b) Time series of the $k$-mode amplitude for $k=1 \cdots 6$ as
obtained from the measurement run shown in (a).
(c) Fourier decomposition of the
time series presented in panel (b). The individual peaks occur
at multiples of the rotation frequency of the container and
represent the different azimuthal modes for $m=1,2,3,...$.
(d) Original signal (black curve) in
comparison with various reconstructed data curves using different
numbers of axial eigenmodes (red curves). The coloured curves show the
individual eigenmodes up to $k=6$ that have been used to reconstruct
the original signal. Here, ${\rm{Re}}=10^4, {\rm{Po}}=0.075$.  
  }
\end{figure}
The decomposition done in Eq.~(\ref{eq::axial_decomposition}) provides
an axial mode $\tilde{u}_k$ in dependence on the axial wavenumber
$k$.  Up to now we have ignored the dependency on the azimuthal coordinate
$\varphi$, because $\varphi$ enters the measurement analysis only
implicitly, since the UDV probes cover the measuring volume once
during one revolution of the cylinder. Since there is a time offset
between the velocity profiles recorded at different angles during a
single revolution, the measured profiles cannot be interpreted as an
instantaneous snapshot of the axial velocity field, as is possible
with the simulation data.  However, in the precession frame of
reference the essential part of kinetic energy is contained in
large-scale components in terms of standing inertial waves so that,
after applying a Fourier decomposition, we are able to disentangle the
mixed dependency on time and on the azimuthal coordinate, because we
know the frequency of the probe.  Therefore, we calculate the Fourier
spectrum in time for each individual $k$-mode according to
\begin{equation}
  \tilde{\tilde{u}}_{k\omega}(r_s)=
    \frac{1}{N_t}\sum\limits_{n=0}^{N_t-1}\tilde{u}_k(r_s,t_n)
     e^{-i\omega t_n}=
  \frac{1}{N_zN_t}\sum\limits_{j=0}^{N_z-1}\sum\limits_{n=0}^{N_t-1}u_z(r_s,z_j,t_n)
  \cdot \sin\left(\frac{\pi z_jk}{H}\right)e^{-i\omega t_n}.
  \label{eq::fft}
\end{equation}
The implementation of the individual steps is shown in
figure~\ref{fig::fft_dst} for the paradigmatic case ${\rm{Re}}=10^4$
and ${\rm{Po}}=0.075$.  Figure~\ref{fig::fft_dst_a} shows a small
section of the measurement series as recorded by a single UDV probe
(the total duration covers approximately 100 rotation periods).
Figure ~\ref{fig::fft_dst_b} shows the decomposition into individual
axial modes (from $k=1$ to $k=6$) according to
Eq.~(\ref{eq::axial_decomposition}). The quasiperiodic behaviour and
the dominance of the first axial mode with $k \propto \sin(\pi z/H)$
can be clearly recognized.  Figure~\ref{fig::fft_dst_c} shows the
Fourier spectrum for the six axial modes from panel (b) and
illustrates that the main contributions come from standing waves in
the precession system, which appear in terms of peaks at integer
multiples of the rotation frequency of the container.  This multiplier
is equivalent to the corresponding azimuthal wavenumber $m$ and in the
following we label the peaks of the modes at integer multiples of the
cylinder frequency with $m$,
i.e. $\tilde{\tilde{u}}_{k\omega=1,2,3...}\widehat{=}\tilde{\tilde{u}}_{km=1,2,3...}$. 
In order to prove that the characterization by means of the standing
inertial modes provides a reasonable representation for the fluid
flow, we calculated the reconstructed signal from the amplitudes of
various numbers of $k$-modes.  The result is shown in
figure~\ref{fig::fft_dst_d}, which presents a single snapshot of an
axial profile (black curve) and the decomposition into several axial
$k$-modes according to Eq.~(\ref{eq::axial_decomposition}) with the
blue curve denoting the dominant contribution for the case $k=1$. The
red curve in turn presents the reconstructed axial profile when using
the first $20$ axial eigenmodes and confirms that the decomposition in
terms of inertial eigenmodes provides a good representation for the
velocity field.

%%%%%%%%%%%%%%%%%%%%%%%%%%%%%%%%%%%%%%%%%%%%%%%%%%%%%%%%%%%%%%%%%%%%%%%%%%%%%

%%\setcitestyle{authoryear,round}
%\bibliography{giesecke_global_state_cylinder_precession}
%%\bibliographystyle{apsrev4-2}

\bibliographystyle{jfm}
\bibliography{giesecke_global_state_cylinder_precession_jfm}

\begin{thebibliography}{89}
\expandafter\ifx\csname natexlab\endcsname\relax\def\natexlab#1{#1}\fi
\def\au#1{#1} \def\ed#1{#1} \def\yr#1{#1}\def\at#1{#1}\def\jt#1{\textit{#1}}
  \def\bt#1{#1}\def\bvol#1{\textbf{#1}} \def\vol#1{#1} \def\pg#1{#1}
  \def\publ#1{#1}\def\arxiv#1{#1}\def\org#1{#1}\def\st#1{\textit{#1}}

\bibitem[{Albrecht} {\em et~al.\/}(2015{\natexlab{{\em a\/}}}){Albrecht},
  {Blackburn}, {Lopez}, {Manasseh} \& {Meunier}]{albrecht2015b}
{\sc \au{{Albrecht}, T.}, \au{{Blackburn}, H.~M.}, \au{{Lopez}, J.~M.},
  \au{{Manasseh}, R.} \& \au{{Meunier}, P.}} \yr{2015{\natexlab{{\em a\/}}}}
  \at{{Triadic resonances in precessing rapidly rotating cylinder flows}}.
  \jt{\jfm}  \bvol{778},  \pg{1}.

\bibitem[Albrecht {\em et~al.\/}(2021)Albrecht, Blackburn, Lopez, Manasseh \&
  Meunier]{albrecht2021}
{\sc \au{Albrecht, Thomas}, \au{Blackburn, Hugh~M.}, \au{Lopez, Juan~M.},
  \au{Manasseh, Richard} \& \au{Meunier, Patrice}} \yr{2021}  \at{On the
  origins of steady streaming in precessing fluids}.  \jt{\jfm}  \bvol{910},
  \pg{A51}.

\bibitem[{Albrecht} {\em et~al.\/}(2015{\natexlab{{\em b\/}}}){Albrecht},
  {Blackburn}, {Meunier}, {Manasseh} \& {Lopez}]{albrecht2015a}
{\sc \au{{Albrecht}, T.}, \au{{Blackburn}, H.~M.}, \au{{Meunier}, P.},
  \au{{Manasseh}, R.} \& \au{{Lopez}, J.~M}} \yr{2015{\natexlab{{\em b\/}}}}
  {\protect{PIV}} of a precessing cylinder flow at large tilt angles.  \bt{In
  {\em Ninth International Symposium on Turbulence and Shear Flow Phenomena
  (TSFP-9)\/}}. Melbourne, Australia.

\bibitem[{Albrecht} {\em et~al.\/}(2016){Albrecht}, {Blackburn}, {Meunier},
  {Manasseh} \& {Lopez}]{albrecht2016}
{\sc \au{{Albrecht}, T.}, \au{{Blackburn}, H.~M.}, \au{{Meunier}, P.},
  \au{{Manasseh}, R.} \& \au{{Lopez}, J.~M.}} \yr{2016}  \at{Experimental and
  numerical investigation of a strongly-forced precessing cylinder flow}.
  \jt{Int. J. Heat Fluid Flow}  \bvol{61},  \pg{68--74}.

\bibitem[Barker(2016)]{barker2016}
{\sc \au{Barker, Adrian~J.}} \yr{2016}  \at{{On turbulence driven by axial
  precession and tidal evolution of the spin–orbit angle of close-in giant
  planets}}.  \jt{\mnras}  \bvol{460}~(3),  \pg{2339--2350}.

\bibitem[{Blackburn} {\em et~al.\/}(2019){Blackburn}, {Lee}, {Albrecht} \&
  {Singh}]{blackburn2019}
{\sc \au{{Blackburn}, H.~M.}, \au{{Lee}, D.}, \au{{Albrecht}, T.} \&
  \au{{Singh}, J.}} \yr{2019}  \at{{Semtex: a spectral element–Fourier solver
  for the incompressible Navier–Stokes equations in cylindrical or Cartesian
  coordinates}}.  \jt{Comput. Phys. Comm.}  \bvol{245},  \pg{106804--1--13}.

\bibitem[{Blackburn} \& {Sherwin}(2004)]{blackburn2004}
{\sc \au{{Blackburn}, H.~M.} \& \au{{Sherwin}, S.~J.}} \yr{2004}
  \at{{Formulation of a Galerkin spectral element-Fourier method for
  three-dimensional incompressible flows in cylindrical geometries}}.  \jt{J.
  Comp. Phys.}  \bvol{197},  \pg{759--778}.

\bibitem[Buffett(2021)]{buffett2021}
{\sc \au{Buffett, B.A.}} \yr{2021}  \at{Conditions for turbulent ekman layers
  in precessionally driven flow}.  \jt{\gji}  \bvol{226}~(1),  \pg{56 – 65}.

\bibitem[Burmann \& Noir(2022)]{buhrmann2022}
{\sc \au{Burmann, Fabian} \& \au{Noir, Jérõme}} \yr{2022}  \at{Experimental
  study of the flows in a non-axisymmetric ellipsoid under precession}.
  \jt{\jfm}  \bvol{932},  \pg{A24}.

\bibitem[{Busse}(1968)]{busse1968}
{\sc \au{{Busse}, F.~H.}} \yr{1968}  \at{{Steady fluid flow in a precessing
  spheroidal shell}}.  \jt{\jfm}  \bvol{33},  \pg{739--751}.

\bibitem[{C{\'e}bron} {\em et~al.\/}(2019){C{\'e}bron}, {Laguerre}, {Noir} \&
  {Schaeffer}]{cebron2019}
{\sc \au{{C{\'e}bron}, D.}, \au{{Laguerre}, R.}, \au{{Noir}, J.} \&
  \au{{Schaeffer}, N.}} \yr{2019}  \at{{Precessing spherical shells: flows,
  dissipation, dynamo and the lunar core}}.  \jt{Geophys. J. Int.}  \bvol{219},
   \pg{S34--S57}.

\bibitem[{Dudley} \& {James}(1989)]{dudley1989}
{\sc \au{{Dudley}, M.~L.} \& \au{{James}, R.~W.}} \yr{1989}
  \at{{Time-dependent kinematic dynamos with stationary flows}}.  \jt{Proc.
  Roy. Soc. Lond. Series A}  \bvol{425}~(1869),  \pg{407--429}.

\bibitem[{Dwyer} {\em et~al.\/}(2011){Dwyer}, {Stevenson} \&
  {Nimmo}]{dwyer2011}
{\sc \au{{Dwyer}, C.~A.}, \au{{Stevenson}, D.~J.} \& \au{{Nimmo}, F.}}
  \yr{2011}  \at{{A long-lived lunar dynamo driven by continuous mechanical
  stirring}}.  \jt{\nat}  \bvol{479},  \pg{212--214}.

\bibitem[{Gans}(1970)]{gans1970}
{\sc \au{{Gans}, R.~F.}} \yr{1970}  \at{{On the precession of a resonant
  cylinder}}.  \jt{\jfm}  \bvol{41},  \pg{865--872}.

\bibitem[{Gans}(1971)]{gans1971}
{\sc \au{{Gans}, R.~F.}} \yr{1971}  \at{{On hydromagnetic precession in a
  cylinder}}.  \jt{\jfm}  \bvol{45},  \pg{111--130}.

\bibitem[{Gao} {\em et~al.\/}(2021){Gao}, {Meunier}, {Le Diz{\`e}s} \&
  {Eloy}]{gao2021}
{\sc \au{{Gao}, D.}, \au{{Meunier}, P.}, \au{{Le Diz{\`e}s}, S.} \& \au{{Eloy},
  C.}} \yr{2021}  \at{{Zonal flow in a resonant precessing cylinder}}.
  \jt{\jfm}  \bvol{923},  \pg{A29}.

\bibitem[Garcia {\em et~al.\/}(2021)Garcia, Giesecke \& Stefani]{garcia2021}
{\sc \au{Garcia, F.}, \au{Giesecke, A.} \& \au{Stefani, F.}} \yr{2021}
  \at{{Modulated rotating waves and triadic resonances in spherical fluid
  systems: The case of magnetized spherical Couette flow}}.  \jt{Phys. Fluids}
  \bvol{33}~(4),  \pg{044105}.

\bibitem[Garcia {\em et~al.\/}(2019)Garcia, Seilmayer, Giesecke \&
  Stefani]{garcia2019}
{\sc \au{Garcia, Ferran}, \au{Seilmayer, Martin}, \au{Giesecke, André} \&
  \au{Stefani, Frank}} \yr{2019}  \at{Modulated rotating waves in the
  magnetised spherical couette system}.  \jt{J. Nonlinear Sci.}  \bvol{29}~(6),
   \pg{2735 – 2759}.

\bibitem[Garcia {\em et~al.\/}(2020)Garcia, Seilmayer, Giesecke \&
  Stefani]{garcia2020}
{\sc \au{Garcia, F.}, \au{Seilmayer, M.}, \au{Giesecke, A.} \& \au{Stefani,
  F.}} \yr{2020}  \at{Four-frequency solution in a magnetohydrodynamic couette
  flow as a consequence of azimuthal symmetry breaking}.  \jt{\prl}
  \bvol{125}~(26),  \pg{264501}.

\bibitem[{Giesecke} {\em et~al.\/}(2015{\natexlab{{\em a\/}}}){Giesecke},
  {Albrecht}, {Gerbeth}, {Gundrum} \& {Stefani}]{giesecke2015a}
{\sc \au{{Giesecke}, A.}, \au{{Albrecht}, T.}, \au{{Gerbeth}, G.},
  \au{{Gundrum}, T.} \& \au{{Stefani}, F.}} \yr{2015{\natexlab{{\em a\/}}}}
  \at{{Numerical simulations for the DRESDYN precession dynamo}}.
  \jt{Magnetohydrodynamics}  \bvol{51}~(2),  \pg{293--302}.

\bibitem[{Giesecke} {\em et~al.\/}(2015{\natexlab{{\em b\/}}}){Giesecke},
  {Albrecht}, {Gundrum}, {Herault} \& {Stefani}]{giesecke2015b}
{\sc \au{{Giesecke}, A.}, \au{{Albrecht}, T.}, \au{{Gundrum}, T.},
  \au{{Herault}, J.} \& \au{{Stefani}, F.}} \yr{2015{\natexlab{{\em b\/}}}}
  \at{{Triadic resonances in nonlinear simulations of a fluid flow in a
  precessing cylinder}}.  \jt{\njp}  \bvol{17}~(11),  \pg{113044}.

\bibitem[{Giesecke} {\em et~al.\/}(2012{\natexlab{{\em a\/}}}){Giesecke},
  {Nore}, {Stefani}, {Gerbeth}, {L{\'e}orat}, {Herreman}, {Luddens} \&
  {Guermond}]{giesecke2012a}
{\sc \au{{Giesecke}, A.}, \au{{Nore}, C.}, \au{{Stefani}, F.}, \au{{Gerbeth},
  G.}, \au{{L{\'e}orat}, J.}, \au{{Herreman}, W.}, \au{{Luddens}, F.} \&
  \au{{Guermond}, J.-L.}} \yr{2012{\natexlab{{\em a\/}}}}  \at{{Influence of
  high-permeability discs in an axisymmetric model of the Cadarache dynamo
  experiment}}.  \jt{\njp}  \bvol{14}~(5),  \pg{053005}.

\bibitem[{Giesecke} {\em et~al.\/}(2012{\natexlab{{\em b\/}}}){Giesecke},
  {Stefani} \& {Burguete}]{giesecke2012b}
{\sc \au{{Giesecke}, André}, \au{{Stefani}, Frank} \& \au{{Burguete}, Javier}}
  \yr{2012{\natexlab{{\em b\/}}}}  \at{Impact of time-dependent nonaxisymmetric
  velocity perturbations on dynamo action of von kármán-like flows}.
  \jt{\pre}  \bvol{86}~(6),  \pg{066303}.

\bibitem[{Giesecke} {\em et~al.\/}(2018){Giesecke}, {Vogt}, {Gundrum} \&
  {Stefani}]{giesecke2018}
{\sc \au{{Giesecke}, Andr{\'e}}, \au{{Vogt}, Tobias}, \au{{Gundrum}, Thomas} \&
  \au{{Stefani}, Frank}} \yr{2018}  \at{{Nonlinear Large Scale Flow in a
  Precessing Cylinder and Its Ability To Drive Dynamo Action}}.  \jt{\prl}
  \bvol{120}~(2),  \pg{024502}.

\bibitem[{Giesecke} {\em et~al.\/}(2019){Giesecke}, {Vogt}, {Gundrum} \&
  {Stefani}]{giesecke2019}
{\sc \au{{Giesecke}, Andr{\'e}}, \au{{Vogt}, Tobias}, \au{{Gundrum}, Thomas} \&
  \au{{Stefani}, Frank}} \yr{2019}  \at{{Kinematic dynamo action of a
  precession-driven flow based on the results of water experiments and
  hydrodynamic simulations.}}  \jt{Geophys. Astrophys. Fluid Dyn.}
  \bvol{113}~(1-2),  \pg{235--255}.

\bibitem[{Giesecke} {\em et~al.\/}(2024){Giesecke}, {Wilbert} \&
  {{\v{S}}imkanin}]{giesecke2024a}
{\sc \au{{Giesecke}, A.}, \au{{Wilbert}, M.} \& \au{{{\v{S}}imkanin}, J.}}
  \yr{2024} Mean magnetic fields and magnetic bursts in mhd simulations of a
  precession-driven dynamo. {\it{in preparation}}.

\bibitem[{Goto} {\em et~al.\/}(2007){Goto}, {Ishii}, {Kida} \&
  {Nishioka}]{goto2007}
{\sc \au{{Goto}, S.}, \au{{Ishii}, N.}, \au{{Kida}, S.} \& \au{{Nishioka}, M.}}
  \yr{2007}  \at{{Turbulence generator using a precessing sphere}}.  \jt{\pof}
  \bvol{19}~(6),  \pg{061705--061705}.

\bibitem[{Goto} {\em et~al.\/}(2014){Goto}, {Matsunaga}, {Fujiwara},
  {Nishioka}, {Kida}, {Yamato} \& {Tsuda}]{goto2014}
{\sc \au{{Goto}, S.}, \au{{Matsunaga}, A.}, \au{{Fujiwara}, M.},
  \au{{Nishioka}, M.}, \au{{Kida}, S.}, \au{{Yamato}, M.} \& \au{{Tsuda}, S.}}
  \yr{2014}  \at{{Turbulence driven by precession in spherical and slightly
  elongated spheroidal cavities}}.  \jt{\pof}  \bvol{26}~(5),  \pg{055107}.

\bibitem[{Greenspan}(1969)]{greenspan1969}
{\sc \au{{Greenspan}, H.~P.}} \yr{1969}  \at{{On the non-linear interaction of
  inertial modes}}.  \jt{\jfm}  \bvol{36},  \pg{257--264}.

\bibitem[{Herault} {\em et~al.\/}(2019){Herault}, {Giesecke}, {Gundrum} \&
  {Stefani}]{herault2019}
{\sc \au{{Herault}, Johann}, \au{{Giesecke}, Andr{\'e}}, \au{{Gundrum}, Thomas}
  \& \au{{Stefani}, Frank}} \yr{2019}  \at{{Instability of precession driven
  Kelvin modes: Evidence of a detuning effect}}.  \jt{Phys. Rev. Fluid.}
  \bvol{4}~(3),  \pg{033901}.

\bibitem[{Herault} {\em et~al.\/}(2015){Herault}, {Gundrum}, {Giesecke} \&
  {Stefani}]{herault2015}
{\sc \au{{Herault}, J.}, \au{{Gundrum}, T.}, \au{{Giesecke}, A.} \&
  \au{{Stefani}, F.}} \yr{2015}  \at{{Subcritical transition to turbulence of a
  precessing flow in a cylindrical vessel}}.  \jt{\pof}  \bvol{27}~(12),
  \pg{124102}.

\bibitem[Horimoto {\em et~al.\/}(2018)Horimoto, Simonet-Davin, Katayama \&
  Goto]{horimoto2018}
{\sc \au{Horimoto, Yasufumi}, \au{Simonet-Davin, Gabriel}, \au{Katayama,
  Atsushi} \& \au{Goto, Susumu}} \yr{2018}  \at{Impact of a small ellipticity
  on the sustainability condition of developed turbulence in a precessing
  spheroid}.  \jt{Phys. Rev. Fluids}  \bvol{3},  \pg{044603}.

\bibitem[{Kerswell}(1993)]{kerswell1993}
{\sc \au{{Kerswell}, R.~R.}} \yr{1993}  \at{{The instability of precessing
  flow}}.  \jt{\gafd}  \bvol{72},  \pg{107--144}.

\bibitem[{Kerswell}(1999)]{kerswell1999}
{\sc \au{{Kerswell}, R.~R.}} \yr{1999}  \at{{Secondary instabilities in rapidly
  rotating fluids: inertial wave breakdown}}.  \jt{\jfm}  \bvol{382},
  \pg{283--306}.

\bibitem[Kida(2020)]{kida2020}
{\sc \au{Kida, Shigeo}} \yr{2020}  \at{Steady flow in a rapidly rotating
  spheroid with weak precession: I}.  \jt{Fluid Dyn. Res.}  \bvol{52}~(1),
  \pg{015513}.

\bibitem[Kida(2021)]{kida2021}
{\sc \au{Kida, Shigeo}} \yr{2021}  \at{Steady flow in a rapidly rotating
  spheroid with weak precession: {\protect{II}}}.  \jt{Fluid Dyn. Res.}
  \bvol{53}~(2),  \pg{025501}.

\bibitem[{Kobine}(1995)]{kobine1995}
{\sc \au{{Kobine}, J.~J.}} \yr{1995}  \at{{Inertial wave dynamics in a rotating
  and precessing cylinder}}.  \jt{\jfm}  \bvol{303},  \pg{233--252}.

\bibitem[{Kobine}(1996)]{kobine1996}
{\sc \au{{Kobine}, J.~J.}} \yr{1996}  \at{{Azimuthal flow associated with
  inertial wave resonance in a precessing cylinder}}.  \jt{\jfm}  \bvol{319},
  \pg{387--406}.

\bibitem[Komoda \& Goto(2019)]{komoda2019}
{\sc \au{Komoda, Ken} \& \au{Goto, Susumu}} \yr{2019}  \at{Three-dimensional
  flow structures of turbulence in precessing spheroids}.  \jt{Phys. Rev.
  Fluids}  \bvol{4},  \pg{014603}.

\bibitem[{Kumar} {\em et~al.\/}(2023){Kumar}, {Pizzi}, {Giesecke},
  {{\v{S}}imkanin}, {Gundrum}, {Ratajczak} \& {Stefani}]{kumar2023}
{\sc \au{{Kumar}, Vivaswat}, \au{{Pizzi}, Federico}, \au{{Giesecke},
  Andr{\'e}}, \au{{{\v{S}}imkanin}, J{\'a}n}, \au{{Gundrum}, Thomas},
  \au{{Ratajczak}, Matthias} \& \au{{Stefani}, Frank}} \yr{2023}  \at{The
  effect of nutation angle on the flow inside a precessing cylinder and its
  dynamo action}.  \jt{\pof}  \bvol{35}~(1),  \pg{014114}.

\bibitem[Lagrange {\em et~al.\/}(2011)Lagrange, Meunier, Nadal \&
  Eloy]{lagrange2011}
{\sc \au{Lagrange, R.}, \au{Meunier, P.}, \au{Nadal, F.} \& \au{Eloy, C.}}
  \yr{2011}  \at{Precessional instability of a fluid cylinder}.  \jt{\jfm}
  \bvol{666},  \pg{104--145}.

\bibitem[Landeau {\em et~al.\/}(2022)Landeau, Fournier, Nataf, C{\'e}bron \&
  Schaeffer]{landeau2022}
{\sc \au{Landeau, Maylis}, \au{Fournier, Alexandre}, \au{Nataf, Henri-Claude},
  \au{C{\'e}bron, David} \& \au{Schaeffer, Nathana{\"e}l}} \yr{2022}
  \at{Sustaining earth’s magnetic dynamo}.  \jt{Nat. Rev. Earth Environ.}
  \bvol{3}~(4),  \pg{255 – 269}.

\bibitem[Le~Bars(2016)]{lebars2016}
{\sc \au{Le~Bars, Michael}} \yr{2016}  \at{Flows driven by libration,
  precession, and tides in planetary cores}.  \jt{Phys. Rev. Fluids}  \bvol{1},
   \pg{060505}.

\bibitem[{Le Bars} {\em et~al.\/}(2015){Le Bars}, {C{\'e}bron} \& {Le
  Gal}]{lebars2015}
{\sc \au{{Le Bars}, M.}, \au{{C{\'e}bron}, D.} \& \au{{Le Gal}, P.}} \yr{2015}
  \at{{Flows Driven by Libration, Precession, and Tides}}.  \jt{Annu. Rev.
  Fluid Mech.}  \bvol{47},  \pg{163--193}.

\bibitem[{Liao} \& {Zhang}(2012)]{liao2012}
{\sc \au{{Liao}, X.} \& \au{{Zhang}, K.}} \yr{2012}  \at{{On flow in weakly
  precessing cylinders: the general asymptotic solution}}.  \jt{\jfm}
  \bvol{709},  \pg{610--621}.

\bibitem[{Lin} {\em et~al.\/}(2015){Lin}, {Marti} \& {Noir}]{lin2015}
{\sc \au{{Lin}, Y.}, \au{{Marti}, P.} \& \au{{Noir}, J.}} \yr{2015}
  \at{{Shear-driven parametric instability in a precessing sphere}}.  \jt{\pof}
   \bvol{27}~(4),  \pg{046601}.

\bibitem[{Lin} {\em et~al.\/}(2016){Lin}, {Marti}, {Noir} \&
  {Jackson}]{lin2016}
{\sc \au{{Lin}, Y.}, \au{{Marti}, P.}, \au{{Noir}, J.} \& \au{{Jackson}, A.}}
  \yr{2016}  \at{{Precession-driven dynamos in a full sphere and the role of
  large scale cyclonic vortices}}.  \jt{\pof}  \bvol{28}~(6),  \pg{066601}.

\bibitem[{Loper}(1975)]{loper1975}
{\sc \au{{Loper}, David~E.}} \yr{1975}  \at{{Torque balance and energy budget
  for the precessionally driven dynamo}}.  \jt{\pepi}  \bvol{11}~(1),
  \pg{43--60}.

\bibitem[{Lopez} \& {Marques}(2016)]{lopez2016}
{\sc \au{{Lopez}, J.~M.} \& \au{{Marques}, F.}} \yr{2016}  \at{{Nonlinear and
  detuning effects of the nutation angle in precessionally forced rotating
  cylinder flow}}.  \jt{\prf}  \bvol{1}~(2),  \pg{023602}.

\bibitem[Lopez \& Marques(2018)]{lopez2018}
{\sc \au{Lopez, Juan~M.} \& \au{Marques, Francisco}} \yr{2018}  \at{Rapidly
  rotating precessing cylinder flows: forced triadic resonances}.  \jt{\jfm}
  \bvol{839},  \pg{239–270}.

\bibitem[{Malkus}(1968)]{malkus1968}
{\sc \au{{Malkus}, W.~V.~R.}} \yr{1968}  \at{{Precession of the Earth as the
  Cause of Geomagnetism}}.  \jt{Science}  \bvol{160},  \pg{259--264}.

\bibitem[{Manasseh}(1992)]{manasseh1992}
{\sc \au{{Manasseh}, R.}} \yr{1992}  \at{{Breakdown regimes of inertia waves in
  a precessing cylinder}}.  \jt{\jfm}  \bvol{243},  \pg{261--296}.

\bibitem[{Manasseh}(1994)]{manasseh1994}
{\sc \au{{Manasseh}, R.}} \yr{1994}  \at{{Distortions of inertia waves in a
  rotating fluid cylinder forced near its fundamental mode resonance}}.
  \jt{\jfm}  \bvol{265},  \pg{345--370}.

\bibitem[{Manasseh}(1996)]{manasseh1996}
{\sc \au{{Manasseh}, R.}} \yr{1996}  \at{{Nonlinear behaviour of contained
  inertia waves}}.  \jt{\jfm}  \bvol{315},  \pg{151--173}.

\bibitem[{Marques} \& {Lopez}(2015)]{marques2015}
{\sc \au{{Marques}, F.} \& \au{{Lopez}, J.~M.}} \yr{2015}  \at{{Precession of a
  rapidly rotating cylinder flow: traverse through resonance}}.  \jt{\jfm}
  \bvol{782},  \pg{63--98}.

\bibitem[McEwan(1970)]{mcewan1970}
{\sc \au{McEwan, A.~D.}} \yr{1970}  \at{{Inertial oscillations in a rotating
  fluid cylinder}}.  \jt{\jfm}  \bvol{40},  \pg{603--640}.

\bibitem[{Meunier}(2020)]{meunier2020}
{\sc \au{{Meunier}, P.}} \yr{2020}  \at{{Geoinspired soft mixers}}.  \jt{\jfm}
  \bvol{903},  \pg{A15}.

\bibitem[{Meunier} {\em et~al.\/}(2008){Meunier}, {Eloy}, {Lagrange} \&
  {Nadal}]{meunier2008}
{\sc \au{{Meunier}, P.}, \au{{Eloy}, C.}, \au{{Lagrange}, R.} \& \au{{Nadal},
  F.}} \yr{2008}  \at{{A rotating fluid cylinder subject to weak precession}}.
  \jt{\jfm}  \bvol{599},  \pg{405--440}.

\bibitem[Moffatt(1970)]{moffatt1970}
{\sc \au{Moffatt, H.~K.}} \yr{1970}  \at{Dynamo action associated with random
  inertial waves in a rotating conducting fluid}.  \jt{\jfm}  \bvol{44}~(4),
  \pg{705--719}.

\bibitem[{Monchaux} {\em et~al.\/}(2007){Monchaux}, {Berhanu}, {Bourgoin},
  {Moulin}, {Odier}, {Pinton}, {Volk}, {Fauve}, {Mordant}, {P{\'e}tr{\'e}lis},
  {Chiffaudel}, {Daviaud}, {Dubrulle}, {Gasquet}, {Mari{\'e}} \&
  {Ravelet}]{monchaux2007}
{\sc \au{{Monchaux}, R.}, \au{{Berhanu}, M.}, \au{{Bourgoin}, M.},
  \au{{Moulin}, M.}, \au{{Odier}, P.}, \au{{Pinton}, J.-F.}, \au{{Volk}, R.},
  \au{{Fauve}, S.}, \au{{Mordant}, N.}, \au{{P{\'e}tr{\'e}lis}, F.},
  \au{{Chiffaudel}, A.}, \au{{Daviaud}, F.}, \au{{Dubrulle}, B.},
  \au{{Gasquet}, C.}, \au{{Mari{\'e}}, L.} \& \au{{Ravelet}, F.}} \yr{2007}
  \at{{Generation of a Magnetic Field by Dynamo Action in a Turbulent Flow of
  Liquid Sodium}}.  \jt{\prl}  \bvol{98}~(4),  \pg{044502}.

\bibitem[{Mouhali} {\em et~al.\/}(2012){Mouhali}, {Lehner}, {L{\'e}orat} \&
  {Vitry}]{mouhali2012}
{\sc \au{{Mouhali}, W.}, \au{{Lehner}, T.}, \au{{L{\'e}orat}, J.} \&
  \au{{Vitry}, R.}} \yr{2012}  \at{{Evidence for a cyclonic regime in a
  precessing cylindrical container}}.  \jt{Exp. Fluids}  \bvol{53},
  \pg{1693--1700}.

\bibitem[{Noir} {\em et~al.\/}(2003){Noir}, {Cardin}, {Jault} \&
  {Masson}]{noir2003}
{\sc \au{{Noir}, J.}, \au{{Cardin}, P.}, \au{{Jault}, D.} \& \au{{Masson},
  J.-P.}} \yr{2003}  \at{{Experimental evidence of non-linear resonance effects
  between retrograde precession and the tilt-over mode within a spheroid}}.
  \jt{\gji}  \bvol{154},  \pg{407--416}.

\bibitem[{Noir} \& {C{\'e}bron}(2013)]{noir2013}
{\sc \au{{Noir}, J.} \& \au{{C{\'e}bron}, D.}} \yr{2013}
  \at{{Precession-driven flows in non-axisymmetric ellipsoids}}.  \jt{\jfm}
  \bvol{737},  \pg{412--439}.

\bibitem[{Olson}(2013)]{olson2013}
{\sc \au{{Olson}, P.}} \yr{2013}  \at{{The new core paradox}}.  \jt{Science}
  \bvol{431--432}~(6157),  \pg{013008}.

\bibitem[{Pizzi}(2023)]{pizziphd}
{\sc \au{{Pizzi}, F.}} \yr{2023}  \at{Numerical studies of a fluid-filled
  precessing cylinder : a framework for the dresdyn precession experiment}.
  Doctoral thesis, BTU Cottbus - Senftenberg.

\bibitem[{Pizzi} {\em et~al.\/}(2021{\natexlab{{\em a\/}}}){Pizzi}, {Giesecke},
  {\v{S}imkanin} \& {Stefani}]{pizzi2021b}
{\sc \au{{Pizzi}, F.}, \au{{Giesecke}, A.}, \au{{\v{S}imkanin}, J.} \&
  \au{{Stefani}, F.}} \yr{2021{\natexlab{{\em a\/}}}}  \at{{Prograde and
  retrograde precession of a fluid -filled cylinder}}.  \jt{New J. Phys.}
  \bvol{23}~(12),  \pg{123016}.

\bibitem[{Pizzi} {\em et~al.\/}(2021{\natexlab{{\em b\/}}}){Pizzi}, {Giesecke}
  \& {Stefani}]{pizzi2021a}
{\sc \au{{Pizzi}, F.}, \au{{Giesecke}, A.} \& \au{{Stefani}, F.}}
  \yr{2021{\natexlab{{\em b\/}}}}  \at{{Ekman boundary layers in a fluid filled
  precessing cylinder}}.  \jt{AIP Advances}  \bvol{11}~(3),  \pg{035023}.

\bibitem[{Pizzi} {\em et~al.\/}(2022){Pizzi}, {Mamatsashvili}, {Barker},
  {Giesecke} \& {Stefani}]{pizzi2022}
{\sc \au{{Pizzi}, F.}, \au{{Mamatsashvili}, G.}, \au{{Barker}, A.J.},
  \au{{Giesecke}, A.} \& \au{{Stefani}, F.}} \yr{2022}  \at{{Interplay between
  geostrophic vortices and inertial waves in precession-driven turbulence}}.
  \jt{\pof}  \bvol{34}~(12),  \pg{125135}.

\bibitem[Poincar{\'e}(1910)]{poincare1910}
{\sc \au{Poincar{\'e}, H.}} \yr{1910}  \at{{Sur la pr{\'e}cession des corps
  d{\'e}formables}}.  \jt{Bulletin Astronomique}  \bvol{27},  \pg{321--356}.

\bibitem[{Ravelet} {\em et~al.\/}(2005){Ravelet}, {Chiffaudel}, {Daviaud} \&
  {L{\'e}orat}]{ravelet2005}
{\sc \au{{Ravelet}, F.}, \au{{Chiffaudel}, A.}, \au{{Daviaud}, F.} \&
  \au{{L{\'e}orat}, J.}} \yr{2005}  \at{{Toward an experimental von
  K{\'a}rm{\'a}n dynamo: Numerical studies for an optimized design}}.
  \jt{\pof}  \bvol{17}~(11),  \pg{117104}.

\bibitem[{Reasor} {\em et~al.\/}(2004){Reasor}, {Montgomery} \&
  {Grasso}]{reasor2004}
{\sc \au{{Reasor}, Paul~D.}, \au{{Montgomery}, Michael~T.} \& \au{{Grasso},
  Lewis~D.}} \yr{2004}  \at{A new look at the problem of tropical cyclones in
  vertical shear flow: Vortex resiliency}.  \jt{J. Atmos. Sci.}  \bvol{61}~(1),
   \pg{3--22}.

\bibitem[{Rochester} {\em et~al.\/}(1975){Rochester}, {Jacobs}, {Smylie} \&
  {Chong}]{rochester1975}
{\sc \au{{Rochester}, M.~G.}, \au{{Jacobs}, J.~A.}, \au{{Smylie}, D.~E.} \&
  \au{{Chong}, K.~F.}} \yr{1975}  \at{{Can precession power the geomagnetic
  dynamo?}}  \jt{Geophysical Journal}  \bvol{43},  \pg{661--678}.

\bibitem[{Rogers} {\em et~al.\/}(2013){Rogers}, {Costello} \&
  {Cooper}]{rogers2013}
{\sc \au{{Rogers}, Jonathan}, \au{{Costello}, Mark} \& \au{{Cooper}, Gene~R.}}
  \yr{2013}  \at{Design considerations for stability of liquid payload
  projectiles}.  \jt{\jsr}  \bvol{50},  \pg{169--178}.

\bibitem[Soward(1975)]{soward1975}
{\sc \au{Soward, A.~M.}} \yr{1975}  \at{Random waves and dynamo action}.
  \jt{\jfm}  \bvol{69}~(1),  \pg{145–177}.

\bibitem[{Stefani} {\em et~al.\/}(2015){Stefani}, {Albrecht}, {Gerbeth},
  {Giesecke}, {Gundrum}, {Herault}, {Nore} \& {Steglich}]{stefani2015}
{\sc \au{{Stefani}, F.}, \au{{Albrecht}, T.}, \au{{Gerbeth}, G.},
  \au{{Giesecke}, A.}, \au{{Gundrum}, T.}, \au{{Herault}, J.}, \au{{Nore}, C.}
  \& \au{{Steglich}, C.}} \yr{2015}  \at{{Towards a precession driven dynamo
  experiment}}.  \jt{Magnetohydrodynamics}  \bvol{51}~(2),  \pg{275--284}.

\bibitem[{Stewartson} \& {Roberts}(1963)]{stewartson1963}
{\sc \au{{Stewartson}, K.} \& \au{{Roberts}, P.~H.}} \yr{1963}  \at{{On the
  motion of a liquid in a spheroidal cavity of a precessing rigid body}}.
  \jt{\jfm}  \bvol{17},  \pg{1--20}.

\bibitem[Tarduno {\em et~al.\/}(2020)Tarduno, Cottrell, Bono, Oda, Davis,
  Fayek, van{\textquoteright}t Erve, Nimmo, Huang, Thern, Fearn, Mitra, Smirnov
  \& Blackman]{tarduno2020}
{\sc \au{Tarduno, J.~A.}, \au{Cottrell, Rory~D.}, \au{Bono, Richard~K.},
  \au{Oda, Hirokuni}, \au{Davis, William~J.}, \au{Fayek, Mostafa},
  \au{van{\textquoteright}t Erve, Olaf}, \au{Nimmo, Francis}, \au{Huang,
  Wentao}, \au{Thern, Eric~R.}, \au{Fearn, Sebastian}, \au{Mitra, Gautam},
  \au{Smirnov, Aleksey~V.} \& \au{Blackman, Eric~G.}} \yr{2020}
  \at{Paleomagnetism indicates that primary magnetite in zircon records a
  strong hadean geodynamo}.  \jt{Proc. Natl. Acad. Sci.}  \bvol{117}~(5),
  \pg{2309--2318}.

\bibitem[Tikoo \& Evans(2022)]{tikoo2022}
{\sc \au{Tikoo, Sonia~M.} \& \au{Evans, Alexander~J.}} \yr{2022}  \at{Dynamos
  in the inner solar system}.  \jt{Annu. Rev. Fluid Mech.}  \bvol{50},  \pg{99
  – 122}.

\bibitem[{Tikoo} {\em et~al.\/}(2017){Tikoo}, {Weiss}, {Shuster}, {Suavet},
  {Wang} \& {Grove}]{tikoo2017}
{\sc \au{{Tikoo}, S.~M.}, \au{{Weiss}, Benjamin~P.}, \au{{Shuster}, David~L.},
  \au{{Suavet}, Cl{\'e}ment}, \au{{Wang}, Huapei} \& \au{{Grove}, Timothy~L.}}
  \yr{2017}  \at{{A two-billion-year history for the lunar dynamo}}.
  \jt{Science Advances}  \bvol{3}~(8),  \pg{e1700207}.

\bibitem[{Tilgner}(1998)]{tilgner1998}
{\sc \au{{Tilgner}, A.}} \yr{1998}  \at{{On models of precession driven core
  flow}}.  \jt{Studia geoph. et geod.}  \bvol{42},  \pg{232--238}.

\bibitem[{Tilgner}(1999)]{tilgner1999}
{\sc \au{{Tilgner}, A.}} \yr{1999}  \at{{Magnetohydrodynamic flow in precessing
  spherical shells}}.  \jt{\jfm}  \bvol{379},  \pg{303--318}.

\bibitem[{Tilgner}(2005)]{tilgner2005}
{\sc \au{{Tilgner}, A.}} \yr{2005}  \at{{Precession driven dynamos}}.
  \jt{\pof}  \bvol{17}~(3),  \pg{034104}.

\bibitem[{Tilgner} \& {Busse}(2001)]{tilgner2001}
{\sc \au{{Tilgner}, A.} \& \au{{Busse}, F.~H.}} \yr{2001}  \at{{Fluid flows in
  precessing spherical shells}}.  \jt{\jfm}  \bvol{426},  \pg{387--396}.

\bibitem[{Vanyo}(1973)]{vanyo1973b}
{\sc \au{{Vanyo}, J.~P.}} \yr{1973}  \at{{An Energy Assessment for Liquids in a
  Filled Precessing Spherical Cavity}}.  \jt{\jam}  \bvol{40},  \pg{851}.

\bibitem[{Vanyo}(1991)]{vanyo1991}
{\sc \au{{Vanyo}, J.~P.}} \yr{1991}  \at{{A geodynamo powered by luni-solar
  precession}}.  \jt{\gafd}  \bvol{59},  \pg{209--234}.

\bibitem[{Vanyo} \& {Likins}(1971)]{vanyo1971}
{\sc \au{{Vanyo}, J.~P.} \& \au{{Likins}, P.~W.}} \yr{1971}  \at{{Measurement
  of Energy Dissipation in a Liquid-Filled, Precessing, Spherical Cavity}}.
  \jt{\jam}  \bvol{38},  \pg{674}.

\bibitem[{Vanyo} \& {Likins}(1973)]{vanyo1973a}
{\sc \au{{Vanyo}, J.~P.} \& \au{{Likins}, P.~W.}} \yr{1973}  \at{{Use of
  Baffles to Suppress Energy Dissipation in Liquid-Filled Precessing
  Cavities}}.  \jt{\jsr}  \bvol{10},  \pg{627}.

\bibitem[Vogt {\em et~al.\/}(2014)Vogt, R{\" a}biger \& Eckert]{vogt2014}
{\sc \au{Vogt, Tobias}, \au{R{\" a}biger, Dirk} \& \au{Eckert, Sven}} \yr{2014}
   \at{Inertial wave dynamics in a rotating liquid metal}.  \jt{\jfm}
  \bvol{753},  \pg{472--498}.

\bibitem[{Weiss} \& {Tikoo}(2014)]{weiss2014}
{\sc \au{{Weiss}, B.~P.} \& \au{{Tikoo}, S.~M.}} \yr{2014}  \at{{The lunar
  dynamo}}.  \jt{Science}  \bvol{346},  \pg{1198}.

\end{thebibliography}

%\expandafter\ifx\csname natexlab\endcsname\relax
%\def\natexlab#1{#1}\fi
%\expandafter\ifx\csname selectlanguage\endcsname\relax
%\def\selectlanguage#1{\relax}\fi

\end{document}